\documentclass[graybox]{svmult}

\usepackage{type1cm}        
\usepackage{makeidx}         
\usepackage{comment}
\usepackage{float}
\usepackage{subfig}
\usepackage[section]{placeins}
\usepackage{graphicx}        
\usepackage{multicol}        
\usepackage[bottom]{footmisc}

\usepackage{url}
\usepackage[colorlinks,allcolors=cyan!70!black]{hyperref}

\usepackage{siunitx}

\usepackage{comment}

\usepackage{algorithm}
\usepackage[noend]{algpseudocode}
\newcommand{\LongState}[1]{\State\hspace{-1.5em}\parbox[t]{\dimexpr\linewidth-\algorithmicindent}{\strut\hspace*{1.3em} #1\strut}}

\usepackage{newtxtext}       %
\usepackage[varvw]{newtxmath}       

\usepackage[capitalise]{cleveref} 

\Crefformat{equation}{Eq.~#2#1#3}
\Crefrangeformat{equation}{Eqs.~#3#1#4--#5#2#6}
\Crefmultiformat{equation}{Eqs.~#2#1#3}{ and~#2#1#3}{, #2#1#3}{ and~#2#1#3}
\Crefformat{figure}{Fig.~#2#1#3}
\Crefrangeformat{figure}{Figs.~#3#1#4--#5#2#6}
\Crefmultiformat{figure}{Figs.~#2#1#3}{ and~#2#1#3}{, #2#1#3}{ and~#2#1#3}
\Crefformat{section}{Sect.~#2#1#3}
\Crefrangeformat{section}{Sects.~#3#1#4--#5#2#6}
\Crefmultiformat{section}{Sects.~#2#1#3}{ and~#2#1#3}{, #2#1#3}{ and~#2#1#3}

\makeindex             

\begin{document}

\title*{ A geometry-dependent, force balance-driven model of \textit{Staphylococcus epidermidis} biofilm cell cluster detachment}

\titlerunning{A geometry-dependent, force-balance driven model of biofilm cell cluster detachment} 
\author{Yuehui Xu\orcidID{0009-0004-1004-9884}, Jasmine A.F. Kreig\orcidID{0000-0002-2976-0949}, Zhuoran Wang\orcidID{0000-0003-2331-5120}, Elizabeth J. Stewart\orcidID{0000-0003-4763-5743}, Rayanne A. Luke$^{\dag}$\orcidID{0000-0002-1636-8438}, and Sarah D. Olson$^{\dag}$\orcidID{0000-0003-3239-0579}}
\authorrunning{Y. Xu, J.A.F. Kreig, Z. Wang, E.J. Stewart, R.A. Luke, and S.D. Olson}

\institute{Yuehui Xu \at Indiana University, 420 University Blvd, Indianapolis, IN 46202, USA, \email{yx28@iu.edu}
\and Jasmine A. F. Kreig \at Los Alamos National Laboratory, Los Alamos, NM 87645, USA, \email{jkreig@lanl.gov}
\and Zhuoran Wang \at University of Kansas, 1460 Jayhawk Blvd, Lawrence, KS 66045, USA, \email{wangzr@ku.edu}
\and Elizabeth J. Stewart \at Worcester Polytechnic Institute, 100 Institute Rd, Worcester, MA 01609, USA, \email{ejstewart@wpi.edu}
\and Rayanne A. Luke$^{\dag}$ \at George Mason University, 4400 University Dr, Fairfax, VA 22030, USA, \email{rluke@gmu.edu}
\and Sarah D. Olson$^{\dag}$ \at Worcester Polytechnic Institute, 100 Institute Rd, Worcester, MA 01609, USA, \email{sdolson@wpi.edu}
\and $^{\dag}$RAL and SDO are co-corresponding authors for this work.
}
%
%

\maketitle

\abstract*{Each chapter should be preceded by an abstract (no more than 200 words) that summarizes the content. The abstract will appear \textit{online} at \url{www.SpringerLink.com} and be available with unrestricted access. This allows unregistered users to read the abstract as a teaser for the complete chapter.
Please use the 'starred' version of the \texttt{abstract} command for typesetting the text of the online abstracts (cf. source file of this chapter template \texttt{abstract}) and include them with the source files of your manuscript. Use the plain \texttt{abstract} command if the abstract is also to appear in the printed version of the book.}

\abstract{ 
Biofilms, bacteria cells surrounded by a self-produced polymeric matrix, are common on medical devices and lead to many hospital infections. The biofilm lifecycle includes disassembly and dispersion, where bacteria clusters detach from the biofilm, circulate in the bloodstream, and potentially colonize secondary infection sites. Existing models often simplify detachment to a function of biofilm thickness or extracellular polymeric substance (EPS) density, without tracking properties of detached clusters that impact their biological fate, including cluster size and morphology. Addressing this gap, our detachment model accounts for drag and adhesion in tagged sections of the biofilm determined by the cluster geometry and local arrangement of bacteria and EPS.  A stickiness parameter controls local EPS adhesion strength, which is modulated to disrupt (or compromise) EPS biomass. We specifically model the detachment of clusters from a \textit{Staphylococcus epidermidis} biofilm grown for 24 hours. Experimental data for biofilm microstructural features are utilized to benchmark the simulated biofilm, which is then subjected to different EPS disruption levels. We examine parameters that influence detached biofilm cell cluster frequency, size, and shape, providing mechanistic insights into how compromised EPS influences detachment dynamics. This integrated modeling framework is a significant advance in the predictive capabilities for biofilm detachment processes.
}

\section{Introduction}
\label{sec:intro}
Bacterial biofilms—surface adherent, structured communities of bacteria encapsulated by self-produced extracellular polymeric substances (EPS) \cite{Jonblat2024}—account for more than 65\% of healthcare-associated infections, approximately 80\% of chronic infections, and 60\% of all human bacterial infections \cite{Assefa}. Biofilms that form on or in medical devices such as catheters, central lines, or feeding tubes are a major cause of such persistent infections \cite{Otto2009}. The protective function of EPS, which can shield embedded bacteria from host immune responses \cite{Alhede20, Fleming01, Stewart20} and hinder antibiotic diffusion \cite{Stewart20}, is a factor that leads to biofilm infections being difficult to eradicate. In addition, biofilms can be up to 100 times less susceptible to antibiotics than planktonic bacteria (individual, free-floating cells) \cite{Alhede20,Stewart20}. 

The final phase of the biofilm lifecycle is disassembly and dispersion, where biofilm cells and cell clusters of bacteria with a subset of the surrounding EPS detach from the parent biofilm to colonize new sites \cite{Bjarnsholt13,Schilcher2020}. In the case of \textit{Staphylococcus epidermidis}, a nonmotile bacteria that frequently causes device-associated infections, the disassembly and dispersion of bacteria from the biofilm can contribute to systemic infections 
or sepsis \cite{Fleming18,WILLE20}. The size and shape of those detached bacterial clusters will impact their 
transport, propensity to attach to different surfaces, ability to evade immune clearance, and recolonization success at additional infection sites. 
For example, simulations of particles in a linear shear blood flow indicate that oblate particles have a higher probability of reaching the vessel wall, while prolate particles exhibit a greater tendency to adhere once contact is made \cite{Vahidkhah15}. In experiments, detached bacterial clusters with a diameter greater than 5 microns are more likely to evade immune clearance   \cite{Alhede20}, with increasing cluster sizes generally having a raised tolerance to antibiotics \cite{Franca16}.
Given this, targeted disruption of biofilms tailored for specific cluster sizes and shapes may offer a viable therapeutic strategy. A range of enzymes (e.g., glycoside hydrolases, proteases, and deoxyribonucleases (DNases)) have been shown to degrade different EPS components and promote detachment \cite{Jonblat2024,Erskine18,Okshevsky,Panlilio,Qin}. Despite these advances, predicting or controlling the properties of the resulting detached biofilm cell clusters  
remains a major challenge.

Numerous 
mathematical models of the biofilm lifecycle, including formation, structure, and growth dynamics \cite{Dzianach, Klapper, Mattei18, Picioreanu04, Wang10}, have been developed. Models such as BacLAB and iDynoMiCS (individual-based Dynamics of Microbial Communities Simulator) \cite{Lardon11,Cockx24} incorporate stochastic representations of bacteria and EPS, coupled with fluid dynamics and solute transport, to offer detailed insights into biofilm development. Very few models benchmark simulated biofilm development with single cell spatial positioning of bacteria since quantitative analysis of biofilm images with cellular level resolution has been limited. Exceptions include Hammond et. al \cite{Hammond14} where parameterization of the model utilized cellular spatial positions of biofilm cells from experiments and a recent agent based model by Li et. al that considered cell ordering \cite{Li24}. Only a small number of models focus specifically on biofilm detachment, further limiting inquiry. Existing frameworks often treat detachment as a simplified function of biofilm thickness, shear stress, or EPS density. For example, Chambless and Stewart modeled detachment probabilities as inversely proportional to biofilm height or local cell density \cite{Chambless}, while Lardon et al. incorporated EPS integrity into erosion terms \cite{Lardon11}. Xavier et al. later extended this approach to include enzymatic degradation of EPS by using iDynoMiCS to model how compromised EPS  (EPS altered by detachment-promoting agents)  affects detachment 
in mature biofilms \cite{xavier2005biofilm,Xavier05,xavier2005framework}. These models focused primarily on long-term biofilm maturation (e.g., 60-day growth to a height of $\SI{500}{\micro\meter}$) and none tracked the location of detachment or geometry of the clusters. 

Other modeling efforts have considered mechanical aspects of detachment. Sudarsan et al. \cite{sudarsan2016simulating} and Hammond et al. \cite{Hammond14} treated biofilms as a spring network and modeled detachment via stress thresholds. Although not specifically developed for biofilms, 
Ting et al. used computational fluid dynamics (CFD) to study the mechanical aspects of detachment of non-spherical particles from a substrate \cite{ting2021impact,ting2022image,ting2023detachment}. This work identified key geometric factors that affect particulate detachment from surfaces, including the aspect ratio and orientation angle, which offers a useful theoretical foundation for studying how cellular aggregates detach from biofilms with different shapes, sizes, and orientations. 
Despite this, there are no integrated models that can predict the detachment of differently shaped clusters from 
biofilms. 

\begin{figure}[h]
\centering
\includegraphics[width=0.99\linewidth]{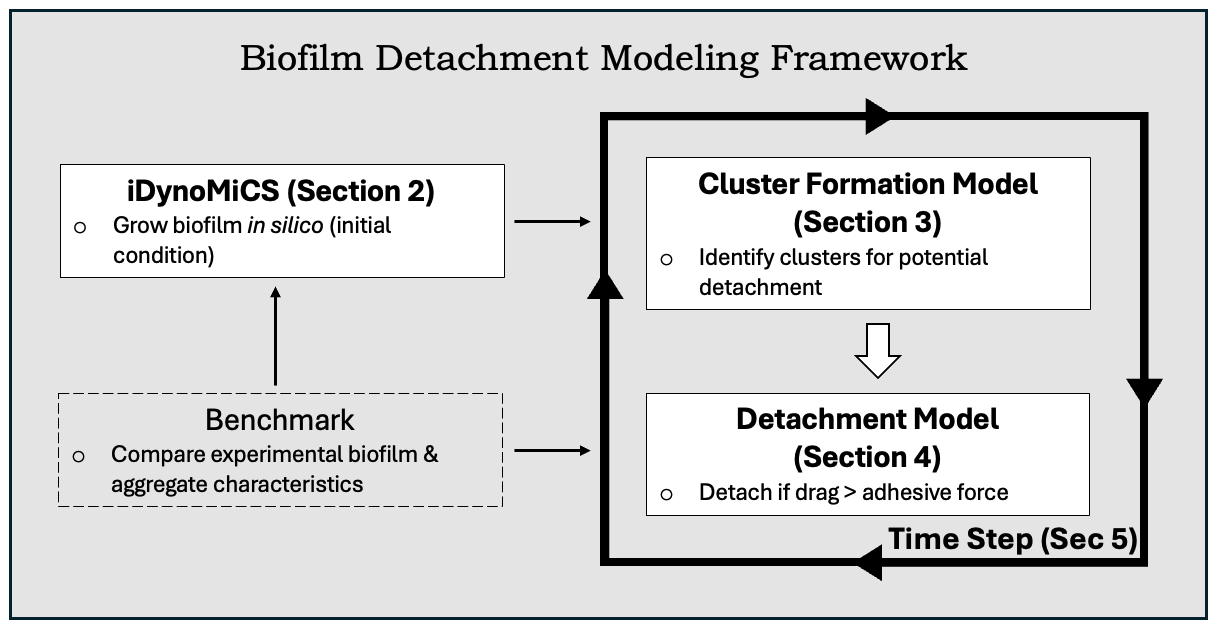}
\caption{Overview of biofilm detachment modeling framework. A \textit{S. epidermidis} biofilm, configured utilizing experimental data for cellular organization and cell density, is created in iDynoMiCS. Our cluster formation and detachment models are run in sequence in each time step (one minute increments), resulting in changes in   the local biofilm topography over the course of one hour.} 
\label{fig:framework}       
\end{figure}

To address this gap, we develop 
equations and an algorithm for the disassembly and break off of cellular aggregates from biofilms that 
exists within a larger model framework (illustrated in Fig. \ref{fig:framework}). We first employ the iDynoMiCS platform to grow a simulated \textit{S. epidermidis} biofilm that is parameterized with experimental data. This process is discussed in Section \ref{sec:idynomics}. The simulated biofilm then acts as the initial condition for our cluster formation model (Section \ref{sec:cluster_model}) and detachment model (Section \ref{sec:detach_model}). 
Distinctive features of our modeling approach include the linking of the relative biomass and spatial positions of the EPS and bacteria with the degree of EPS disruption via a stickiness parameter, and utilization of geometric factors to identify which groups of cells would potentially detach together.   Final detachment of a cluster is governed by a balance between drag and adhesive moments, where the latter utilizes an interaction potential that captures the weakening of cluster-bacteria and cluster-EPS interactions in a region around the cluster. We 
simulate cluster detachment over one hour (Section \ref{sec:results}); in each time step, bacteria and EPS agents are grouped together in the clustering algorithm and a subset are identified to detach. As a result, the local topography and neighborhoods of EPS and bacteria change at the top of the biofilm with each time step.  
To simulate different experimental conditions and disruption strategies, we vary iDynoMiCS parameter combinations and EPS disruptor levels across both 2-d and 3-d biofilms.  The influence of model parameters on frequency, size, and
shape of detached clusters are studied throughout Section \ref{sec:results}. We compare our model results to experiments  in Section \ref{sec:discussion}, along with a discussion of model limitations and extensions.

\section{Biofilm Growth Model}
\label{sec:idynomics}
We utilize iDynoMiCs \cite{Lardon11},  an open-source software developed under the  General Public License (GPL)-like CEA CNRS INRIA Logiciel Libre (CeCILL) license, to develop a representative  24-hour-old \textit{S. epidermidis} biofilm. The details of iDynoMiCs have been published previously \cite{Lardon11,Cockx24} so we only provide a brief summary of our setup here. We then discuss parameter choices and benchmark our \textit{in silico} biofilms with reported experimental ranges of \textit{S. epidermidis} biofilm height, cellular counts, cellular arrangements, and EPS content after 24 hours of growth.

\subsection{Overview of iDynoMiCS}
The agents are the immobile spherical \textit{S. epidermidis} bacteria ($B$) and the EPS ($E$). We assume that there is a single population of \textit{S. epidermidis} bacteria, which are heterotrophs,  i.e., they cannot produce their own food and therefore must obtain nutrients from their environment. The age, biomass, and location  
of each bacteria agent, $i=1,\ldots,N_B^t$, at each time $t$ is tracked. When cells reach a specified radius (set by the parameter \texttt{divRadiusB}), the biomass is divided equally and a shoving algorithm based on the local pressure and biomass determines the location of the two new cells 
(close to the original cell location). Even though EPS is a very complex structure with different sub-components \cite{Karygianni20}, here it is only tracked in two ways. Bacteria agents produce EPS which is first added to the capsule around the bacteria cell, accumulating up to a specified volume fraction of the bacteria agent $B$ (defined by the parameter  \texttt{epsMax}). Additional EPS produced by the $i-th$ bacteria agent is then distributed evenly to nearby EPS agents. If there are no EPS agents near a bacteria cell,  one is created immediately adjacent to the bacteria. Once an EPS agent reaches a critical radius (set by the parameter \texttt{divRadiusE}), it is equally divided into two EPS agents with location adjacent to the existing EPS determined via the shoving algorithm.  As with the bacteria, the radius, biomass, and location of each EPS agent $E$ is also tracked. 

With iDynoMiCS, we simulate 2-d or 3-d biofilms by initializing $N_B^0$ bacteria cells with random locations at the bottom of the domain (attached to the substratum with a no flux boundary condition). On the sides of the domain we have periodic boundary conditions, which is reasonable when representing a small section of a larger biofilm domain. 
The agents are solved at discrete time steps, but the growth of bacteria cells, cell division, and EPS production are all coupled to solute concentrations \cite{Lardon11,Jo22}, which are continuous variables. These solutes include oxygen and carbon (expressed as chemical oxygen demand or COD). In experiments, biofilms are often grown in a flow cell where nutrient-rich growth medium is flowing at a slow, constant velocity \cite{packard2026biophysical}. In iDynoMiCS, the solute concentrations are dynamic and satisfy a reaction-diffusion equation with appropriate sinks and sources in the diffusive layer (region throughout the biofilm plus a boundary layer above the biofilm set by the parameter \texttt{boundaryLayer}). The area above the top of the boundary layer is considered a bulk layer with constant concentration, representing a steady flow of nutrient-rich growth medium well above the biofilm. The growth of bacteria agents and the production of EPS are coupled via Monod kinetics to the concentration of solutes, which then reduces the solute concentration locally within the biofilm. EPS hydrolysis, or breakdown, is assumed to occur at a constant rate (first order kinetics determined by the parameter \texttt{muMax}). Additionally, at each time step there is erosion where very small regions at the top of the biofilm break off, quadratically related to the height of the biofilm and scaled with parameter \texttt{kDet}. An overview of the processes executed at each time step is given in  Algorithm \ref{Alg:iDyn}.

\begin{programcode}{Alg. 2.1: Overview of time step iterations in iDynoMiCS}
\begin{algorithmic}
 \State \textbf{Environment submodels:}
  \State \quad Determine local sinks/sources based on location of agents in the biofilm.
  \State \quad Solve mass balance equations (reaction-diffusion) in the biofilm and boundary layer to determine new solute concentrations. 
  
\State \textbf{Agent submodels:}
\While{agent time step $<$ global time step}
 \For{each agent}
    \LongState{Compute bacteria agent growth, division, and death, which are coupled to local solute concentrations.}
    \LongState{Update bacteria agent locations via the shoving algorithm (in the case of collisions or bacteria cell division).}
    \LongState{Compute EPS secreted by each bacteria agent, which are coupled to local solute concentrations.} 
    \LongState{Determine portion of EPS secreted that goes to bacteria capsule and individual EPS agents, tracking growth of biomass and radius.}
    \LongState{Compute EPS agent growth, division, and death.}
    \LongState{Update EPS agent locations via the shoving algorithm (in the case of collisions or EPS divisions).}
    \LongState{Apply erosion to the top of the biofilm.}
    \EndFor
  \EndWhile
\end{algorithmic}
\label{Alg:iDyn}
\end{programcode}

\subsection{Parameters in iDynoMiCS}
\label{sec:iDynoParam}
To set the parameters for our simulated \textit{S. epidermidis} biofilm, we first start with the standard parameter set given in the iDynoMiCS tutorial \cite{Lardon11}. 
The resolution of the grids for the agents and solutes are updated to be in line with the radius of the bacteria agent. \textit{S. epidermidis} biofilm growth is heterogeneous,  with average heights of $\sim$8-200 $\SI{}{\micro\meter}$ after 24-hours of growth \cite{Jonblat2024,packard2026biophysical,Cerca2012,Stewart13}. Thus, we choose to only simulate a domain within the lower end of the reported height range with a  maximum height of $\SI{33}{\micro\meter}$. The diffusive layer includes the biofilm region and a boundary layer $\SI{6}{\micro\meter}$ above the biofilm, with a steady (well-mixed) bulk layer above this boundary layer.  
A summary of all parameters in the iDynoMiCS model are  in Table \ref{tab:iDynomics_parameters}-\ref{tab:iDynomics_parameters_2}. We note that we did not  vary  many of the parameters and reactions in iDynoMiCS as initial parameters resulted in biofilms with characteristics representative of experimentally grown \textit{S. epidermidis} biofilms. 

Biofilm height and EPS composition exhibit a high level of heterogeneity \cite{Jonblat2024,Karygianni20,Stewart17}, which we aim to capture in our \textit{in silico} biofilms. With that in mind, we create four Parameter Sets (summarized in Table \ref{tab:iDynomics_param_sets}) that alter two iDynoMiCs parameters to create distinct biofilms. To have biofilms of different heights, we modify \texttt{kDet}, which controls the rate of erosion at the top of the biofilm. We note that increased erosion can mimic experimental conditions with a higher flow rate of the medium above the biofilm \cite{Wang22}. We also vary \texttt{YBeps}, which determines EPS production by bacteria. Higher values of this parameter correspond to higher EPS biomass, leading to increased biofilm height and tighter adherence among agents in the biofilm. 

\begin{table}[h]
\begin{tabular}{p{1.5cm}p{2.3cm}p{2cm}p{4.4cm}}
\hline\noalign{\smallskip}
Parameter Set & \texttt{kDet} (1/($\SI{}{\micro\meter}$ h)) & \texttt{YBeps} (unitless) &  Physical Scenario \\
\noalign{\smallskip}\svhline\noalign{\smallskip}
1 & $1 \times 10^{-4}$ & 1 & Low erosion \& EPS yield \\
2 & $1 \times 10^{-3}$ & 1 & Moderate erosion \& low EPS yield\\
3 & $1 \times 10^{-3}$ & 1.3 & Moderate erosion \& EPS yield\\
4 & $1 \times 10^{-3}$ & 1.5 & Moderate erosion \& high EPS yield\\
\noalign{\smallskip}\hline\noalign{\smallskip}
\end{tabular}
    \caption{Parameter Sets used in iDynoMiCS (2-d and 3-d). Here, \texttt{kDet} is the detachment coefficient for agents on the top of the biofilm and \texttt{YBeps} sets the yield of EPS from bacteria agents in the biofilm, based on Monod kinetics that are coupled to the local solute concentrations. Lower erosion creates taller biofilms whereas increased EPS biomass leads to more adherent agents in taller biofilms.}
    \label{tab:iDynomics_param_sets}
\end{table}

\begin{figure}
    \centering
    \includegraphics[width=0.99\linewidth]{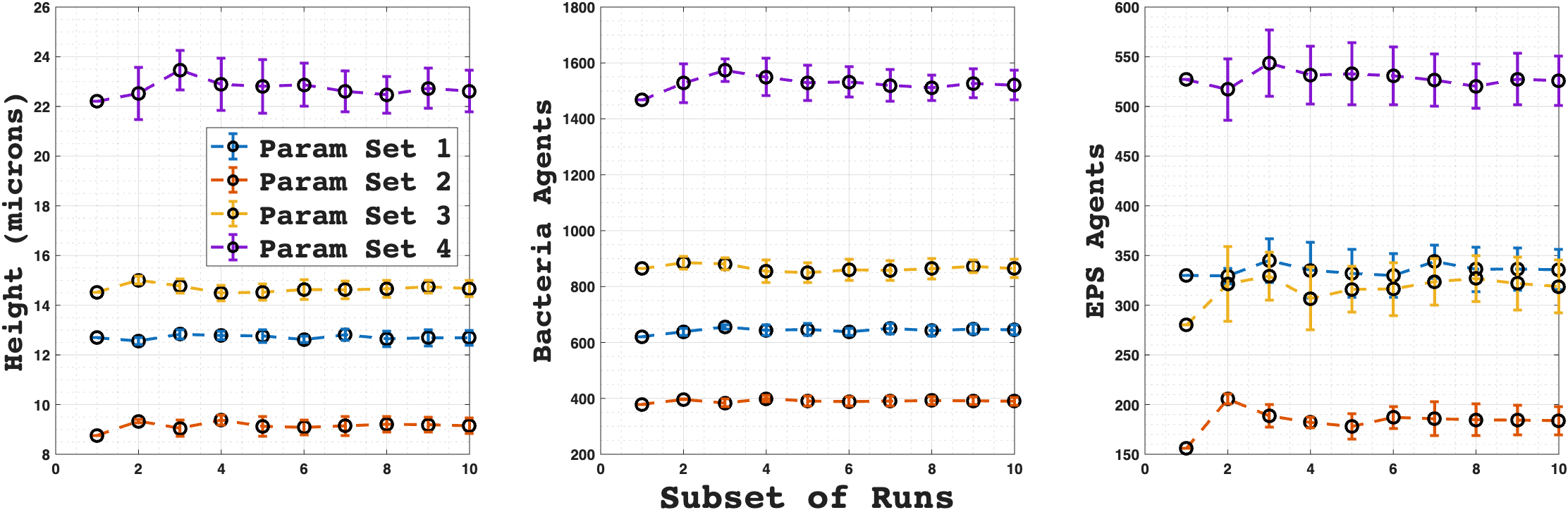}\\
    (a) Average biofilm height \hspace{1cm}(b) Average bacteria agents\hspace{1cm} (c) Average EPS agents \\
    \caption{iDynoMiCS results for a 24-hour \textit{S. epidermidis} biofilm in 2-d utilizing 4 different Parameter Sets (Table \ref{tab:iDynomics_param_sets}). (a) Biofilm height (average over the highest bacteria agents in 1 $\SI{}{\micro\meter}$ intervals), (b) number of bacteria agents, and (c) number of EPS agents for a subset of simulations is shown. The subset of runs corresponds to randomly choosing 1 to 10 runs (stochastic simulations). The average and standard deviation are shown with respect to the runs.}
    \label{fig:RunSetAvgStd}
\end{figure}

There is stochasticity in the iDynoMiCS simulations (\texttt{CV} parameters in Table \ref{tab:iDynomics_parameters}) for time to agent division and mass in each newly divided agent. For example, a bacteria agent $B$ will divide when its radius is 0.65+0.2$\mathcal{R}_B$ where $\mathcal{R}_B$ is drawn from a uniform distribution on $[0,1]$. Here, the random seed is set by a random.State protocol file in iDynoMiCS. Due to this stochasticity, there is aleatory uncertainty \cite{Alden13} in how time of cell division and other factors will impact total cell counts and biofilm height. We define a single stochastic simulation with fixed parameters in iDynoMiCS as a run. As shown in Fig.~\ref{fig:RunSetAvgStd}, the average biofilm height, number of bacteria agents, and number of EPS agents changed only by a small percentage after 5 runs. Even though these metrics stabilize after 5 runs, the individual locations of agents, size of agents, and biomass of agents does vary greatly due to the stochasticity in timing of cell divisions. As a result, we chose to utilize 10 runs (Run $=1,\dots,10$) for a given parameter set in 2-d to capture the full variation (and 5 runs in 3-d due to increased computational time). 

\subsection{\textit{In Silico} Biofilm Properties}
\begin{figure}[h]
    \centering
    \subfloat[][Average biofilm height]{\includegraphics[width=0.45\linewidth]{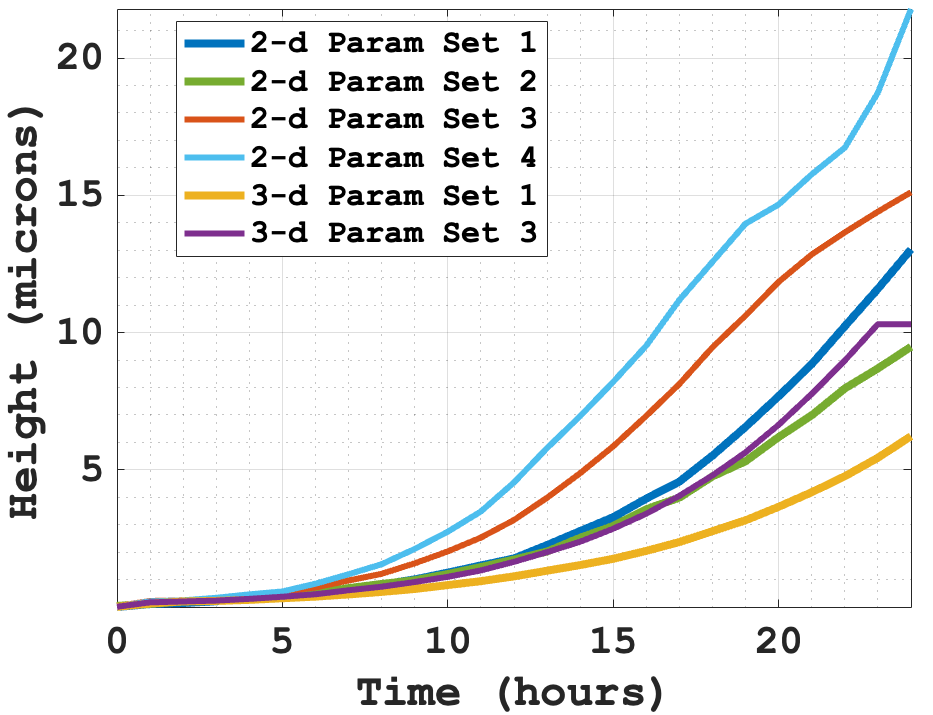}}
    \subfloat[][3-d biofilm height at 24 hours]{\includegraphics[width=0.45\linewidth]{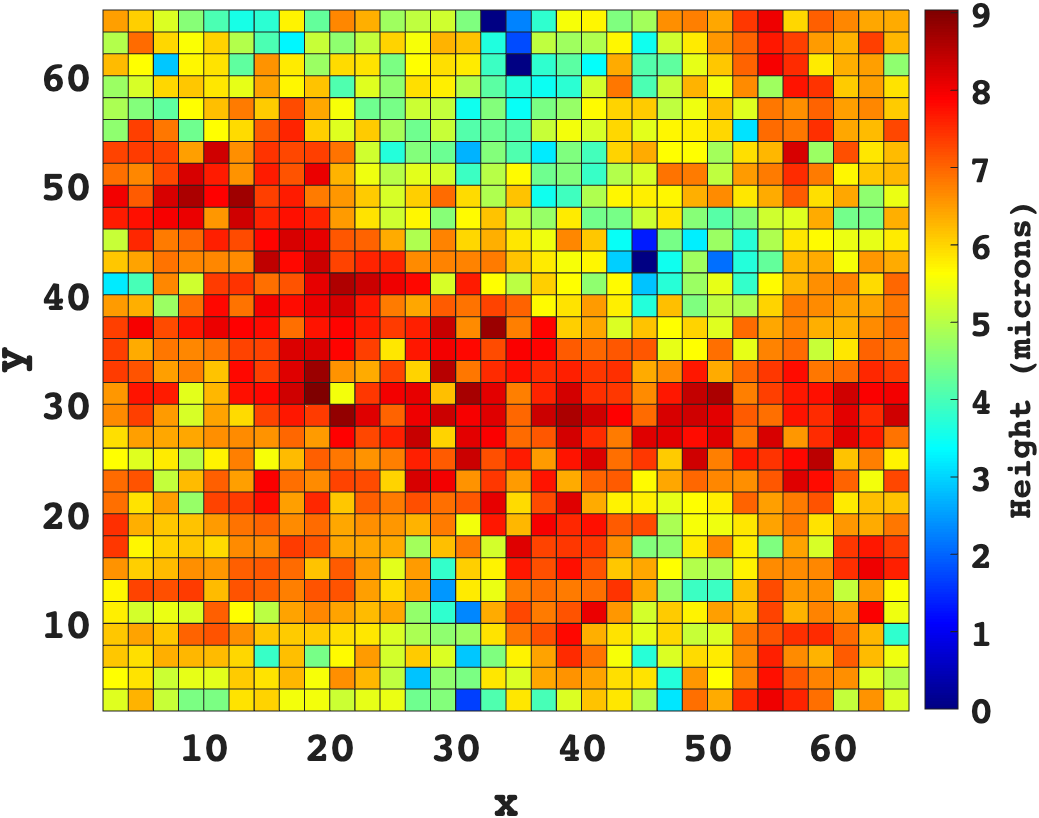}}\\
    \subfloat[][Cell density over time]{\includegraphics[width=0.45\linewidth]{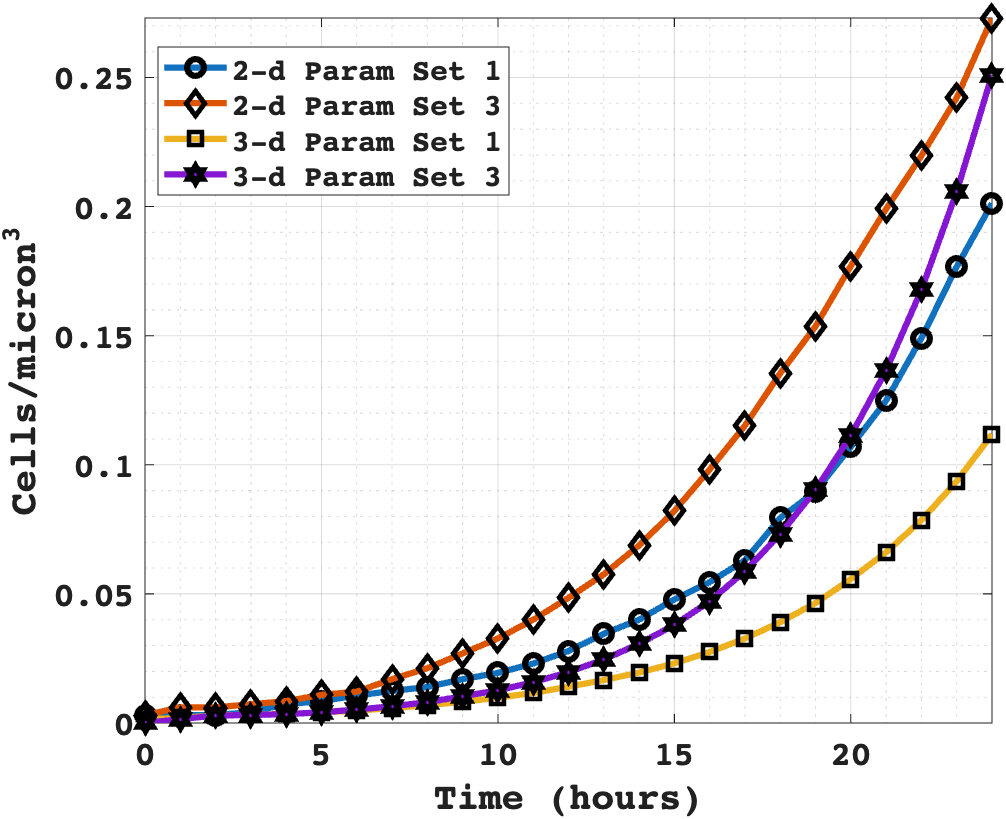}}
    \subfloat[][Bacteria clusters : total bacteria]{\includegraphics[width=0.45\linewidth]{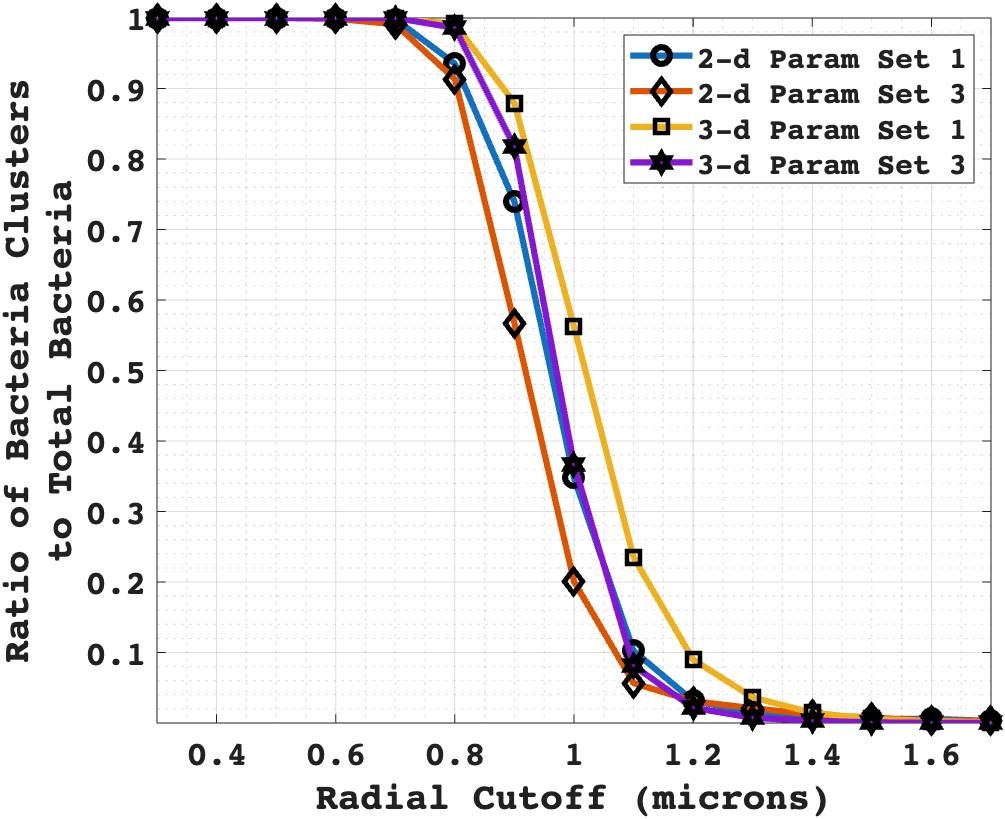}}
    \caption{24 hours of \textit{S. epidermidis} biofilm growth that are simulated by iDynoMiCS in a single run (Run=1). (a) Average biofilm height over 1 $\SI{}{\micro\meter}$ intervals for 2-d and 4 $\SI{}{\micro\meter^2}$ for 3-d.   (b) Surface plot showing the height of the 3-d biofilim at 24 hours utilizing Parameter Set 1.  (c) Density of cells over time (with equivalent initial cell densities in 2-d and 3-d). (d) Quantifying cellular organization by the ratio of bacteria clusters to total bacteria as a function of different radial cutoffs in 2-d and 3-d.}
    \label{fig:iDynoValidate}
\end{figure}

We created four different \textit{in silico} biofilms and we now examine their attributes to ensure they fall within reported experimental ranges collected from reports of \textit{S. epidermidis} biofilms grown for 24-hours across three strains (RP62A \cite{packard2026biophysical,Stewart13,Ganesan13}, CIP 444 \cite{Jonblat2024}, 9142 \cite{Cerca2012}) with varying growth conditions.    
We examine biofilm height, cellular number density, and cellular organization over time for different parameter sets in both 2-d and 3-d (summarized in Fig. \ref{fig:iDynoValidate}). We also 
benchmark EPS biomass across \textit{in silico} and experimental biofilms.

We first look at the average biofilm height over time (Fig.\ref{fig:iDynoValidate}a), calculated as the average over the highest bacteria agents in 1 $\SI{}{\micro\meter}$ intervals for 2-d (or 4 $\SI{}{\micro\meter^2}$ for 3-d).  
At 24 hours, 2-d heights range from 9.48-21.8 $\SI{}{\micro\meter}$,
which aligns with some experimental reports of average \textit{S. epidermidis} heights that ranged from $\sim$8-52 $\SI{}{\micro\meter}$ \cite{Jonblat2024,packard2026biophysical,Cerca2012}. We note that biofilm heights can reach 200 $\SI{}{\micro\meter}$ \cite{Stewart13}, but we are focusing on shorter biofilms or shallower regions in a biofilm. 
Due to the additional computational complexity for runs in 3-d, we focus on Parameter Sets 1 and 3 for this analysis. For Parameter Set 1, the 3-d version has an average height of 6.23 $\SI{}{\micro\meter}$ whereas the 2-d biofilm with the same parameters has a height of 13.03 $\SI{}{\micro\meter}$. Differences between 2-d and 3-d are expected; 2-d biofilm simulations have been shown to grow more quickly in terms of total biomass per unit surface area \cite{Picioreanu04,Clegg17}. We observe that Parameter Set 1 has a lower average biofilm height at 24 hours in comparison to Parameter Set 3; this is consistent across runs in 2-d and 3-d. Of course, these values are averages. Biofilms are incredibly heterogeneous; an illustrative example of the topography is shown in Fig. \ref{fig:iDynoValidate}b. This figure shows the maximum height of the 3-d biofilm in each of the 4 $\SI{}{\micro\meter^2}$ intervals at 24 hours, clearly displaying the variation and roughness of the biofilm.  

To understand the microstructure of the biofilm, we next examine cell density. Fig.~\ref{fig:iDynoValidate}c displays cell densities across time for Parameter Sets 1 and 3 for both 2-d and 3-d. Our simulations have a range of 0.05-0.3 cells/$\SI{}{\micro\meter^3}$ in the time range of 20-24 hours. Experimentally, \textit{S. epidermidis} has been reported to have an average local number density of 0.19$\pm$0.03 cells/$\SI{}{\micro\meter^3}$  with  densities ranging from 0.02 - 0.41 cells/$\SI{}{\micro\meter^3}$  \cite{Stewart13,Ganesan13}; thus, our model is in alignment with experimental cellular number densities. 
 
Next, we analyze cellular organization. To this end, we utilize a density-based clustering algorithm (DBSCAN) in MATLAB \cite{dbscan} to group the bacteria agents into clusters. We have two parameters for DBSCAN: the neighborhood search radius threshold, $\epsilon$, and the minimum number of agents to form a cluster, $n_\text{min}$. If an entity has at least $n_\text{min}$ neighbors, then it will be considered a core point; if an agent has fewer than $n_\text{min}$ neighbors, then it will be considered a border point. We set $n_{min}=1$ and explore $\epsilon$ in the range of 0.3 to 1.7 $\SI{}{\micro\meter}$, tracking the ratio of the number of bacteria clusters identified to the total number of bacteria. This is shown in Fig.~\ref{fig:iDynoValidate}d; smaller radial cutoffs $\epsilon$ correspond to clusters of a single bacteria by itself, so the ratio tends to 1. Analogously, for larger search radii $\epsilon$, there are generally other bacteria in that radius and thus the algorithm determines a small number of massive clusters, so the ratio tends to 0. Our bacterial cellular clustering results are similar to those of \textit{S. epidermidis} biofilms grown experimentally for 24 hours in a flow cell, where the ratio of bacteria clusters to total bacteria converged to 0 with a radial cutoff greater than 1.5 $\SI{}{\micro\meter}$ \cite{Stewart13}.

Finally, we compare simulated EPS biomass. 
Experimental biofilms grown to 24 hours with \textit{S. epidermidis}   reported that   $\sim$52\% of the biofilm was b-polysaccharides (biofilm-associated exopolysaccharides)   \cite{Jonblat2024}. Polysaccharide intercelluar adhesin, PIA, is one of the main components of the b-polysaccharides and a different study measured $\sim1.56\times10^4$ molecules of PIA per bacteria cell (in a region of $\sim$5.1 $\SI{}{\micro\meter^3}$) for a 24 hour \textit{S. epidermidis} biofilm   \cite{Ganesan13}. From our simulated biofilm in iDynoMiCS, we have EPS biomass ($\SI{}{fg/\micro\meter^3}$) of EPS agents and capsule EPS biomass ($\SI{}{fg/\micro\meter^3}$) associated with the bacteria agents. Averaging EPS biomass in a region of 5.1 $\SI{}{\micro\meter^3}$ around each bacteria (circular region in 2-d and spherical region in 3-d), 
we arrive at simulated values of 4.60$\times10^4$ molecules in 2-d and 2.00$\times10^4$ molecules in 3-d for Parameter Set 1 and Run=1. Based on this, we determine that our simulated EPS biomass per cell is on the same order of magnitude as experimental data, noting that PIA is only a percentage of the total EPS biomass.

\section{Cluster Formation Model}
\label{sec:cluster_model}
To identify agents or a cluster of agents that are likely to detach from the biofilm in an aggregate, we consider specific mechanisms of biofilm detachment and explicitly model the adhesion strength provided by EPS through a stickiness parameter. 
At the top of the biofilm layer, fluid flow at and above the biofilm  creates a shear stress that leads to 
 erosion, the continuous removal of small clusters \cite{Petrova}, 
 or less common sloughing events, where  large portions of the biofilm detach \cite{Kaplan}. Existing models typically correlate erosion and sloughing with biofilm height \cite{Xavier05}; we also adopt this approach. EPS disruption can be induced by increased production of matrix-degrading enzymes triggered by an environmental change (e.g. pH) \cite{Fleming01,Petrova} or human interventions, such as 
 application of EPS disruptors that target specific components of the EPS \cite{packard2026biophysical,Wang23}. We account for detachment due to EPS disruption by allowing EPS biomass to exist in one of two states: healthy or compromised. Healthy EPS biomass ($m^H$) is stickier, equating to a higher local adhesion strength. Compromised EPS biomass ($m^{H*}$) arises from the application of a matrix disruptor or other environmental factors which locally decrease the attachment strength of EPS. 

We model a generic scenario in which there is a constant level of compromised EPS throughout the biofilm. This could represent a disruptor that is applied and rapidly diffuses throughout the entire biofilm or a constant level of increased matrix-degrading enzymes throughout the biofilm.  
The EPS biomass in each EPS and bacteria agent (from the capsule) at 24 hours, denoted $m_E^{24}$ and $m_B^{24}$, respectively, are obtained from our \textit{in silico} biofilm.  The total biomass of each agent $i$ is  divided into healthy EPS biomass, $m_{A_i}^{\text{H}}$, and compromised EPS biomass, $m_{A_i}^{\text{H*}}$, as 
\begin{equation}
    m_{A_i} =m_{A_i}^{\text{H}} + m_{A_i}^{\text{H*}}=(1-r_\text{H*})m_{A_i}^{24} + r_\text{H*}m_{A_i}^{24},
\end{equation}
for agents $A=B,E$. Here, $r_\text{H*}=m^\text{H*}_{A_i}/m_{A_i}^{24}$ is the fixed disruption level of the EPS ($0\leq r_\text{H*}\leq 1$) corresponding to the ratio of an agent's compromised EPS biomass to the total EPS biomass.

\subsection{EPS Attachment Strength and Tagging Probability}
We define the local EPS attachment strength, a nondimensional stickiness factor of the $i-th$ agent $A=B,E$  as 
\begin{equation}
\label{eq:attachment_strength}
    \beta_{A_i} =
    \left(1-r_\text{H*}\right)
    \left( \frac{m_{A_i}^{\text{H}}}{\max(m_{A}^{\text{H}})}+ \frac{n^{E}_{A_i}}{\max(n^{E}_{A})}\right).
\end{equation}
The first term of the product accounts for the relative fraction of healthy biomass. The limit $r_\text{H*}\to 1$ corresponds to completely compromised EPS and therefore $\beta\to 0$ (no stickiness). The second term of the product is a sum of two ratios. The first quotient compares the healthy EPS biomass of the agent to the maximum amount of healthy EPS biomass over all agents currently in the biofilm ($\max(m^{H}_{A})$). The second fraction operates similarly, focusing on the number of EPS neighbors ($n^{E}_{A_i}$) within a specified region ($d_{\beta}$ of $A_i$). Here, $0\leq\beta\leq 2$, and stickiness decreases when the local healthy biomass decreases (i.e. the amount of compromised EPS increases) or when the number of nearby EPS agents decreases.  

Prior to identifying clusters of agents that may detach, we first calculate the probability of removal for each agent currently in the biofilm. This probability encompasses three concepts. To capture 
erosion, this probability depends quadratically on the local height $h_{A_i}$ of the $i-th$ agent $A$ in the biofilm as measured relative to the substrate. Neighbor density influences the probability because densely packed agents that have many bacteria agents are less likely to detach. Finally, local attachment strength accounting for compromised EPS biomass, $\beta_{A_i}$ (Eq.~\ref{eq:attachment_strength}), also impacts the probability of detachment. Accounting for each idea, we calculate the probability of an EPS agent detaching as
\begin{equation}
P_{E_i} = \left(\frac{h_{E_i}}{\max(h_{E})}\right)^2\left(1-\frac{n_{E_i}^{B}}{\max(n^{B}_E)}\right)\exp\left(-\beta_{E_i}\right),
\label{eq:prob_eps}
\end{equation}
where $n^{B}_{E_i}$ is the number of bacterial agents within $d_\beta$ of the $i-th$ EPS agent $E_i$ and the maximum values in the denominators are over all EPS agents in the biofilm. We model the probability of a bacteria agent detaching in a similar manner: 
 \begin{equation}
P_{B_i} = \left(\frac{h_{B_i}}{\max(h_{B})}\right)^2\left(1-\frac{n_{B_i}^{B}}{\max(n^{B}_B)}\right)\exp\left(-\beta_{B_i}\right)\left(\frac{\min(\rho_{B})}{\rho_{B_i}}\right),
\label{eq:prob_bact}
\end{equation}
where $n_{B_i}^B$ is the number of bacterial agents in the neighboring area of agent $B_i$. Here, minimum and maximum values are being taken over all bacteria agents currently in the biofilm.
 In Eq. \eqref{eq:prob_bact} we also account for the bacteria agent's density, $\rho_{B_i}$, since each bacteria agent is surrounded by an EPS capsule.

\subsection{Cluster Formation Algorithm}
To identify agents that are likely to detach, we craft an algorithm (Alg. \ref{Alg:tag_add_screen}) that operates in four phases: \textit{Tagging}, \textit{Adding}, \textit{Screening}, and \textit{Clustering}. Agents are first selected during the \textit{Tagging} phase if their calculated detachment probabilities (Eqs.~\ref{eq:prob_eps}-\ref{eq:prob_bact}) for EPS and bacteria, exceed thresholds $p_E$ and $p_B$, respectively. We let $T_{A}$ be the subset of all agents $A=B,E$ that are tagged based on initial probability calculations. 

During the \textit{Adding} phase, we determine the total number of tagged agents (bacteria and EPS) that are in a radial region $d_n$ of each agent that is currently not tagged. That is, we calculate 
$N_{A_i}^\mathrm{add}=N_{A_i}^{B,\mathrm{add}}+N_{A_i}^{E,\mathrm{add}}$ for all agents ($A_i\in A\setminus T_A$). Given the thresholds, $n_{A}^\mathrm{add}$, selected agents $A_i$ are added to the tagged set $T_A$ only when $N^\mathrm{add}_{A_i}\geq n^\mathrm{add}_A$. 

We further refine the set of tagged agents in the \textit{Screening} phase by removing entities from $T_A$ that are unlikely to detach or form a cluster. Agents located on the surface of the biofilm have potential to detach as single agents whereas we assume those deep within the biofilm will not detach. We use alphaShape in MATLAB \cite{alphashape} to determine the biofilm boundary which is constrained to be within \SI{3}{\micro\meter} of the maximum height of current agents. Agents that lie within this boundary are retained in the tagged set \( T_A \). 
For each agent $A_i$ that remains  tagged,
\( A_i \in T_A \), we determine the number of tagged bacterial and tagged EPS agents within the screening neighborhood ($d_n$), denoted by \( N_{A_i}^{B,\mathrm{scr}} \) and \( N_{A_i}^{E,\mathrm{scr}} \), respectively. A tagged bacterial agent \( B_i \) is removed from \( T_B \) if \( N_{B_i}^{B,\mathrm{scr}} \leq n^{B,\mathrm{scr}}_B \) or \( N_{B_i}^{E,\mathrm{scr}} \leq n^{E,\mathrm{scr}}_B \). Similarly, a tagged EPS agent \( E_i \) is removed from \( T_E \) if \( N_{E_i}^{E,\mathrm{scr}} \leq n^{E,\mathrm{scr}}_E \) or \( N_{E_i}^{B,\mathrm{scr}} \leq n^{B,\mathrm{scr}}_E \). Here, \( n^{B,\mathrm{scr}}_B \), \( n^{E,\mathrm{scr}}_B \), \( n^{B,\mathrm{scr}}_E \), and \( n^{E,\mathrm{scr}}_E \) denote threshold values for the number of neighboring tagged agents used in the \textit{Screening} process. 

Finally, in the \textit{Clustering} phase, the geometric arrangement of the bacteria and EPS determines the clusters that are likely to detach together (calculated with \\ DBSCAN in MATLAB \cite{dbscan}). We use the eigenvalues and eigenvectors of the covariance matrix of agent locations to find the major and minor axes of an ellipse fitted to that cluster in 2-d (or axes of an ellipsoid in 3-d). The DBSCAN algorithm treats agents that are tagged but have no neighbors as noise, instead we consider them to be representative of a single-agent break off and retain them for further analysis. 

\begin{programcode}{Alg. 3.2: Cluster Formation Algorithm\label{Alg:tag_add_screen}}
\begin{algorithmic}
    \State \textbf{\textit{Tagging}}
    \For{each agent \( A_i \)}
        \State Compute detachment probability \( P_{A_i} \).
        \If{\( P_{A_i} > p_A \)}
            \State Add \( A_i \) to the tagged agent set \( T_A \).
        \EndIf
    \EndFor

    \State \textbf{\textit{Adding}}
    \For{each untagged agent \( A_i \in A \setminus T_A \)}
        \State Compute the number of neighboring tagged agents, denoted by \( N_{A_i}^{\mathrm{add}} \).
        \If{\( N_{A_i}^{\mathrm{add}} \geq n_A^{\mathrm{add}} \)}
            \State Add \( A_i \) to the tagged agent set \( T_A \).
        \EndIf
    \EndFor

    \State \textbf{\textit{Screening}}
    \For{each tagged agent \( A_i \in T_A \)}
        \If{\( A_i \) is located on the top surface of the biofilm}
            \State Keep $A_i$ in the tagged agent set $T_A$.
        \Else
            \State Compute the number of neighboring tagged bacterial agents \( N_{A_i}^{B,\mathrm{scr}} \).
            \State Compute the number of neighboring tagged EPS agents \( N_{A_i}^{E,\mathrm{scr}} \).
            \If{\( N_{A_i}^{B,\mathrm{scr}} \leq n_A^{B,\mathrm{scr}} \) or \(N_{A_i}^{E,\mathrm{scr}} \leq n_A^{E,\mathrm{scr}} \)}
                \State Remove \( A_i \) from the tagged agent set \( T_A \).
            \Else
                \State Keep $A_i$ in the tagged agent set $T_A$.
            \EndIf
        \EndIf
    \EndFor

    \State \textbf{\textit{Clustering}}
    \State Apply DBSCAN to all tagged agents \( T_B \cup T_E \) to identify clusters.
    \For{each identified cluster}
        \LongState{Compute the covariance matrix of agent locations and use its eigenvalues and eigenvectors to determine an ellipse (2-d) or ellipsoid (3-d) representing the cluster shape.}
    \EndFor
    \State Label agents not assigned to any cluster as single-agent break offs.
\end{algorithmic}
\end{programcode}

\subsection{Cluster Formation Parameters and Example}
In the following examples, we use the parameter values summarized in Table \ref{tab:cluster_form_params} in our cluster formation model, but first, we discuss specific parameter value choices here. Based on the agent size from iDynoMiCS (refer to Table \ref{tab:iDynomics_parameters}), we use \SI{1.5}{\micro\meter} as the radius of the agent neighborhood in our model, which applies to $d_\beta$, $d_n$, and $\epsilon$. 
To account for inherent stochasticity, we introduce noise ($\mathcal{R}_\tau$) into the probability thresholds using independent random number generators in MATLAB. Specifically, for a given seed \( s \), the random number generator is initialized as \texttt{rng(2000*s + 4)}, $s\in\{1,2,\dots,10\}$ \cite{rng}.  By drawing $\mathcal R_\tau$ from a uniform distribution on $[0,1]$, 
we can add stochasticity to the probability thresholds for bacteria and EPS agents by setting $p_B = 0.25+0.2\mathcal R_\tau$ and $p_E = 0.35+0.2\mathcal R_\tau$.  Seeds introduce randomness into probability threshold values as opposed to runs, which introduce randomness in the timing of when an agent divides~(Section~\ref{sec:iDynoParam}).

The probability, distance, and neighborhood thresholds for \textit{Tagging}, \textit{Adding}, and \textit{Screening} are calibrated using experimental data and simulation estimates; aggregates with diameter of 5-200 $\SI{}{\micro\meter}$ are observed in chronic infections and $\sim$60\% of aggregates broken off in biofilm simulations were of very small diameter \cite{xavier2005biofilm}. 
Since we expect some small clusters, we set the minimum number of agents in a cluster to $n_{\text{min}}=2$ in DBSCAN. The choice of the disrupted EPS ratio $r_\text{H*}$ will be discussed in Section~\ref{sec:results}.

\begin{table}[h]
\renewcommand{\arraystretch}{1.3}
\begin{tabular}{p{1.4cm}p{7.8cm}p{1.8cm}}
\hline\noalign{\smallskip}
Symbol & Description & Value  \\
\noalign{\smallskip}\svhline\noalign{\smallskip}
$d_{\beta}$ & Radius of region that agent searches for nearby agents ($\SI{}{\micro\meter}$) &  1.5 \\ 
$d_{n}$ & Radius of region that agent searches for tagged neighbors ($\SI{}{\micro\meter}$) &  1.5 \\ 
$p_B$ & Probability threshold for bacteria cells & $[0.25,0.45]$ \\ 
$p_E$ & Probability threshold for EPS entities & $[0.35,0.55]$\\ 
$n^\mathrm{add}_B$ & Neighbor threshold to add bacteria agent to tagged bacteria & $2$\\
$n^\mathrm{add}_E$ & Neighbor threshold to add EPS agent to tagged EPS & $2$\\
$n^{B,\mathrm{scr}}_B$ & Bacteria neighbor threshold to untag bacteria agent \ & $2$\\
$n^{E,\mathrm{scr}}_B$ & EPS neighbor threshold to untag bacteria agent \ & $2$\\
$n^{B,\mathrm{scr}}_E$ & Bacteria neighbor threshold to untag EPS agent \ & $2$\\
$n^{E,\mathrm{scr}}_E$ & EPS neighbor threshold to untag EPS agent \ & $2$\\
$\epsilon$ & DBSCAN neighborhood search radius threshold ($\SI{}{\micro\meter}$) & 1.5\\ 
$n_{\min}$ & Minimum number of agents needed to form cluster in DBSCAN & 2\\
{$r_\text{H*}$} & {Disrupted EPS ratio} & 0-1 \\
\noalign{\smallskip}\hline\noalign{\smallskip}
\end{tabular}
   \caption{Parameters and values used for the cluster formation model. 
   }
    \label{tab:cluster_form_params}
    \renewcommand{\arraystretch}{1}
\end{table}

\begin{figure}[h]
    \centering
    \subfloat[Tagging]{\includegraphics[width=0.45\textwidth]{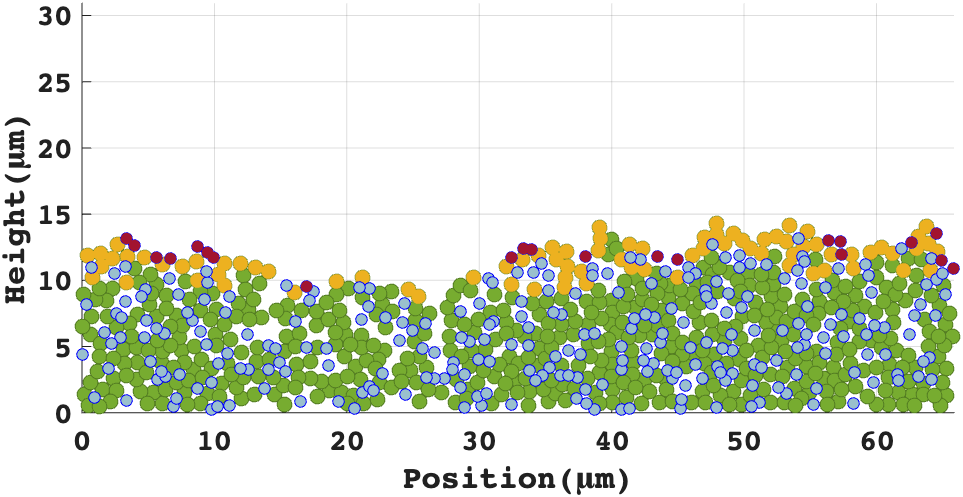}\label{fig:Tagging}}
    \hfill
    \subfloat[Adding]{\includegraphics[width=0.45\textwidth]{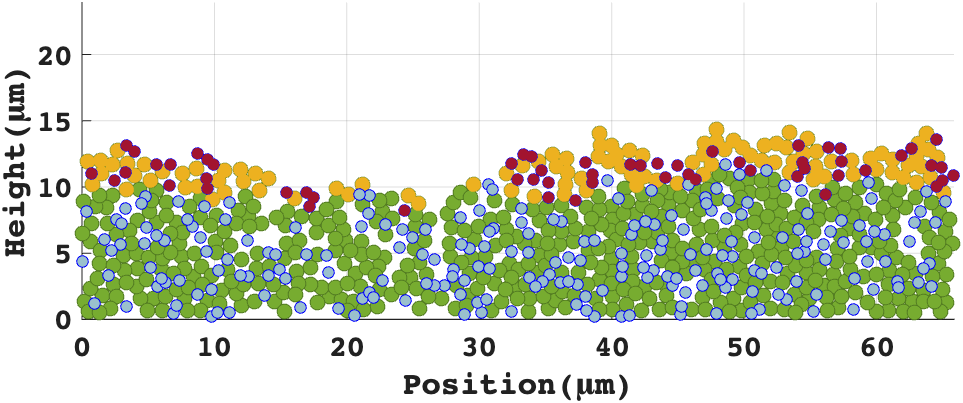}\label{fig:Adding}}
    
    \vspace{0.1cm}
    
    \subfloat[Screening]{\includegraphics[width=0.45\textwidth]{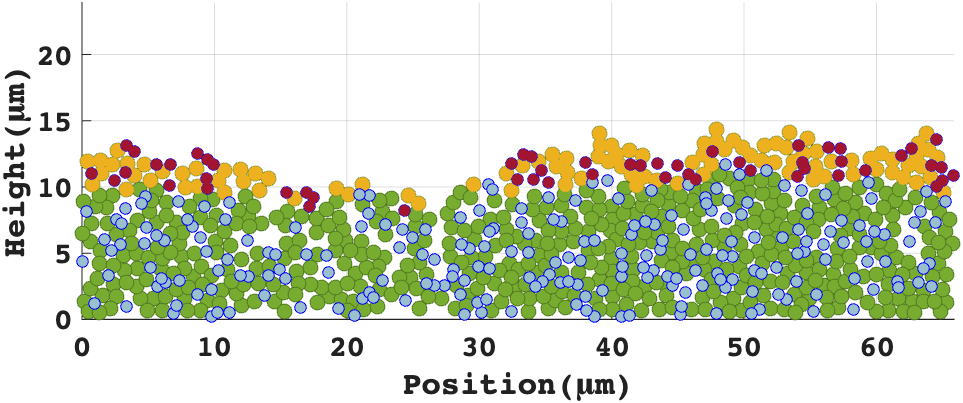}\label{fig:Screening}}
    \hfill
    \subfloat[Clustering]{\includegraphics[width=0.45\textwidth]{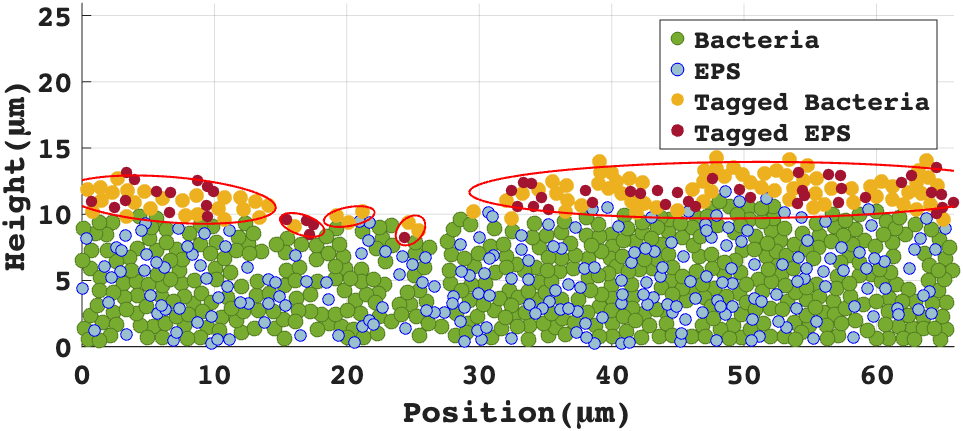}\label{fig:Clustering}}
    
    \caption{Example biofilm undergoing Algorithm \ref{Alg:tag_add_screen} using Parameter Set 1 (see Table \ref{tab:iDynomics_param_sets}), Run 6, Seed 3, and $r_\text{H*}=0.95$ at time step 2. Each of the four phases of the algorithm are illustrated: (a) Tagging, (b) Adding, (c) Screening, and (d) Clustering. The biofilm is compromised of bacteria agents (green), EPS agents (blue), tagged bacteria agents become (orange), and tagged EPS agents (red).  
    }
    \label{fig:Algo1}
\end{figure}

To illustrate the types of results we get from our cluster formation model, we present an example simulated biofilm 
undergoing the various phases of Alg. \ref{Alg:tag_add_screen} in Fig.~\ref{fig:Algo1}. Agents with high detachment probabilities are identified during the \textit{Tagging} phase (Fig. \ref{fig:Tagging}). Amid the \textit{Adding} phase, previously untagged agents that are surrounded by tagged agents are added to the tagged set (Fig. \ref{fig:Adding}). For example, a few agents near (40,13) were initially untagged but transition to tagged during this phase. The \textit{Screening} phase allows agents that are deep inside the biofilm and not surrounded by tagged agents to become untagged, as illustrated with the agent near (56,10). Finally, in the \textit{Clustering} phase (Fig. \ref{fig:Clustering}), all remaining tagged agents are collected and used to form clusters.

\begin{figure}[h]
    \centering
    \subfloat[][Before clustering]{\includegraphics[width=0.5\linewidth]{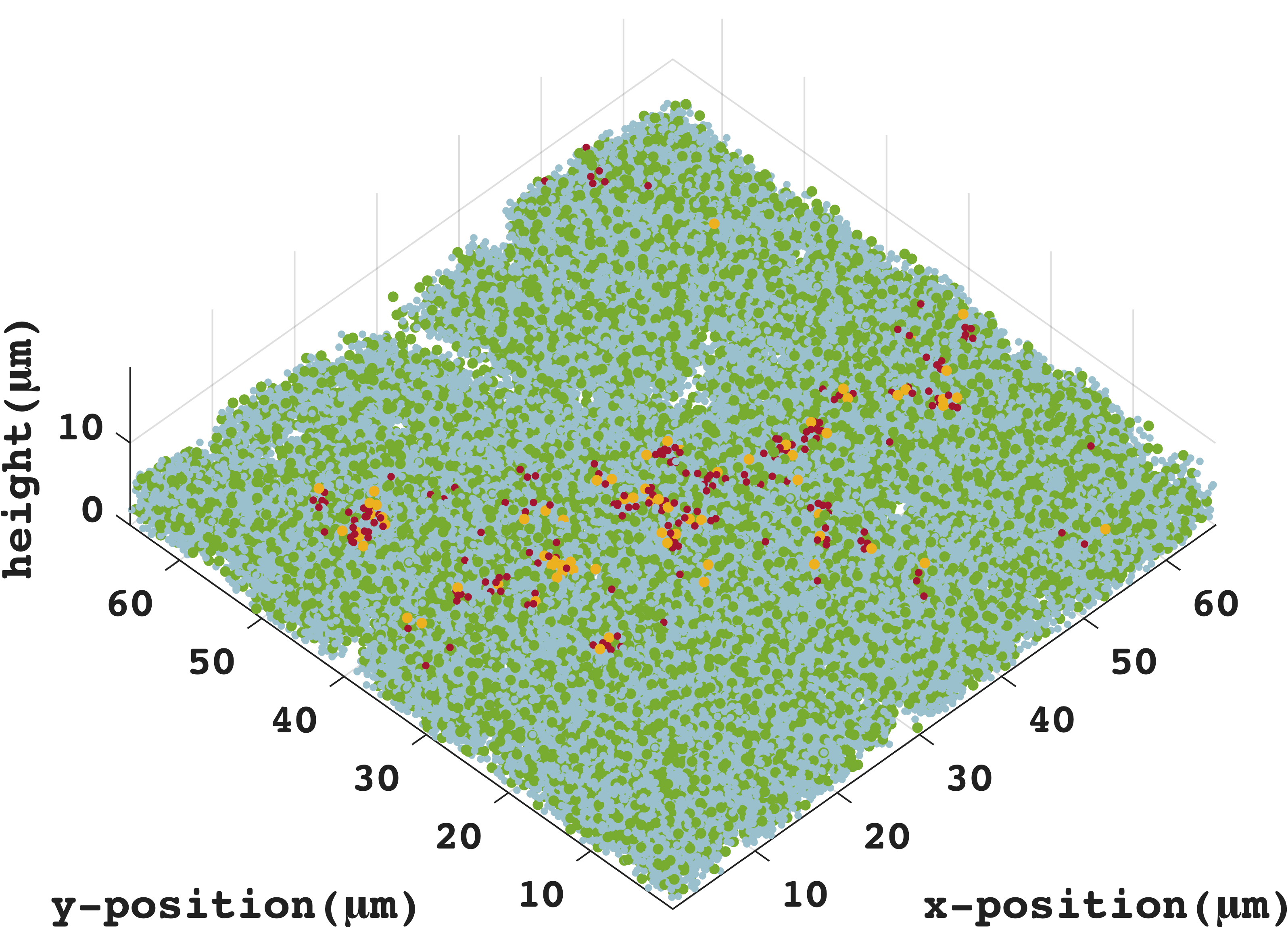}\label{fig:3d_cluster_before}}
    \subfloat[][After clustering]{\includegraphics[width=0.5\linewidth]{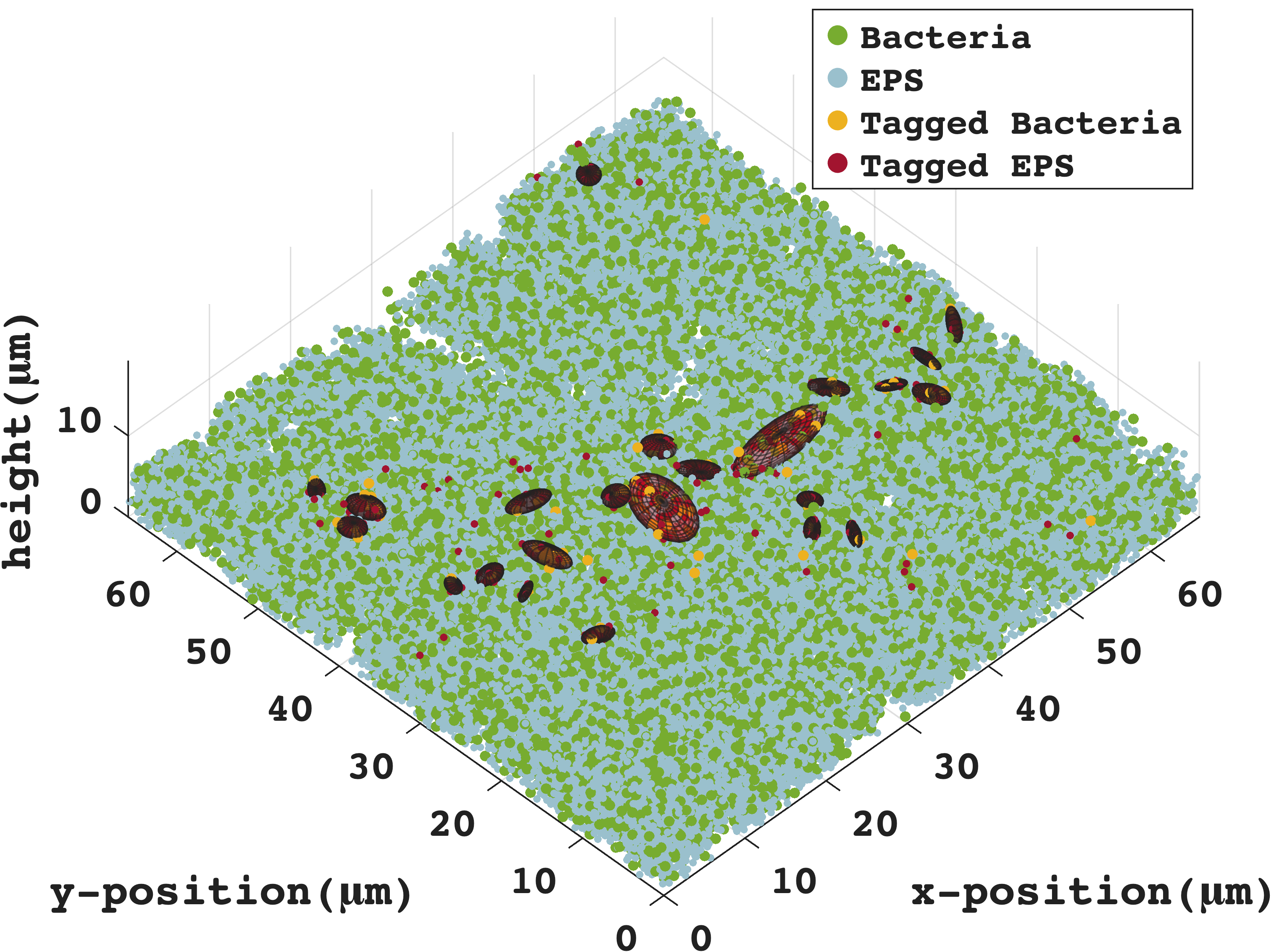}\label{fig:3d_cluster_after}}
    \caption{Example for clustering in 3-d setting with Parameter Set 1 (Table \ref{tab:iDynomics_param_sets}), Run 1, Seed 6,  and $r_\text{H*}=0.65$ at the first time step. 
    See Supplementary Video S1 for a 3-d rotation of this clustered configuration and  \ref{supp:vid_1} for a description.}
    \label{fig:3d_cluster_example}
\end{figure}

As previously mentioned, we also run our model in 3-d and provide an example in Fig. \ref{fig:3d_cluster_example}. The majority of the phases remain the same when moving from 2-d to 3-d with the exception of \textit{Clustering}. Rather than identify clusters as ellipses, we describe their shapes with ellipsoids. This is illustrated in Fig. \ref{fig:3d_cluster_after} where clusters are represented as ellipsoids of differing sizes and asphericities.

\section{Cluster Detachment Model}
\label{sec:detach_model}
Algorithm \ref{Alg:tag_add_screen} identifies clusters of agents that are likely to detach and we now create a model to determine whether detachment events occur. We borrow a rolling detachment inequality from Ting et al.  \cite{ting2021impact,ting2022image,ting2023detachment}. Detachment occurs if 
\begin{equation}
    M_d  > M_a,
    \label{eq:detach}
\end{equation} 
where $M_d$  is the drag moment and $M_a$ is the adhesive moment; both are measured in N$\cdot$m. To make a detachment decision for each cluster identified, 
both $M_d$ and $M_a$ must be calculated. Below we describe how we compute $M_d$ and $M_a$, which takes the 2-d  characteristics of biofilm clusters as inputs. When applying this to 3-d clusters, the clusters are first  projected  onto the 2-d plane that represents  a cross-section through the component of the biofilm that is perpendicular to both the  direction of the flow and the height of the flow cell (the flow and biofilm position are in the $x$-direction, and the height is in the $z$-direction). This is a reasonable approximation of 3-d dynamics that captures the essence of the relevant quantities for detachment. 

\subsection{Drag Moment}
The drag moment  in Eq. \eqref{eq:detach} is given by 
\begin{equation}
    M_d = \ell_d F_d  f_d(\alpha) f_{d, \alpha}(\alpha, \varphi), \label{eq:dragmoment}
\end{equation}
where  $\ell_d$ is the drag lever arm (m), $F_d$ is the drag force  (N), $f_d(\alpha)$ is the aspect ratio-dependent drag moment shape factor (unitless), and $f_{d, \alpha}(\alpha, \varphi)$ is the aspect ratio and orientation-dependent shape factor (unitless) \cite{ting2023detachment}. The drag lever arm $\ell_d$ is defined as half the
ellipsoid height in \cite{ting2021impact}. We adapt this to be half the height (in the direction normal to the fluid flow) of the portion of the cluster that is at the biofilm-fluid interface. 
This takes into account the fact that some clusters are completely within the biofilm and thus will be less likely to detach in the current time step. More specifically, we first identify the boundaries of each individual cluster and of the entire top surface of the biofilm. The intersection of these two boundaries corresponds to the portion of the cluster that is at the biofilm-fluid interface. 
The height is then defined as the maximum vertical distance within this common boundary region. The aspect ratio $\alpha$ describes the proportion of the semi-minor to semi-major axis of the ellipse; $\alpha = 1$ indicates a perfect circle in 2-d. In the 3-d setting, the ratio of the semi-major axis to the larger of the two semi-minor axes is considered. The orientation angle $\varphi$ is measured from the vertical; $\varphi = 0^{\circ}$ indicates that the ellipsoid
makes a right angle with the flow direction and $\varphi = 90^{\circ}$ that the ellipsoid
and flow direction are parallel. We separately discuss each component of this product that defines the drag moment. 

We depart from the strategy taken by \cite{ting2023detachment} for the specific scenario of a single spheroid sitting on a flat substrate. In Eq. \eqref{eq:dragmoment} we utilize the more general form of the force due to drag, 
\begin{equation} 
F_d = \frac{1}{2} \rho v^2 C_d A_c.
\end{equation}
Here, $\rho$ is the fluid density, $v$ is the fluid velocity, $C_d$ is the drag coefficient, and  $A_c$ is the cross-sectional area, which, in our 2-d formulation, is represented as a 1-d version by a scaled version of the drag lever height, $2\ell_d$. 
Formulas for ellipsoids as functions of the Reynolds number $Re$ and other cluster shape characteristics exist. Such models are determined from the results of CFD experiments; see Fig. 2 in \cite{maramizonouz2022drag} for an overview.  Few such empirical models are valid below $Re < 0.1$; we are modeling a low shear stress, laminar flow setting with a low Reynolds number.  
We select an option that is relevant for such a scenario following \cite{ganser1993rational,leith1987drag} given by
\begin{equation}
    C_d = \frac{24 K_S}{Re} \left[ 1 + 0.1118 \left( \frac{Re K_N}{K_S} \right)^{0.6567} \right] + \frac{0.4305 K_N}{ \left( 1 + \frac{3305}{Re K_N/K_S} \right)}.
\end{equation}
Here,
\begin{equation}
    K_N = 10^{1.8148(-\log \Phi_W)^{0.5743}}, \qquad K_S = \frac{1}{3} + \frac{2}{3 \sqrt{\Phi_W}}, \qquad \Phi_W = \frac{A_{\rm sph}}{A_p},
\end{equation}
where $K_N$ and $K_S$ are the Newton's and  Stokes' shape factors and  $\Phi_W$ is the sphericity given in terms of the surface area of the volume equivalent sphere, $A_{\rm sph}$, and the cluster surface area, $A_p$. These latter two quantities  are calculated as $A_{\rm sph} = 4 \pi (a^2 b)^{2/3}$ and $A_p = \pi a b$ using the semi-major and semi-minor axes lengths $a$ and $b$.  Note that other empirical models for $C_d$ may be selected  for settings involving larger Reynolds numbers. The drag moment model variables and parameters used in our model are summarized in Table~\ref{table:drag_force}.

\begin{table}[h]
\begin{tabular}{p{2.3cm}p{6.6cm}p{2.5cm}}
\hline\noalign{\smallskip}
Symbol & Description & Value  \\
\noalign{\smallskip}\svhline\noalign{\smallskip}
$M_d$ & Drag moment (J or N$\cdot$m) & - \\
$F_d$ & Drag force (N) & -\\  
$\ell_d$ & Drag lever arm (m) & - \\
$\varphi$ & Orientation angle (degrees) & - \\
$f_d(\alpha)$ & Aspect-ratio dependent drag moment shape factor (-) & - \\
$f_{d, \alpha} (\alpha, \varphi)$ & Aspect ratio and orientation-dependent shape factor (-) & - \\
$C_d$ & Drag coefficient (-) & - \\
$A_c$ & Cross-sectional area [2-d version: height] (m) & - \\ 
$K_N$ & Newton's shape factor (-) & - \\
$K_S$ & Stokes' shape factor (-) & - \\
$\Phi_W$ & Sphericity (-) & - \\
$\rho$ & Fluid density (kg/m$^3$) &  997 \\
$v$ & Fluid velocity (m/s) & $1.9 \times 10^{-3}$  \cite{packard2026biophysical} \\
$Re$ & Reynolds number (-) & $6.73\times 10^{-3}$  \cite{packard2026biophysical} \\
\noalign{\smallskip}\hline\noalign{\smallskip}
\end{tabular}
\caption{ Variables and parameters for the drag moment model. 
}
\label{table:drag_force}
\end{table}

Utilizing a curve fit to CFD results, Ting et al. \cite{ting2021impact,ting2023detachment} determined functional forms for the shape factors in Eq. \eqref{eq:dragmoment}. 
The aspect ratio-dependent drag moment shape factor $f_d(\alpha)$ \cite{ting2021impact} is given as
\begin{equation}
    f_d(\alpha) = \frac{1.296 \alpha^2 + 0.1509 \alpha + 0.03718}{\alpha^2 + 0.0843 \alpha + 0.0002284},
\end{equation}
which we directly utilize. The aspect ratio and orientation angle-dependent moment shape factor $f_{d, \varphi}(\alpha, \varphi)$ as derived in \cite{ting2023detachment} had six parameters that were different for each aspect ratio $\alpha$. 

\begin{figure}[h]
\centering
\subfloat[][Drag moment shape factor]{\includegraphics[width=0.5\linewidth]{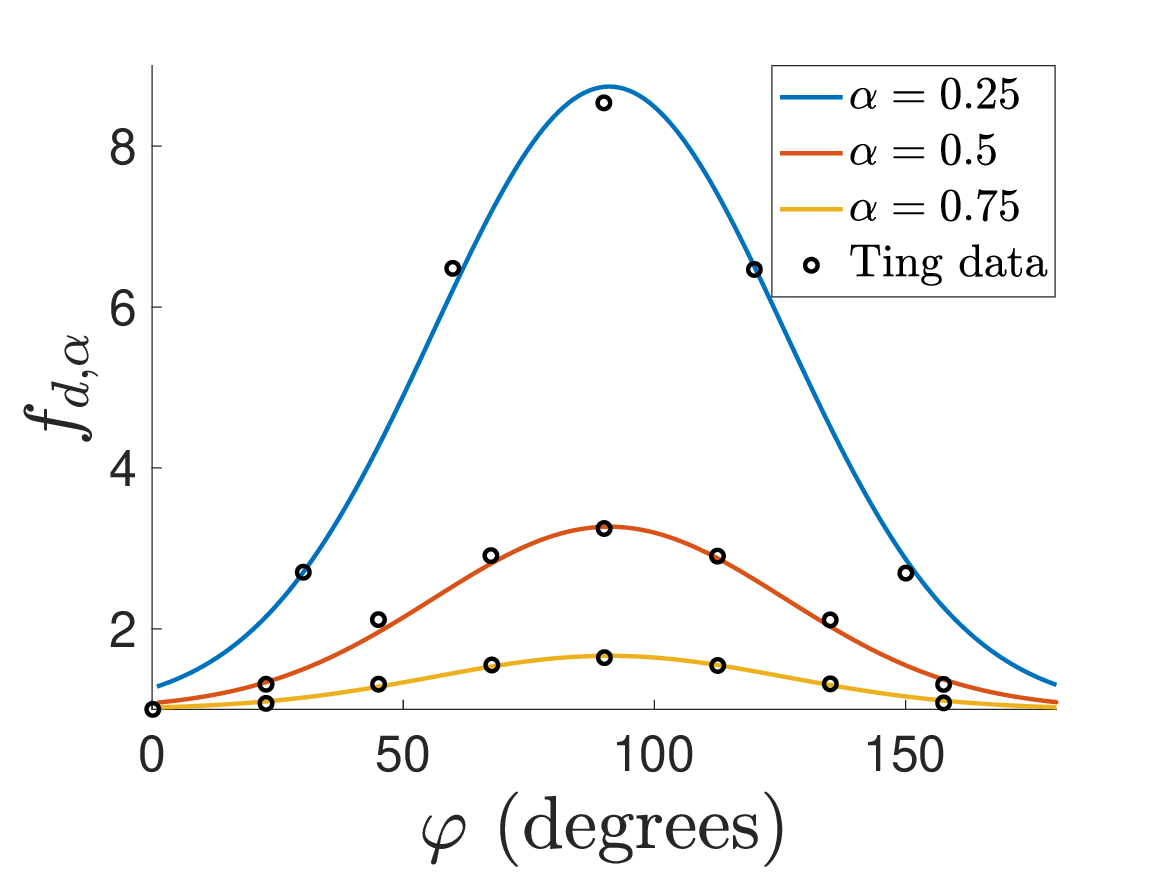}\label{fig:fMalpha}}
\subfloat[][Adhesive moment shape factor]{\includegraphics[width=0.5\linewidth]{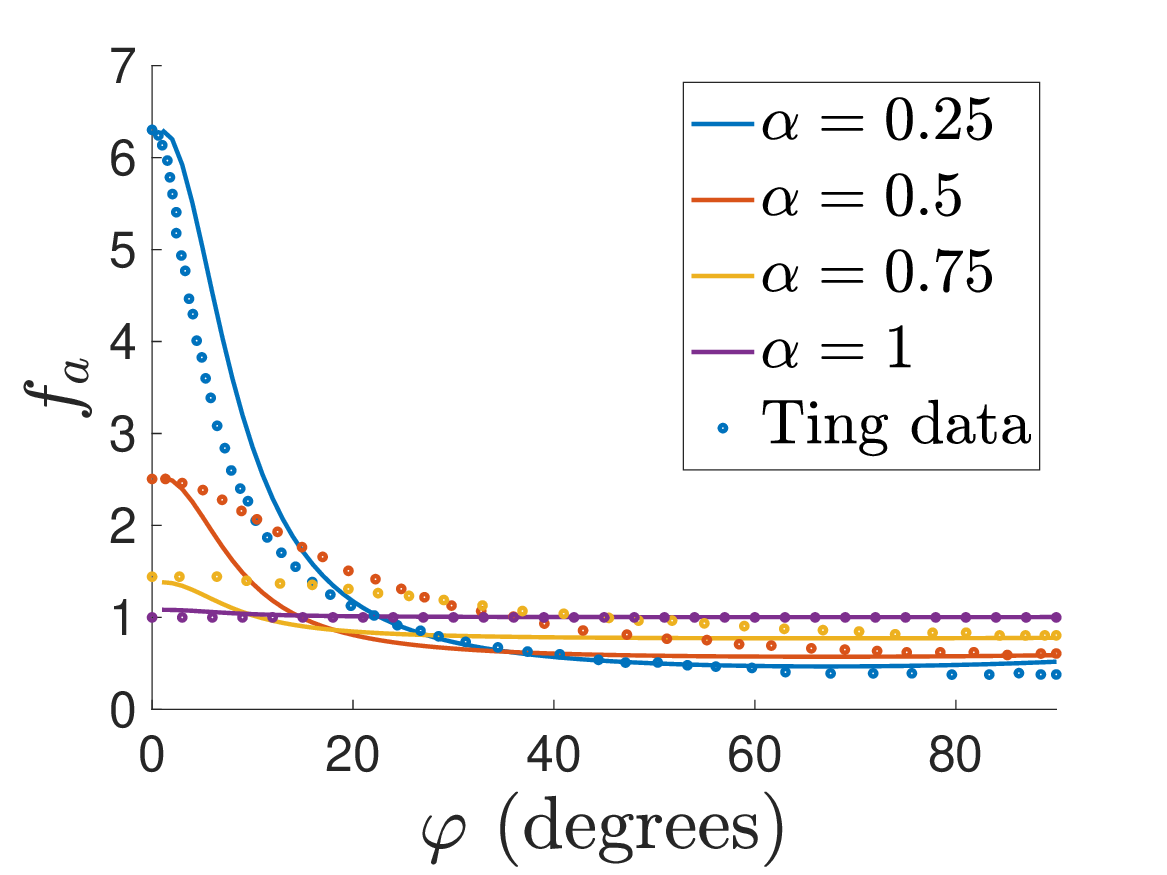}\label{fig:f_a}}
    \caption{Fit of Eq. \eqref{eq:fmalpha} in (a) and Eq. \eqref{eq:f_a} in (b) to the  data from Ting et al. in \cite{ting2023detachment}.}
    \label{fig:fMalpha_f_a}
\end{figure}

We do not directly utilize $f_{d, \alpha}$ from \cite{ting2023detachment}, rather we make use of their data to determine a functional form  
that is written explicitly as a function of $\alpha$ and $\varphi$. Using the MATLAB code \verb|grabit.m| \cite{grabit}, we extract data from Fig. 8 of \cite{ting2023detachment} and first curve fit each of the three curves for $f_{d, \alpha}$ in the figure (for different $\alpha$) with a translated Gaussian:
\begin{equation}
 f_{d, \alpha}(\varphi) =   a_1 \exp\left\{ - \left(\frac{\varphi - b_1}{c_1} \right)^2 \right\} + 1.
\end{equation}
We find that for each curve (corresponding to a different $\alpha$), the parameters $b_1$ and $c_1$ are nearly equal, so we fix $b_1 = 90$ and $c_1 = 49.5$. That is,  $f_{d, \alpha}$ depends on  $\varphi$ in such a way that $f_{d,\alpha}$ is maximal when $\varphi = 90^{\circ}$ and decreases as $\varphi$ deviates from $\varphi = 90^{\circ}$. The parameters $a_1$ seem to decrease exponentially with $\alpha$. Thus, we perform an exponential curve fit to $a_1$ as a function of $\alpha$, adding in the pair $(\alpha, a_1) = (1, 0)$ so that when $\alpha = 1$, the shape factor is uniformly 1. This yields
\begin{equation}
 f_{d, \alpha}(\alpha, \varphi) =   a_1(\alpha) \exp\left\{ - \left(\frac{\varphi - b_1}{c_1} \right)^2 \right\} + 1 = d_1 e^{f_1 \alpha} \exp\left\{ - \left(\frac{\varphi - b_1}{c_1} \right)^2 \right\} + 1,
 \label{eq:fmalpha}
\end{equation}
where $d_1 = 26.4$ and $f_1 = -4.91.$ The fit is shown in Fig. \ref{fig:fMalpha}.
The fit for $f_{d, \alpha}$ is not as good as in Ting's Fig. 8 (although slight errors may be attributed to the use of \verb|grabit.m|). However, it is useful for the functional form of $f_{d,\alpha}(\alpha,\varphi)$ to take $\alpha$ into account and we utilize Eq. \eqref{eq:fmalpha} in the calculation of the drag moment in Eq. \eqref{eq:dragmoment}.

\subsection{Adhesion Moment}
To compute an adhesive moment in Eq. \eqref{eq:detach} for a cluster of agents 
at or near the biofilm surface, we modify ideas from Hartmann et al. 
\cite{hartmann2019emergence} who use concepts from Gay-Berne potential theory for non-identical ellipsoids \cite{cleaver1996extension}. The adhesive moment is given by
\begin{equation}
    M_a = U_i f_a(\alpha, \varphi), \label{eq:adhesivemoment}
\end{equation}
where $U_i$ is the sum of the cluster-agent interaction potentials between all neighboring agents and cluster $i$ (described in more detail below) and $f_a(\alpha, \varphi)$ is the adhesive shape factor. The authors \cite{hartmann2019emergence} model the cell-cell interaction potential between cells $i$ and $j$, which we extend to cluster-agent interaction (cluster $i$ and agent $j$), by
\begin{equation}
    U_{ij} = \epsilon_0  \left[ \underbrace{\exp \left( - \frac{(r_{ij}/\sigma_{0,i})^2}{\lambda_r^2} \right)}_{\text{cluster-agent repulsion}} + \underbrace{\frac{\nu}{1 + \exp \left( \frac{\rho_a - r_{ij}/\sigma_{0,i}}{\lambda_a} \right)}}_{\text{cluster-agent attraction}} \right],
    \label{eq:interaction_potential}
\end{equation}
where $U_{ij}$ is the cluster-agent interaction potential between cluster $i$ and agent $j$, $r_{ij}$ is the distance from the center of cluster $i$ to the center of agent $j$, $b_i, b_j$ are the  semi-minor axes of cluster $i$ and agent $j$, respectively, $\lambda_r$ is the repulsion width, $\nu$ is the attraction strength, $\rho_a$ is the attraction shift, $\lambda_a$ is the attraction width, $\epsilon_0$ is the strength of repulsion, and  $\sigma_{0,i} = \sqrt{b_i^2 + b_j^2}$.  For a single agent, we take  $b_j$ to be the agent radius. Also note that in 3-d, the larger of the two semi-minor axis is used as input into the formulas. The variables and parameters are summarized in Table \ref{table:adhesion_force}. We differentiate between bacteria agents and EPS agents by considering different attraction strengths and strengths of repulsion. We note that the authors \cite{hartmann2019emergence} used a range of $5 \times 10^{-19} - 5 \times 10^{-15}$ J  for $\epsilon_0 $ and an attraction strength of  $0.65 \times 10^{-20}$ J when fitting to experimental data. We extend the upper bound of the repulsion strength range and increase the attraction strength in our simulations, in part because \textit{S. epidermidis} bacteria cells have a radii that is approximately double of that studied by \cite{hartmann2019emergence}.

\begin{table}[h]
\begin{tabular}{p{1.2cm}p{6.1cm}p{4.1cm}}
\hline\noalign{\smallskip}
Symbol & Description & Value  \\
\noalign{\smallskip}\svhline\noalign{\smallskip}
$M_a$ & Adhesive moment (J or N$\cdot$ $\SI{}{\micro \meter}$) & - \\
$U_{ij}$ & Cluster-agent interaction potential  between cluster $i$ and agent $j$  (J or N$\cdot$ $\SI{}{\micro \meter}$) & -\\ 
$r_{ij}$ & Distance from cluster $i$ center to agent $j$ center  ($\SI{}{\micro \meter}$) &  -\\ 
$b_i,b_j$ & Semi-minor axes of cluster $i$ and agent $j$  ($\SI{}{\micro \meter}$) & -\\  
$\lambda_r$ & Repulsion width ($\SI{}{\micro\meter}$) & $1.5\times1.16$ (B),  $1.16^*$ (E) \\ 
$\nu$ & Attraction strength (J or N$\cdot$ $\SI{}{\micro \meter}$) &  $0.65\times 10^{-10}$ (B), $0.325 \times 10^{-7}$ (E)\\  
$\rho_a$ & Attraction shift ($\SI{}{\micro\meter}$) & $2.0^*$ \\  
$\lambda_a$ & Attraction width ($\SI{}{\micro\meter}$)  & 1.5 $\times$ 0.11 (B), $0.11^*$ (E) \\  
$\epsilon_0$ & Strength of repulsion (J or N$\cdot$ $\SI{}{\micro \meter}$) &  $2 \times 10^{-12}$ (B),  $2 \times 10^{-10}$ (E) \\  
$\kappa$ & Neighborhood parameter ($\SI{}{\micro\meter}$) & 1.5 \\
\noalign{\smallskip}\hline\noalign{\smallskip}
\end{tabular}
\caption{Variables and parameters for the adhesive moment model. Here, B and E are used to distinguish values that vary between bacteria and EPS. Parameter's values and ranges are taken or adapted from \cite{hartmann2019emergence} and those marked with $^*$ correspond to a      ``typical overlap factor'' (for \cite{hartmann2019emergence}) of $\sigma =0.7$ $\SI{}{\micro\meter}$, which corresponds to a sphere with mean cell volume of $\SI{0.4}{\micro\meter^3}$.}
\label{table:adhesion_force}
\end{table}

We identify a $U_i$ for each cluster as the sum of $U_{ij}$ for all pairwise cluster-agent interactions involving agents surrounding the cluster in some neighborhood. To define the neighborhood, we begin by defining the cluster boundary. In 2-d, the ellipsoidal cluster boundary is given by 
\begin{equation}
    \frac{(\cos (\varphi) (x - x_0) + \sin(\varphi) (y - y_0))^2}{a^2} + \frac{\sin(\varphi)(x - x_0) - \cos(\varphi) (y - y_0))^2 }{b^2 } = 1,
    \label{eq:cluster_boundary}
\end{equation}
where $(x_0, y_0)$ is the cluster center and $a, b$ are the cluster semi-major and semi-minor axes. We define a larger ellipse using $a + \kappa$ and $b + \kappa$ as the semi-major and semi-minor axes, where $\kappa$ is a neighborhood parameter. We evaluate the left-hand-side of Eq. \eqref{eq:cluster_boundary} for all bacteria and EPS agent coordinates to determine whether they are part of the cluster (if the left-hand-side returns a value of at most 1), and perform a similar calculation to determine whether they fall inside the cluster neighborhood. The agents that lie inside the neighborhood but outside the cluster are considered neighbors, and we compute Eq. \eqref{eq:interaction_potential} for each to arrive at the sum $U_i$.

We also 
need $f_a$, the adhesive shape factor, to determine the adhesive moment in Eq. \eqref{eq:adhesivemoment}. This shape factor is shown in \cite{ting2023detachment} for $0 \leq \varphi \leq 90$ (degrees) and for $\alpha = 0.25, 0.5, 0.75, 1$, from CFD results. To write $f_a$ explicitly as a function of $\alpha$ and $\varphi$, we again use \verb|grabit.m| \cite{grabit}, extracting data from Fig. 9 of \cite{ting2023detachment} and fit it with the function:
\begin{equation}
    f_a(\alpha, \varphi) = \left[ h_1 \exp \left( \frac{j_1 \alpha }{\alpha - k_1} \right) + 1 - \alpha \right] (1 + \ell_1 \varphi^2)^{-1} + \alpha.
    \label{eq:f_a}
\end{equation}
The functional form of $f_a(\alpha, \varphi)$ is chosen as such: we propose that $f_a = g(\alpha)h(\varphi) + \alpha$, noting that the value of $f_a$ for each $\alpha$ in Fig. 9 of \cite{ting2023detachment} decays to approximately $\alpha$ by $\varphi = 90^{\circ}$. We note that the maximum value of $f_a$ (at $\varphi = 0^{\circ}$) for each $\alpha$ seems to decay exponentially with $\alpha$ to 1; this is modeled by the term in the square brackets in Eq. \eqref{eq:f_a}. We then note that $f_a$ decays with $\varphi$; this is modeled with the second piece of the functional form of $f_a$. We first fit for $g(\alpha) - \alpha$ using the data at $\varphi = 0^{\circ}$ to determine the coefficients $h_1, j_1,$ and $k_1$. Fixing those constants, we then fit $f_a$ to the data to determine $\ell_1$. The fit is shown in Fig. \ref{fig:f_a}.

Our curve fit of the data in Fig. 9 of \cite{ting2023detachment} with our Eq. \eqref{eq:f_a} yields $h_1 = 16.5, j_1 = 23.7, k_1 = 5.47$, and $\ell_1 = 0.0155$. The fit has more variation from that of \cite{ting2023detachment} than for the drag moment shape factor, but 
the fit still provides a satisfactory, functional representation of the dynamics.

\subsection{Cluster Detachment Decision Example}

To finalize our model, we tune parameter values for the adhesive force  
and strength of repulsion (see Table \ref{table:adhesion_force}) until our model predicts detachment for identified clusters that we deem likely to detach. This corresponds to clusters with $\varphi$ near 0 and with a substantial fraction 
 of the cluster 
 that is at the biofilm-media interface.  We then fix these parameter values in the final model in which biofilm cluster detachment is simulated over time, which is described in the next section.

\begin{figure}[h]
    \centering
    \subfloat[][]{\includegraphics[width=0.48\linewidth]{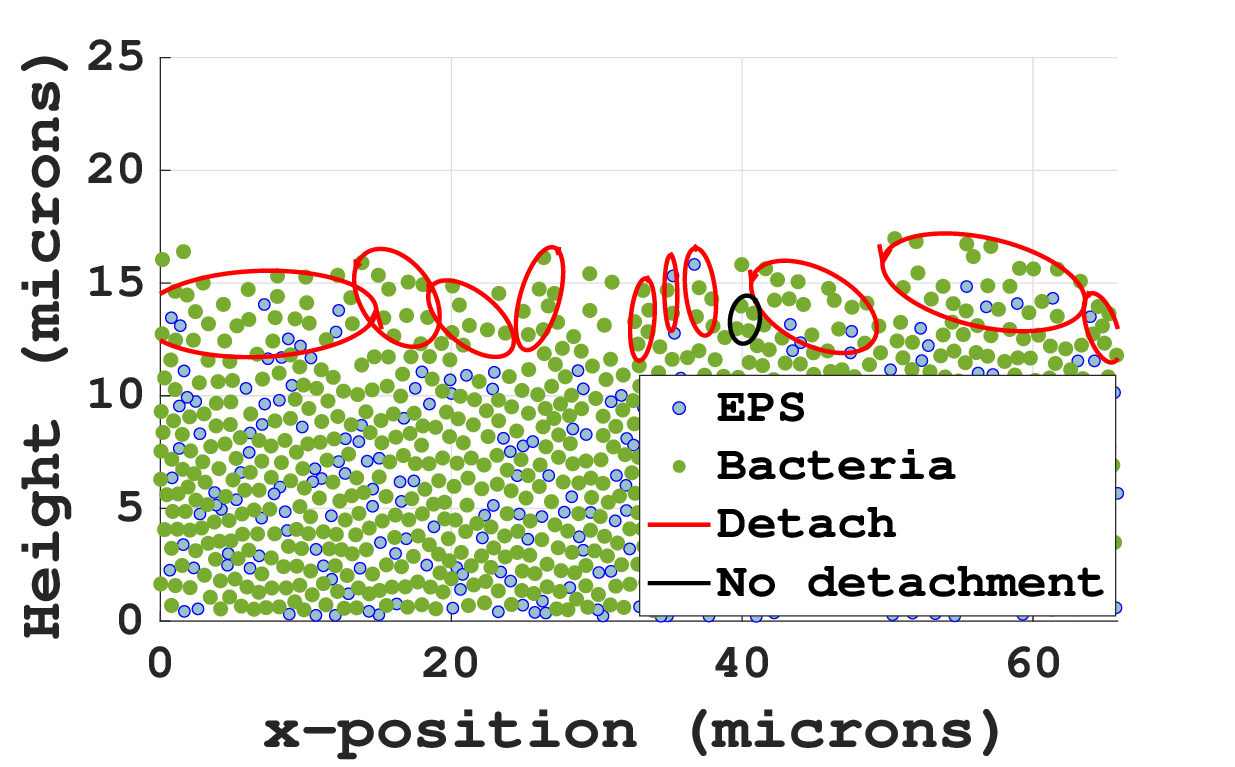}
    \label{fig:detach_1}} \hspace{1mm}
      \subfloat[][]{\includegraphics[width=0.48\linewidth]{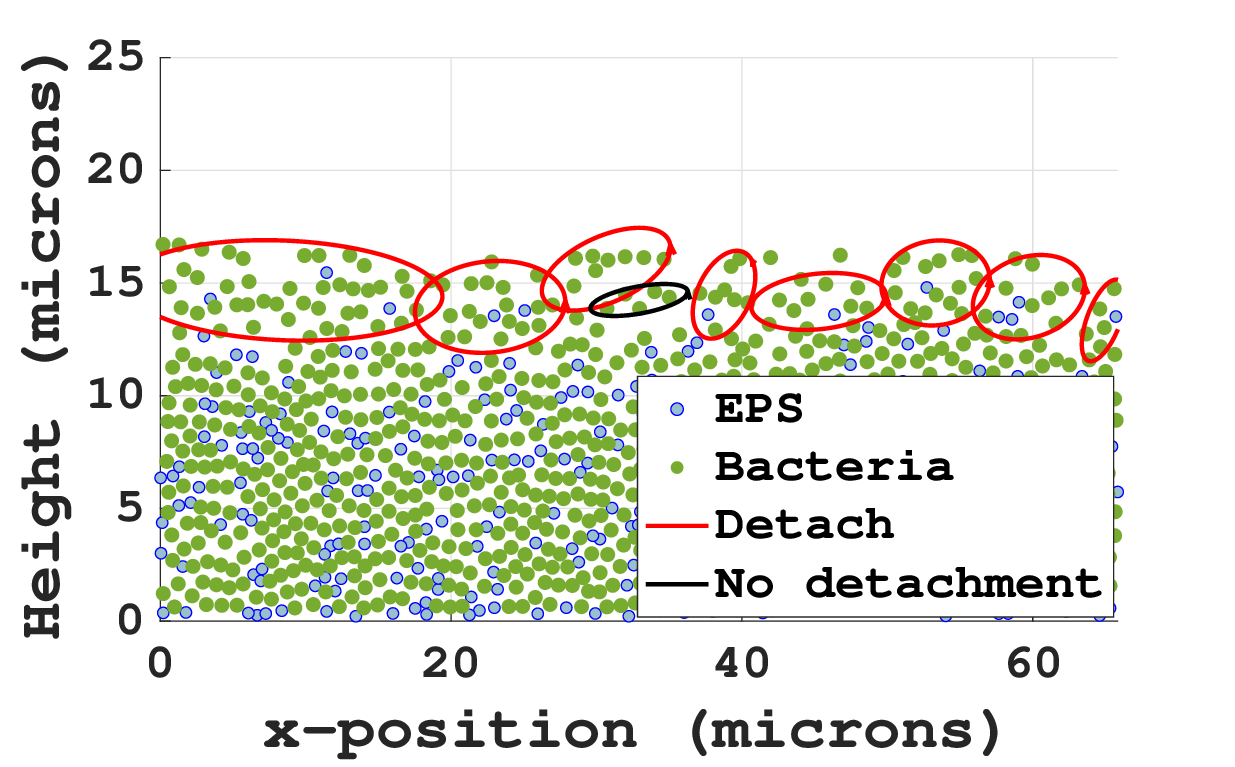}
      \label{fig:detach_2}}
    \caption{Detachment decisions for example biofilms generated by Parameter Set 4, (a) Run 4 and (b) Run 9 , $r_{\text{H}^*} = 0.5$, Seed 2, and time step 0. Bacteria agents are colored green and EPS agents are colored blue. Clusters are indicated by ellipses. A red ellipse indicates detachment of that cluster (satisfying Eq.~\eqref{eq:detach}) and a black ellipse indicates no detachment.}
    \label{fig:detach}
\end{figure}

In Fig. \ref{fig:detach}, we show two example biofilms that highlight the elliptical clusters and their detachment decisions. The examples chosen are illustrative of the cluster detachment decision model as they show at least one cluster not tagged for detachment. (There are example situations, not shown, where either all clusters are or are not selected for detachment.) The examples demonstrate that the detachment decisions can vary across Parameter sets, runs, and disrupted EPS ratio, $r_{\text{H}^*}$. 
In Fig. \ref{fig:detach_1}, we see that of all clusters potentially tagged for detachment, the only cluster not selected for detachment is the cluster that is mostly buried in the biofilm; that is, it has a large cluster consisting of many agents above it. In particular, three clusters tagged for detachment are oriented nearly vertical (around $x = 30$ $\SI{}{\micro\meter}$). In Fig. \ref{fig:detach_2}, the single cluster not tagged for detachment is simultaneously below another cluster and has a positive but large orientation angle, corresponding to a larger adhesive force that prevents detachment.

\section{Biofilm Detachment Over Time}
\label{sec:results}
\begin{figure}[h]
    \centering
    \includegraphics[width=1\linewidth]{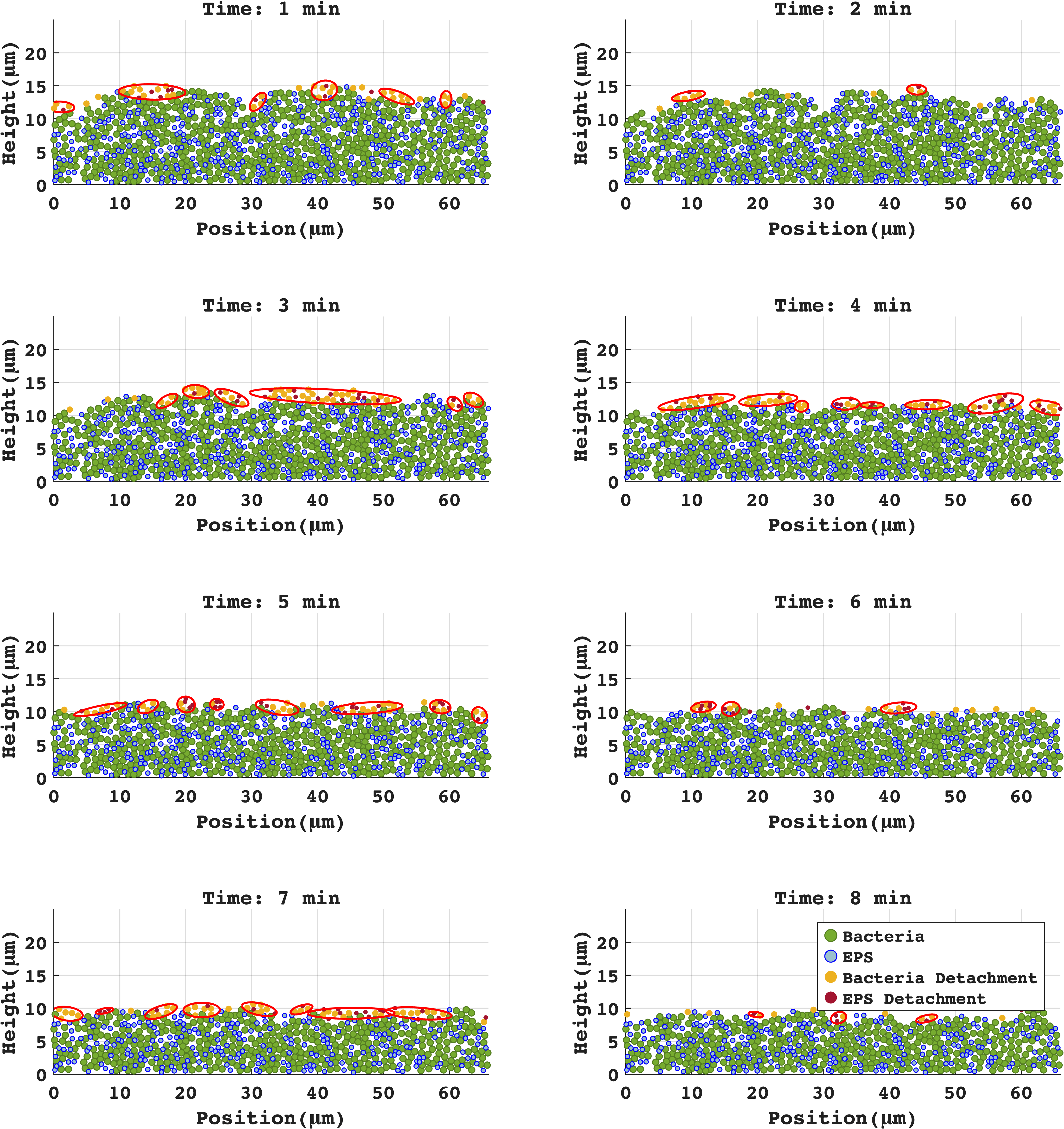}
    \caption{Example of the cluster formation and detachment models applied in sequence in a time loop for the first eight minutes. The initial biofilm is from Parameter Set 1, Run 1, Seed 1, with  $r_\text{H*}=0.5$. 
    Clusters that will detach 
    are identified as red ellipses.  All 60 time steps can be viewed in Supplementary Video~S2  (see \ref{vid:2dprocess} for a description). 
    }
    \label{fig:TimeStep}
\end{figure}

We now discuss how we simulate biofilm detachment over one hour. As an initial condition, we develop 24 hour biofilms (Section \ref{sec:idynomics}) using the iDynoMiCS parameters described in Table \ref{tab:iDynomics_parameters} and 
Table~\ref{tab:iDynomics_param_sets}. To then simulate detachment over one hour with compromised EPS, we apply our cluster formation and detachment models in a time loop where each iteration represents one minute. 
At each iteration, we run Algorithm \ref{Alg:tag_add_screen} (Section \ref{sec:cluster_model}) to identify potential clusters of cells that may detach together. We then determine whether identified clusters break off via the detachment criterion (Section \ref{sec:detach_model}). 
Single agents without any agents in the neighboring area and above the biofilm-fluid interface are also  
detached. 
To illustrate, we show the first eight time steps (one minute each) in Fig. \ref{fig:TimeStep}. Before the next time step, 
corresponding agents marked for detachment are removed from the biofilm. 
In the subsections that follow, we study the influence of biofilm parameters on frequency, size, and shape of detached clusters.

\subsection{Results for 2-d Biofilms}
\label{sec:2d_results}
Under various EPS disruption regimes, we simulate 2-d biofilms  for one hour. We characterize the clusters that detach via average height of the biofilm at time of detachment, cluster size, cluster linearity, and cluster orientation angle. After considering each characteristic independently, we then investigate the time-dependence of and interdependence amongst characteristics. We also quantitatively categorize cluster size (e.g., small, medium, and large) and examine relationships between these cluster sizes and characteristics. Finally, we confirm the robustness of results across parameter sets. 

\begin{figure}[h]
\centering
\subfloat[][Avg. height at which clusters detach]{
\includegraphics[width=0.49\linewidth,height=1.5in]{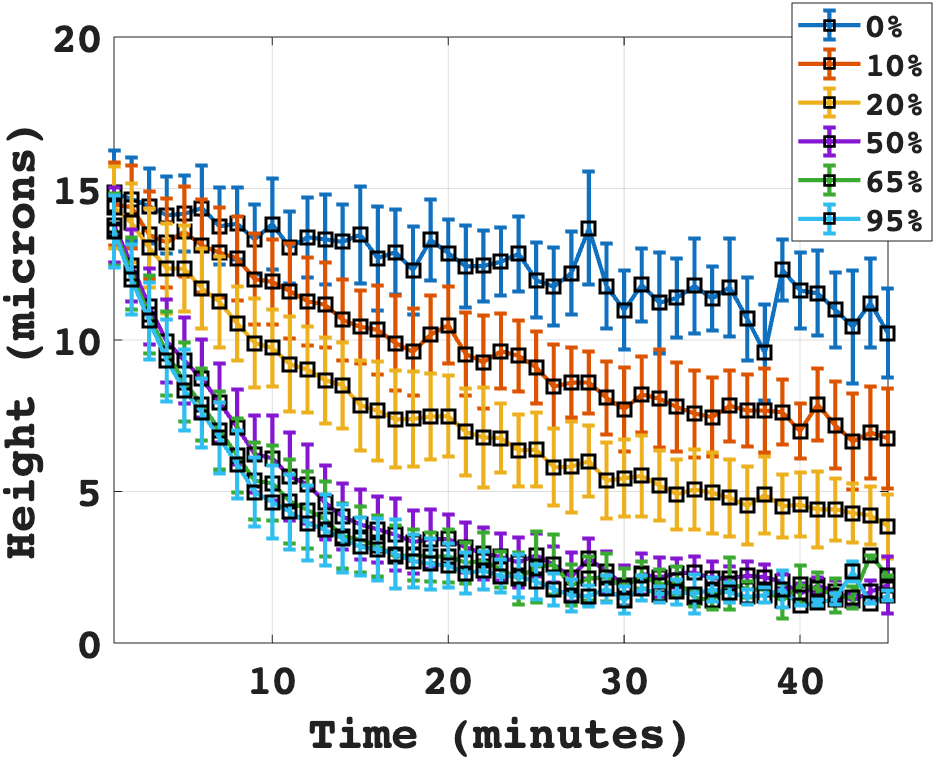}} \hspace{3mm}
\subfloat[][Cluster size at detachment]{\includegraphics[width=0.47\linewidth,height=1.5in]{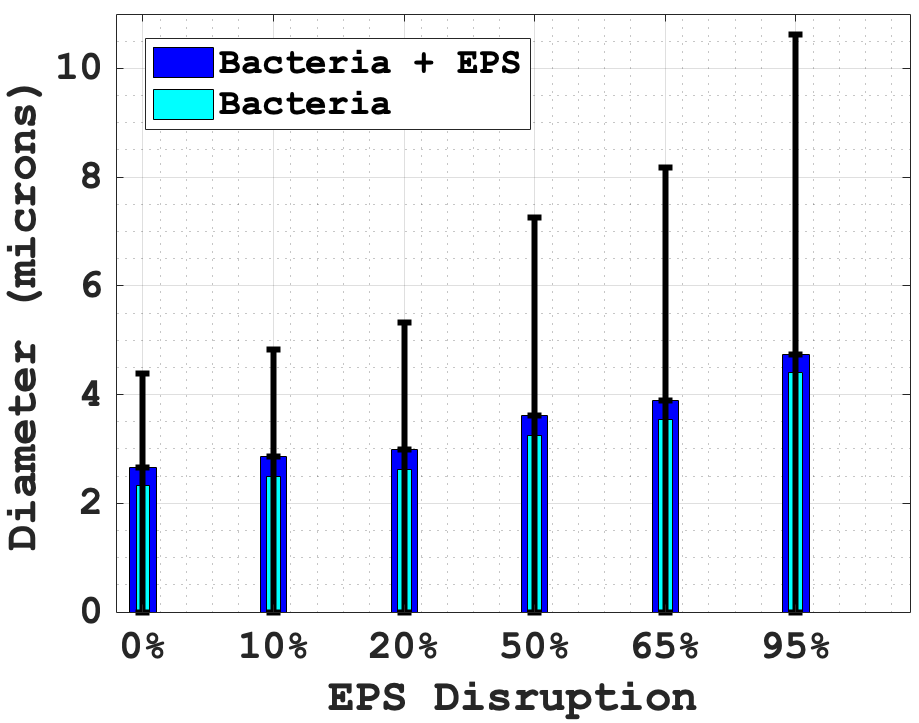}}\\ 
\subfloat[][Cluster linearity ]{\includegraphics[width=0.48\linewidth,height=1.5in]{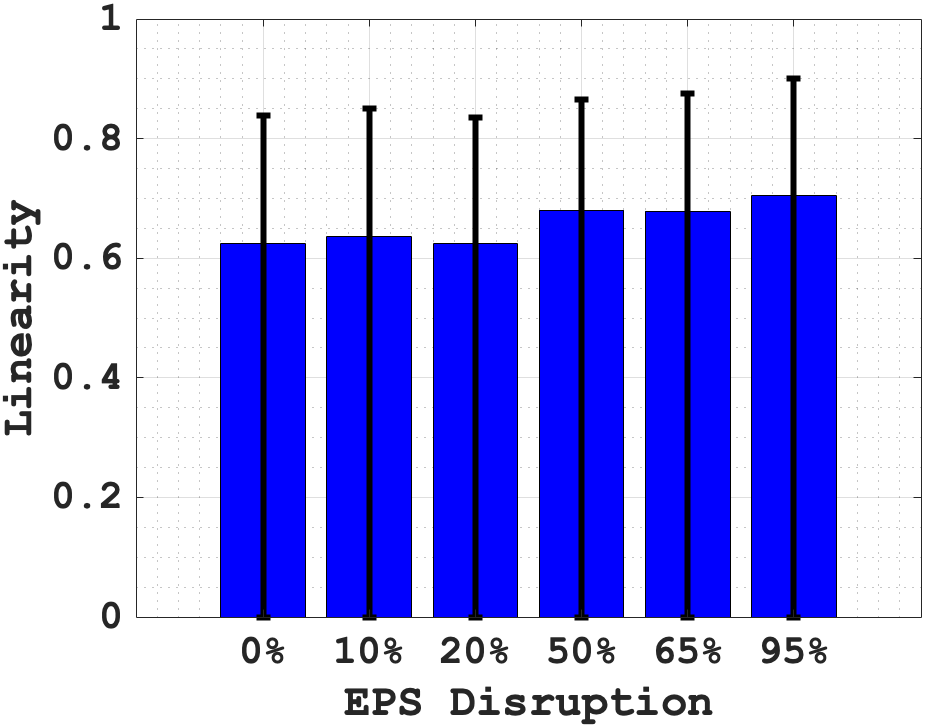}} \hspace{3mm}
\subfloat[][Cluster orientation angle]{\includegraphics[width=0.48\linewidth,height=1.5in]{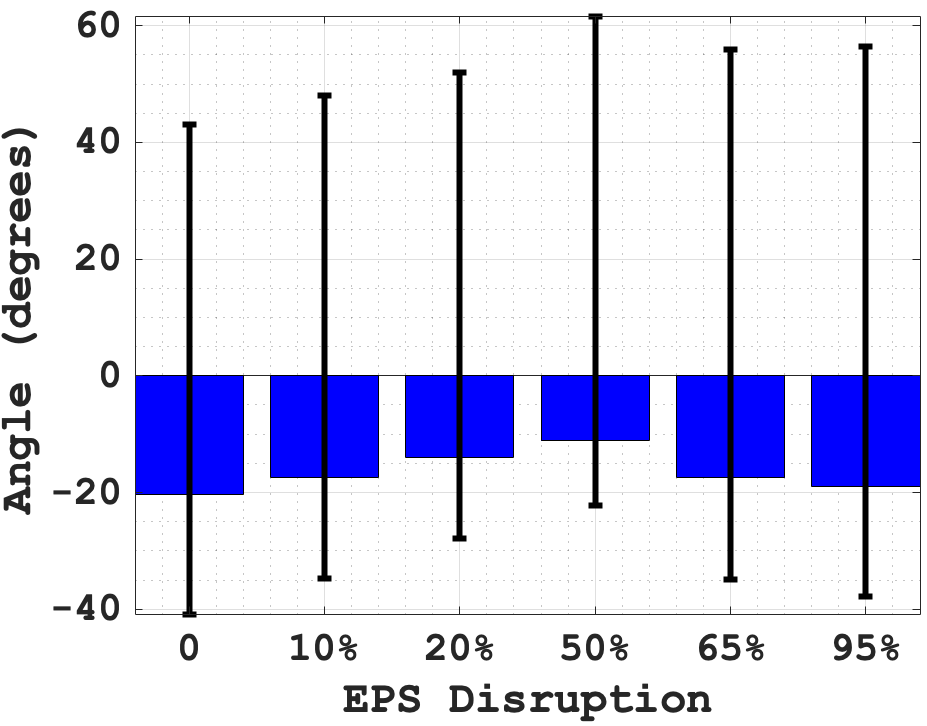}}
\caption{Characterizing clusters that detach from 2-d biofilms with different percentages of disrupted EPS (0\%, 10\%, 20\%, 50\%, 65\%, and 95\%). (a) Average height of biofilm at which clusters break off as a function of time. Detached clusters are studied across (b) maximum Euclidean diameter of either bacteria agents only or both bacteria and EPS agents, (c) linearity, and (d) angle relative to the $y$-axis. In (b)-(d), a different number of cell clusters detach at each time point (total clusters are 3008 for 0\%, 3965 for 10\%, 5159 for 20\%, 8090 for 50\%, 8680 for 65\%, and 7924 for 95\%). All averages and standard deviations are taken over 100 simulations (10 runs each with 10 seeds); in (b)-(d), the statistics comprise all time points.} 
\label{fig:2dSummary}
\end{figure}

We first examine the average height of the biofilm when detachment (single agents or clusters) occurs; this information is tracked over time and summarized in Fig.~\ref{fig:2dSummary}a. The different levels of disruption correspond to how much EPS is compromised (e.g., 20\% disruption corresponds to $r_\text{H*}=0.2$ in attachment strength $\beta$, Eq.~\eqref{eq:attachment_strength}). Disruption levels impact the ability of an agent or cluster to detach by influencing the local stickiness. In all cases, we start with an average  height of $\SI{14.41}{\micro\meter}$ at time zero. As EPS disruption increases, we observe both a rapid decrease in the average biofilm height and a negative correlation with detachment height. In the case of no disruptor (0\%), the mechanisms of sloughing and erosion result in a decrease in biofilm height, 
with a final average biofilm height of 10.22 $\SI{}{\micro\meter}$, a 29\% decrease. In comparison, a 20\% disruption has an additional 14.21\% decrease in average biofilm height while 65\% disruption yields a further 51.27\% decrease.  The biofilm height begins to level off and reach a steady state around 30 minutes for higher disruptor levels (65\% and 95\%); minimal additional decrease is observed.

We next interrogate the size of clusters with 2 or more agents that detach from the biofilm. 
The maximum Euclidean diameter of a cluster, $\mathfrak{D}_E$,  is the largest distance between two agents. Averaging over time and simulations, we observe that $\mathfrak{D}_E$ does increase with increased disruption (Fig.~\ref{fig:2dSummary}b). This trend is clearly observed; EPS disruption of 20\% and 65\% correspond to a 12\% and 46\% increase in cluster size, respectively, when only utilizing bacteria agent locations in the cluster. Since experiments often fluorescently label the bacteria \cite{packard2026biophysical}, we report $\mathfrak{D}_E$ for bacteria only, which does account for EPS between bacteria within the cluster. As expected, cluster size does increase when calculating $\mathfrak{D}_E$ using EPS and bacteria agents as EPS agents on the outer part of the cluster increase the diameter. The average size of detached clusters is less than $\SI{5}{\micro\meter}$ across all disruptor levels which increases the likelihood of successful phagocytosis of the cell clusters 
\cite{Alhede20,Vahidkhah15}. However, since the standard deviation is very large, there will be clusters larger than $\SI{5}{\micro\meter}$ with greater frequency as EPS disruption increases.

To describe the shape of multi-agent clusters, we fit an ellipse with major axis $a$ and minor axis $b$ to the point cloud of agents. Linearity $\mathcal{L}$ is then calculated as
 \begin{equation}
     \mathcal{L}=\frac{a-b}{a},
     \label{eq:linearity}
 \end{equation}
where $\mathcal{L}\to 0$ when the cluster is circular, having nearly equivalent major and minor axes. Similarly, $\mathcal{L}\to1$ when there is only one primary direction that describes most of the variance. As observed in Fig.~\ref{fig:2dSummary}c, averaging over time and all simulations reveals only a small increase in overall average linearity from the case of no disruptor to 95\% of the EPS disrupted ($\mathcal{L}=$0.62 for 0\% and 0.71 for 95\%). 

Next, we investigate cluster orientation angle of detached clusters. Unlike with experiments, in simulations, we can track the height of where a cluster broke off, as well as the orientation of that cluster in the biofilm. We examine cluster orientation with respect to the $y$-axis, which is represented via the angle $\varphi$ as described in Section \ref{sec:detach_model}. A positive angle is oriented towards the direction of the flow whereas a negative angle is oriented away from the direction of the flow. On average, most clusters are oriented away from the flow (Fig.~\ref{fig:2dSummary}d). Furthermore, there is a nonmonotonic change in the average angle of cell clusters, reaching a maximum at 50\% disruption. EPS disruption at 0\% and 95\%  have similar average angles of -20.4$^{\circ}$ and -17.3$^{\circ}$, respectively.

\begin{figure}
    \centering
\includegraphics[width=0.99\linewidth,height=1.75in]{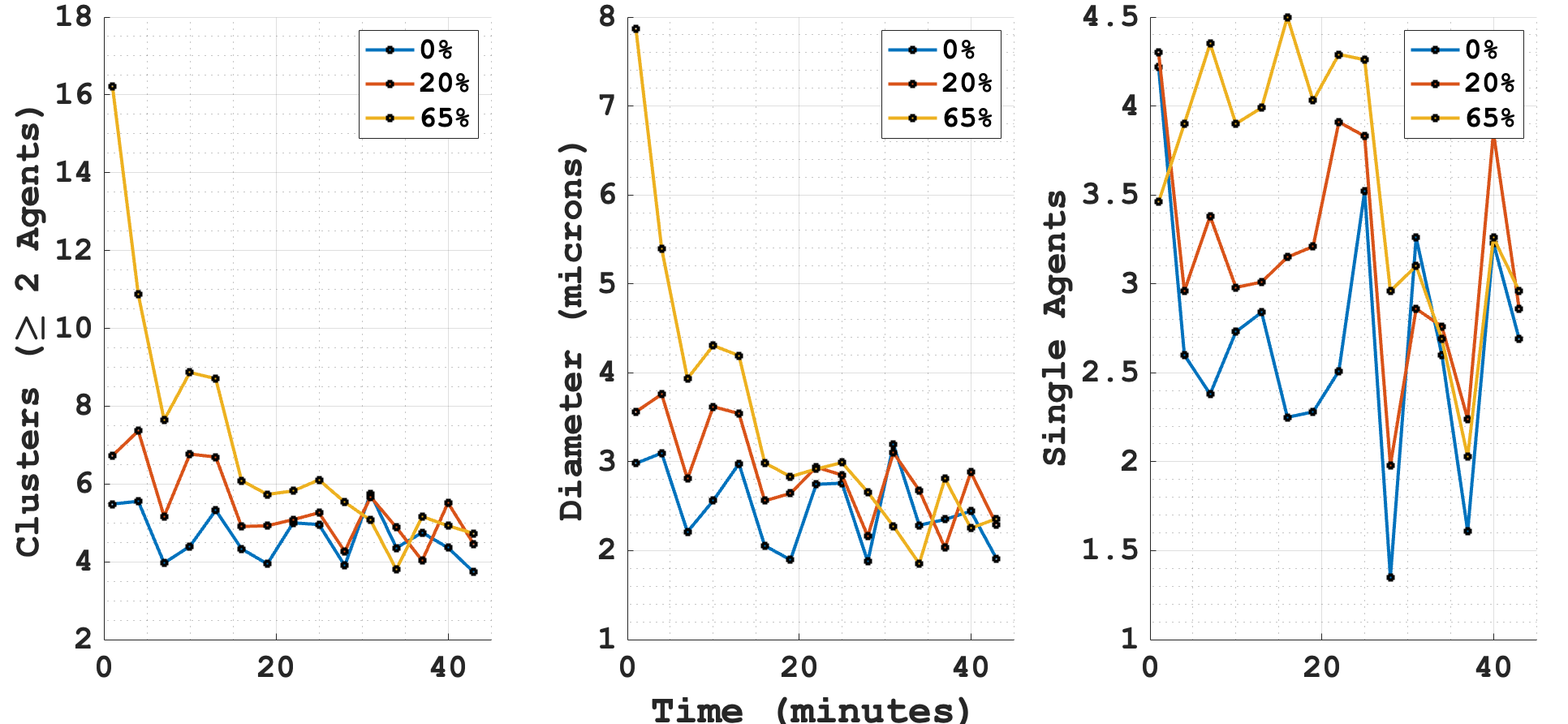} \\
\hspace{.2in} (a) Number of clusters \hspace{0.65in} (b) Cluster size \hspace{0.5in} (c) Single agent detachment
    \caption{Time dependence of clusters and single agents breaking off from the 2-d biofilm. The (a) number of clusters that detach, (b) cluster diameter, and (c) number of single agents that detach are all shown over time. Each point is an average over 100 simulations (10 runs each with 10 seeds). Results are shown for EPS disruptions at 0\%, 20\%, and 65\%.}
    \label{fig:TimeCluster}
\end{figure}

We now determine if there is any time-dependence to cluster geometry and frequency of detachment 
across three EPS disruption levels. 
The total number of clusters ($\geq 2$ agents) that detach decreases over time (Fig. \ref{fig:TimeCluster}a), eventually reaching an oscillatory steady state that is very similar across disruptor levels. The time to reach this state correlates with the time at which the average biofilm height stabilizes (Fig.\ref{fig:2dSummary}a). We see a similar pattern with cluster size (Fig.\ref{fig:TimeCluster}b): initially larger cluster sizes that decrease over time to an oscillatory steady state, 2-3 $\SI{}{\micro\meter}$ in this case. We note that the initial biofilm was grown for 24 hours with only standard erosion being applied and the EPS was not compromised (100\% healthy EPS). When we start the cluster formation and detachment algorithm at time zero, we assume that the EPS disruption is constant throughout the biofilm and occurs instantaneously. This is likely why we have larger clusters and more clusters detaching in the first few minutes at higher disruptor levels. We also track the detachment of single agents over time (Fig. \ref{fig:TimeCluster}c). This oscillates in time and has a lower initial value for the highest disruption level (in comparison to Fig.~\ref{fig:TimeCluster}a,b). 

\begin{figure}
\centering
\includegraphics[width=0.99\linewidth]{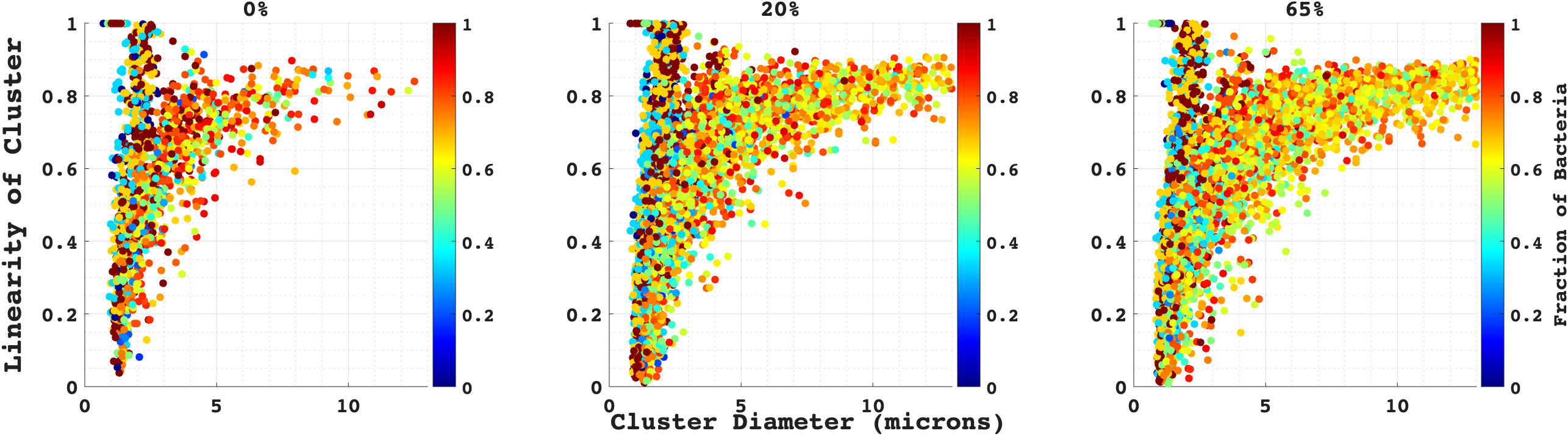} \\
(a) 0\% disruption \hspace{0.5in} (b) 20\% disruption \hspace{0.5in} (c) 65\% disruption
\caption{Linearity of clusters that detach as a function of cluster diameter for 2-d biofilms. Results are shown for three different levels of EPS disruption: (a) 0\%, (b) 20\%, and (c) 65\%. Each shaded circle represents a detached cluster across all time points and 100 simulations. The circle is colored by the fraction of bacteria agents in the cluster.}
\label{fig:LinDiamBacFrac}
\end{figure}

We next evaluate the correlation among cluster characteristics. We calculate the Pearson correlation coefficient using complete time series data from all 100 simulations (10 runs each with 10 seeds). Cluster diameter and linearity are the only characteristics to display a linear relationship. For all disruptors in these cases, correlation was in the range of -0.0334 to 0.0414, compared to correlation values of 0.4016, 0.4429, and 0.5349 for 0\%, 20\%, and 65\% EPS disruption between cluster diameter and linearity. To investigate this relationship further, we plot linearity with respect to diameter of detached clusters (Fig. \ref{fig:LinDiamBacFrac}). We observe that larger clusters have a high linearity (around 0.8) while smaller clusters can attain a wider possible range of values. This trend is consistent across 0\%, 20\% and 65\% EPS disruption levels; at 65\% disruption, clusters of diameter greater than 10   $\SI{}{\micro\meter}$ have a slightly larger range of linearity (0.7-1 versus 0.75-9 for 0\%). When considering cluster composition, we see that the fraction of bacteria in smaller clusters can vary greatly while larger clusters tend to have high fractions of bacteria. We note that at 65\% EPS disruption (Fig. \ref{fig:LinDiamBacFrac}c) more large clusters detach and these clusters are bacteria-dominant (bacteria fraction $>0.5$).

\begin{figure}
    \centering
     \subfloat[][Linearity]{\includegraphics[width=0.5\linewidth]{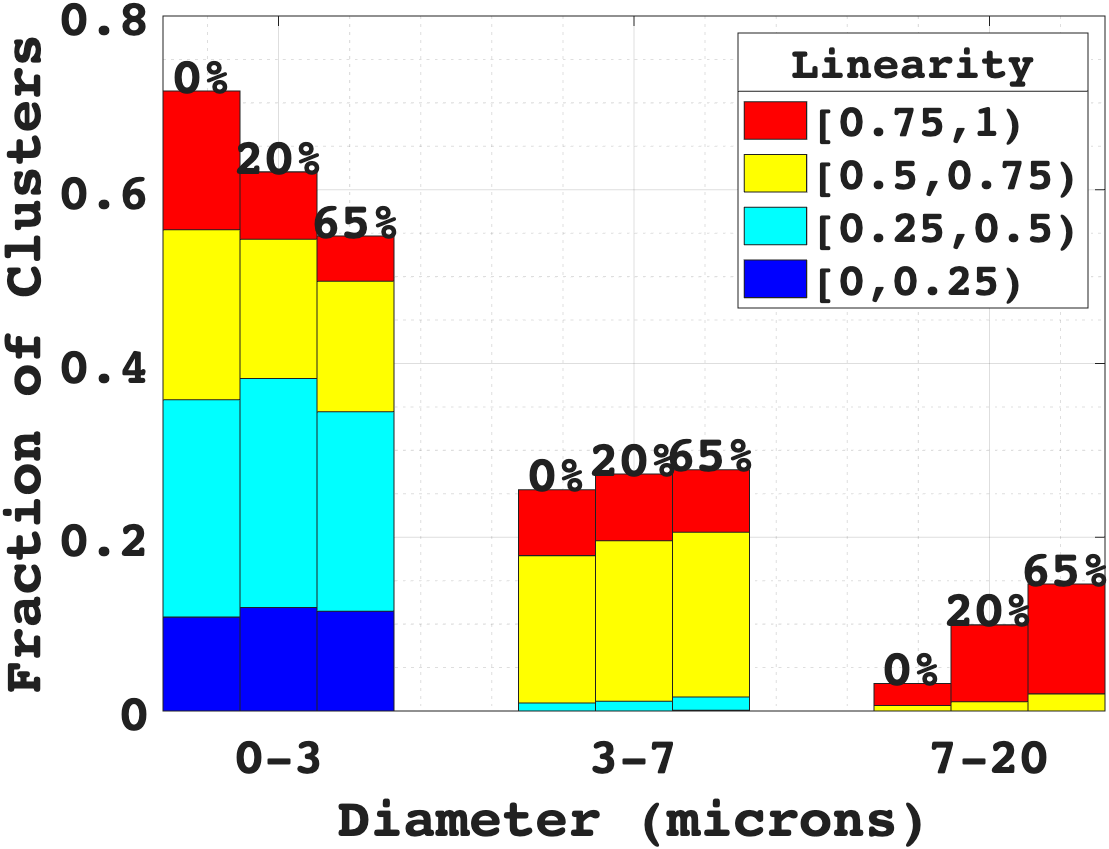}}
\subfloat[][Angle relative to the $y$-axis]{\includegraphics[width=0.5\linewidth]{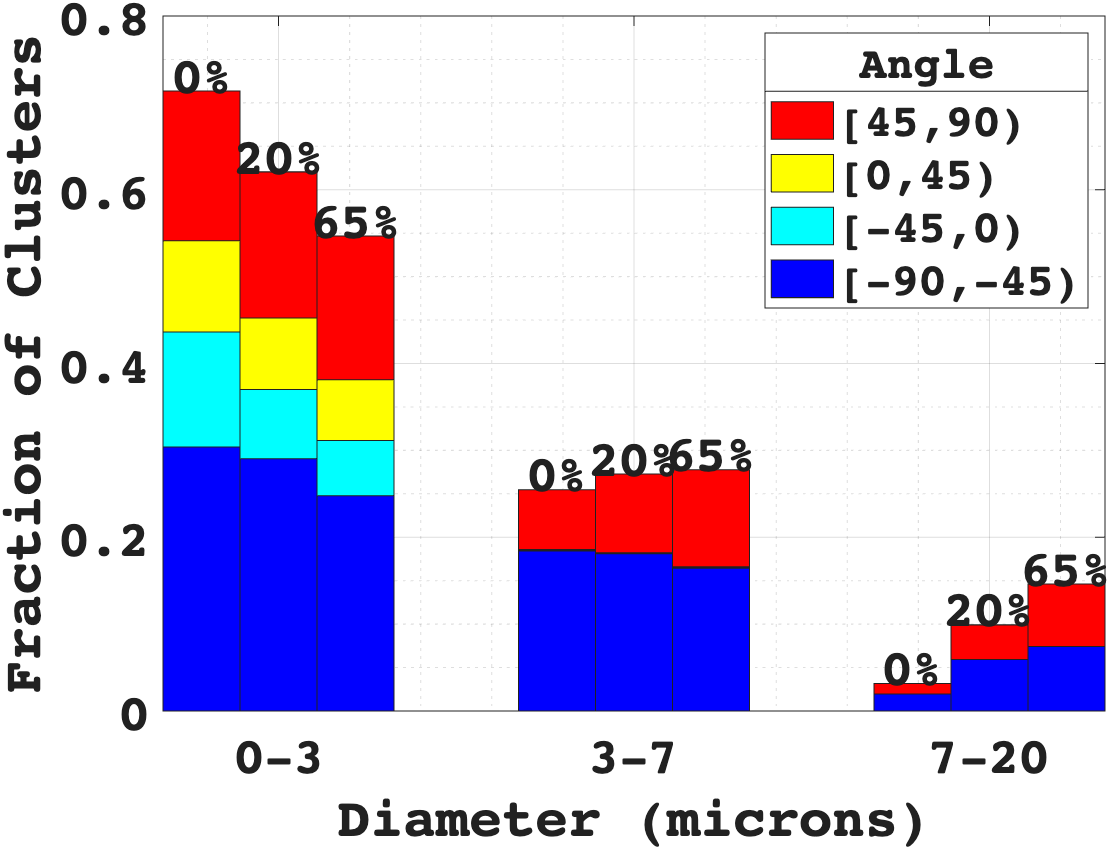}} \\ 
    \caption{Properties of detached clusters across cluster size groups for 2-d biofilms. Clusters are small, medium, or large (maximum Euclidean diameter ranges of 0-3, 3-7, or 7-20 $\SI{}{\micro\meter}$). Relative fraction of clusters is shown across (a) linearity and (b) angle relative to the $y$-axis. The three bars in each size group correspond to the three EPS disruptor levels of 0\%, 20\%, and 65\%, respectively.}
    \label{fig:PercentDiamLin}
\end{figure}

We further scrutinize the relationship between cluster linearity and diameter by examining these characteristics across qualitative size groups (Fig \ref{fig:PercentDiamLin}). We sort clusters into three categories: small (0-3 $\SI{}{\micro\meter}$), medium (3-7 $\SI{}{\micro\meter}$), and large (7-20 $\SI{}{\micro\meter}$). About $25\%$ of small clusters have a linearity $\mathcal{L}\in[0.25,0.5)$ for 0\%, 20\% and 65\% EPS disruption. Medium size clusters account for about $26\%$ of clusters that detach. Of that, the majority (about 65\%) of medium sized clusters have linearity in the range of [0.5,0.75). A higher disruptor level increases the number of larger clusters that detach, the majority of which have linearity $\mathcal{L}\in[0.75,1)$. In terms of detachment angle (Fig. \ref{fig:PercentDiamLin}b), we clearly see that larger clusters detach at an angle of [-90$^{\circ}$,-45$^{\circ}$), with a smaller fraction in the [45$^{\circ}$,90$^{\circ}$) range.  Neither medium nor large clusters detach at shallower angles (between [-45$^{\circ}$, 45$^{\circ}$)). In contrast, smaller clusters may detach at a range of angles, however detachment is more common at a larger angle.

\begin{figure}
    \centering
    \subfloat[][Average Biofilm Height]{\includegraphics[width=0.49\linewidth]{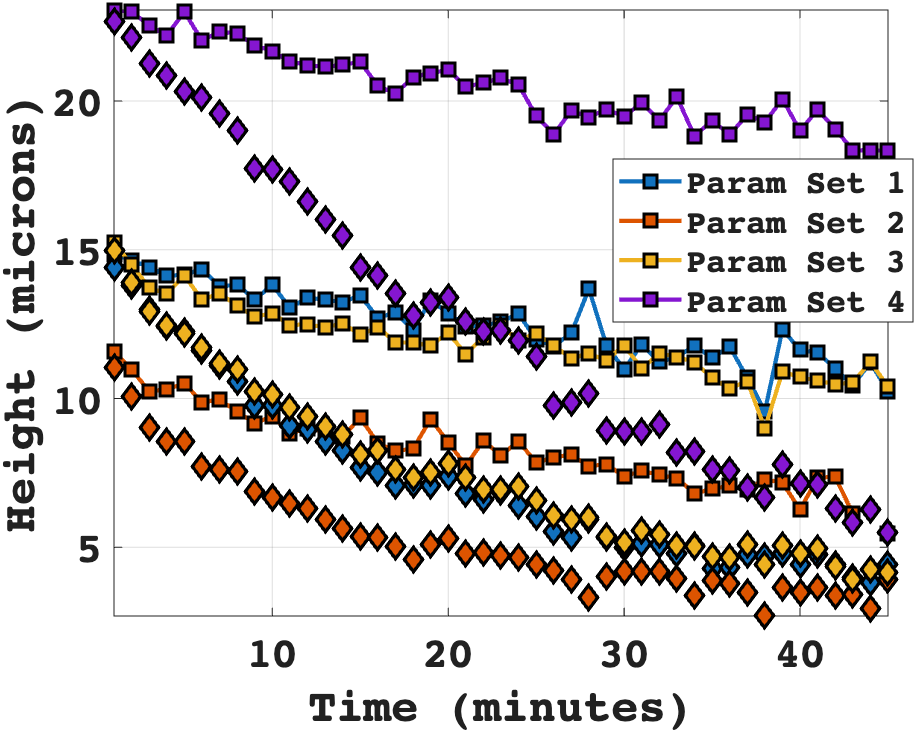}}
    \subfloat[][Cluster Ratio]{
    \includegraphics[width=0.49\linewidth]{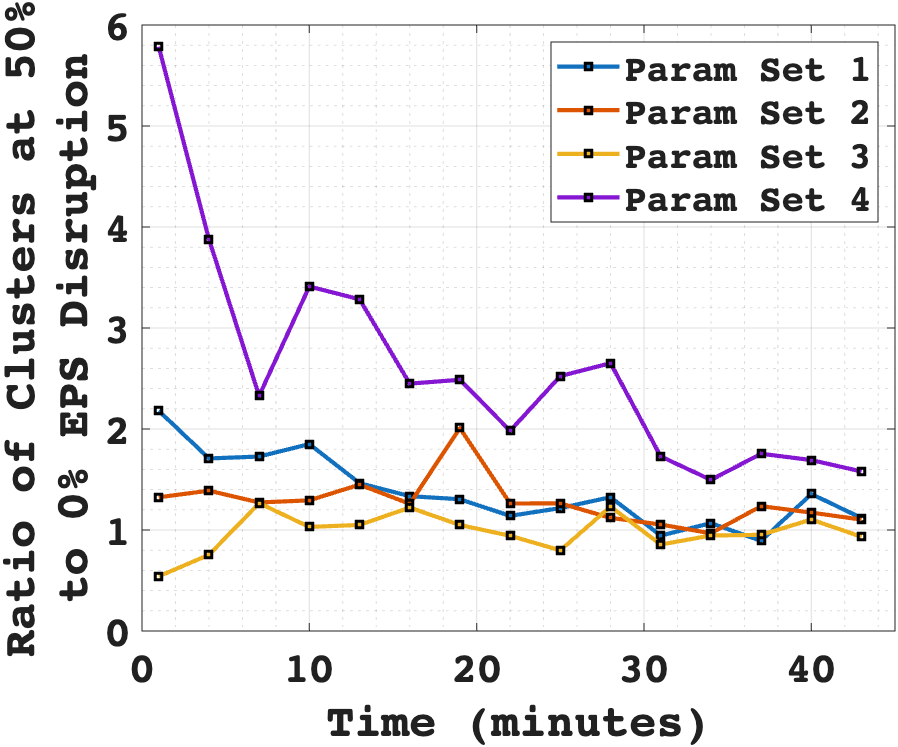}}\\
    \subfloat[][Linearity]{\includegraphics[width=0.49\linewidth]{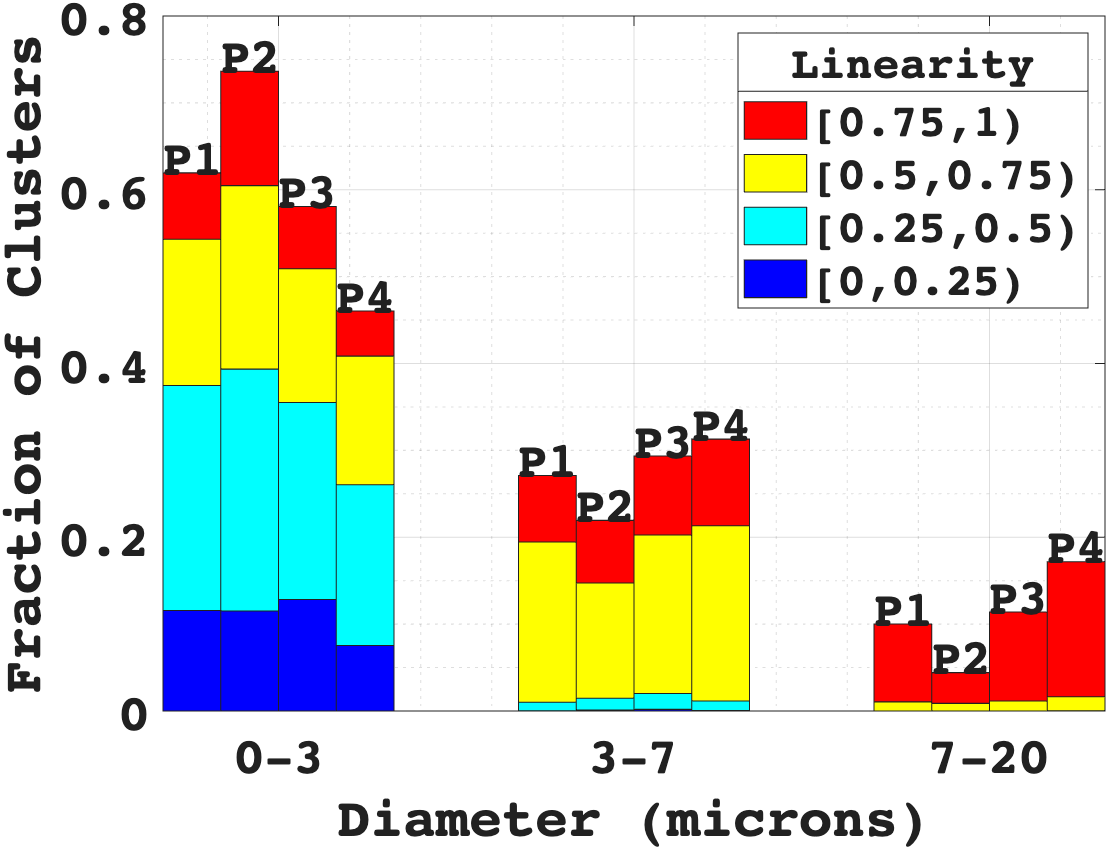}}
    \subfloat[][Angle relative to the $y$-axis]{
    \includegraphics[width=0.49\linewidth]{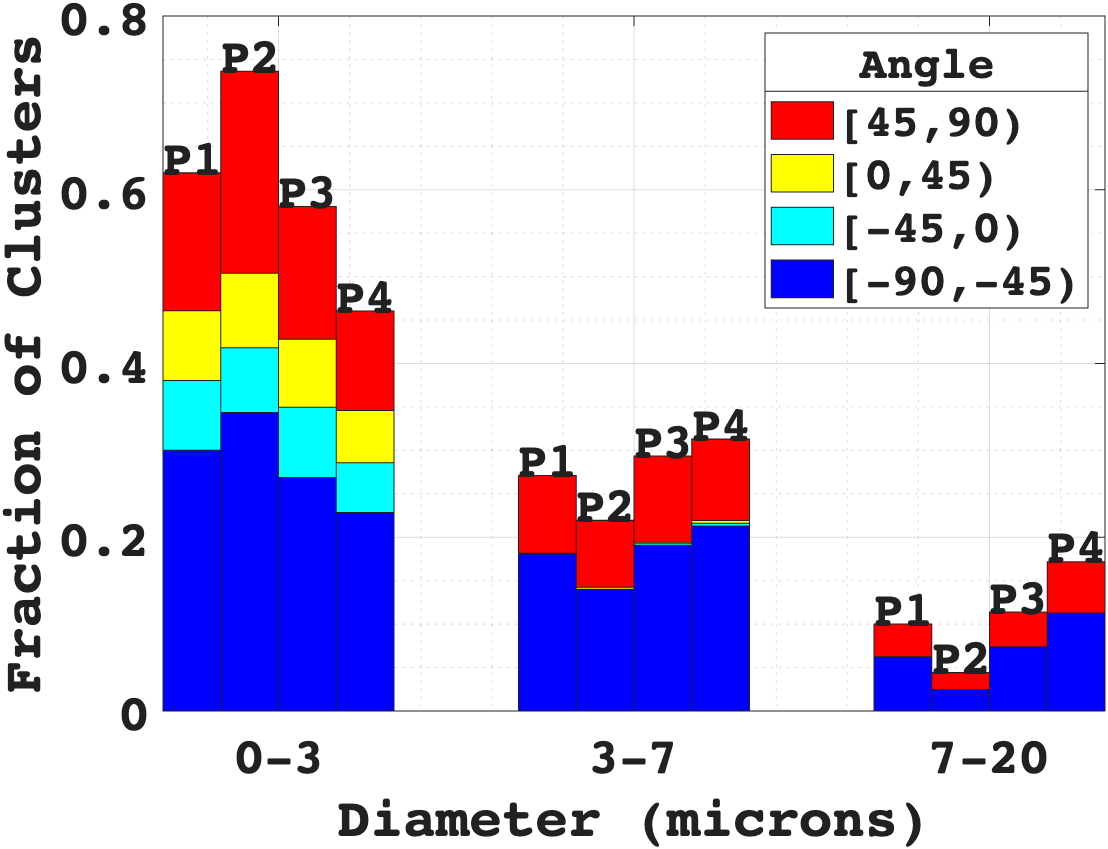}}
    \caption{Cluster properties for 2-d biofilms using Parameter Sets 1-4 (Table \ref{tab:iDynomics_param_sets}). (a) Average biofilm height when clusters detach with respect to time, colored by parameter set. No EPS disruption (0\%) is denoted with a $\square$ and \begin{Large}$\diamond$\end{Large} denotes 50\% EPS disruption. (b) Ratio of the average number of clusters ($\geq 2$ cells) breaking off in the 50\% EPS disruption to the 0\% EPS disruption case. Relative fractions of clusters are shown across (c) linearity and (d) angle relative to the $y$-axis. In (c)-(d), clusters are binned by size as small, medium, or large (maximum Euclidean diameter ranges of 0-3, 3-7, or 7-20 $\SI{}{\micro\meter}$).}
    \label{fig:2dparamcomp}
\end{figure}

Finally, we explore the robustness of trends across different parameter sets since all previous results pertain to a simulated biofilm that uses Parameter Set 1. Once more, we begin by plotting the average biofilm height at cluster detachment over time for each parameter set and two EPS disruption levels (0\% and 50\%). As seen in Fig. \ref{fig:2dparamcomp}a, all parameter sets with 50\% disruption display a large reduction in biofilm height. The tallest biofilm initially (Parameter Set 4) had a 20.47\% decrease in height for 0\% EPS disruption and 75.8\% decrease for 50\% disruption. Analogously, Parameter Set 1 exhibits a 31.16\% decrease in height for 0\% EPS disruption and 69.29\% for 50\% disruption. Parameter Sets 1 and 3 simulate 24-hour biofilms with different properties (1=low erosion and EPS yield whereas 3=moderate erosion and EPS yield), yet the resultant height time series closely mimic each other. Next, we study the impact that biofilm height has on the number of clusters that detach over time. We do this by calculating the ratio of the number of clusters that detach under 50\% EPS disruption to 0\% disruption over time (Fig. \ref{fig:2dparamcomp}b). The majority of parameter sets (1, 2, 4) have more clusters detach from biofilms under 50\% disruption (ratios $>1$). The average ratio of clusters that break off per minute stabilizes around 1 for Parameter Sets 1, 2, and 3 at the same time the biofilm height stabilizes (see Fig.~\ref{fig:2dparamcomp}a). Parameter Set 4 biofilm height is decreasing throughout the 45 minutes; thus, we observe that the ratio of clusters is also decreasing on average. We also analyze cluster diameter, linearity, and angle relative to the $y$-axis over all parameter sets. In Fig.\ref{fig:2dparamcomp}c-d, we observe that the tallest biofilm (Parameter Set 4) has the lowest fraction of small cell clusters and the highest fraction of larger cell clusters. The opposite is true for the shortest biofilm (Parameter Set 2), which has the largest amount of smaller clusters and the smallest fraction of larger clusters. The trends for linearity are consistent; larger clusters are more linear whereas smaller clusters have a range of linearity values (Fig.\ref{fig:2dparamcomp}c). Furthermore, cluster orientation is stable across parameter sets; larger clusters are generally oriented further from the $y$-axis.

\subsection{Results for 3-d Biofilms}
\label{sec:3d_results}
\begin{figure}[h]
    \centering
    \subfloat[][Average Biofilm Height]{\includegraphics[width=0.5\linewidth]{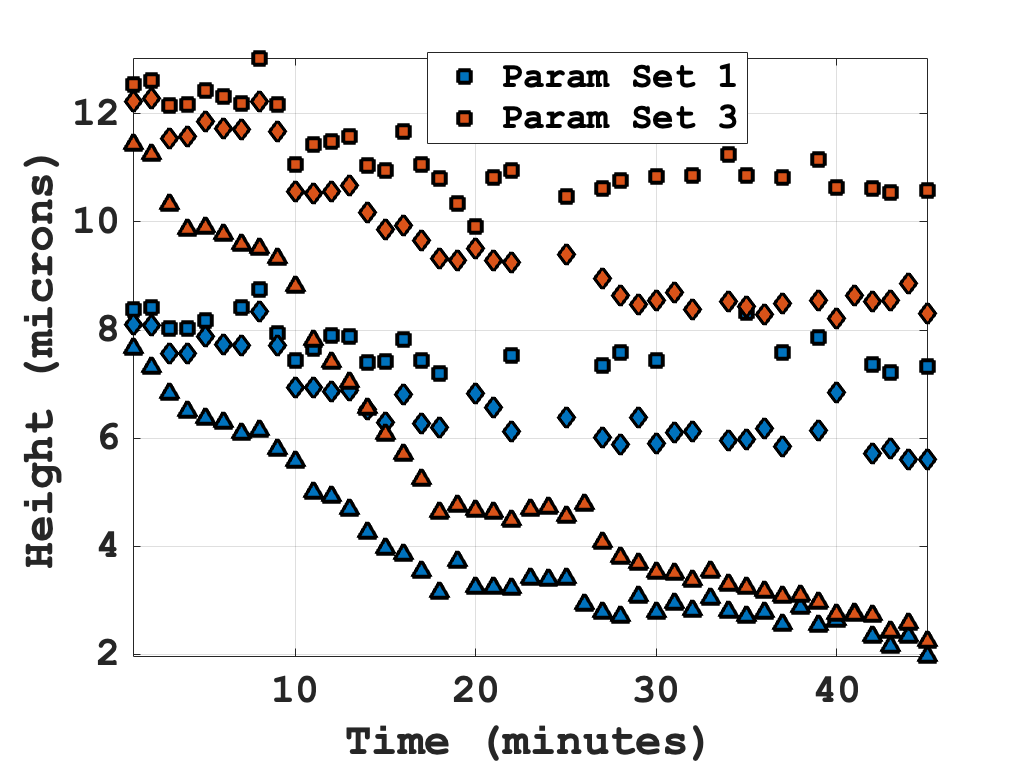}}
    \subfloat[][Shape Anisotropy]{\includegraphics[width=0.5\linewidth]{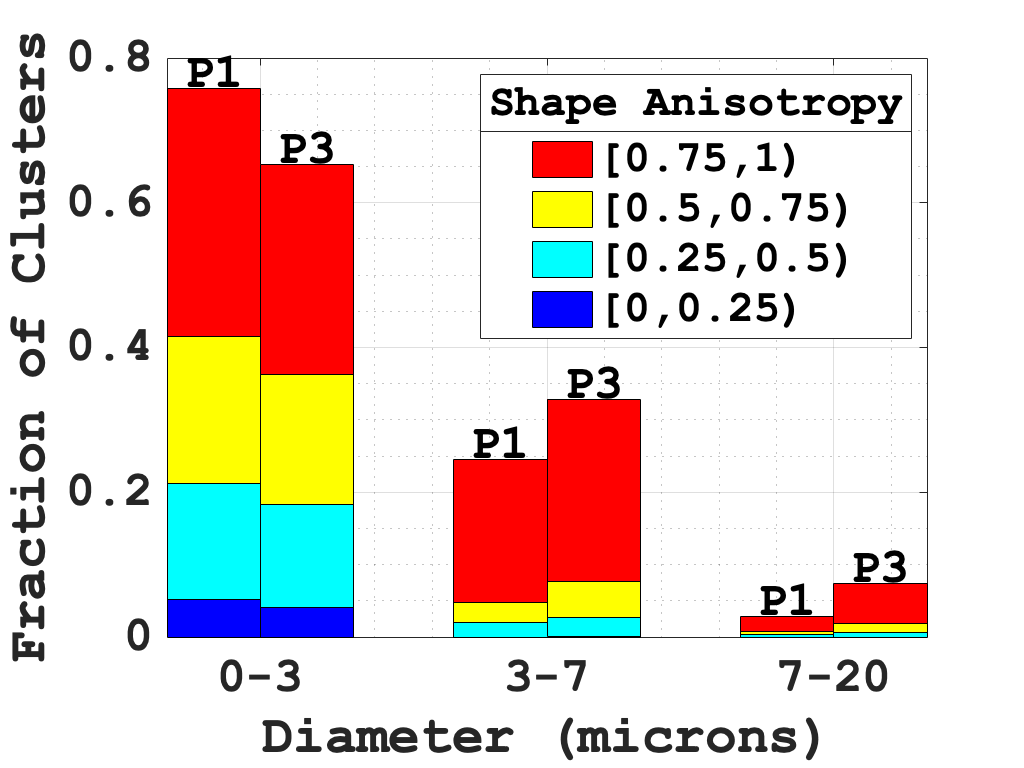}} \\
    \subfloat[][Angle relative to the $y$-axis]{\includegraphics[width=0.5\linewidth]{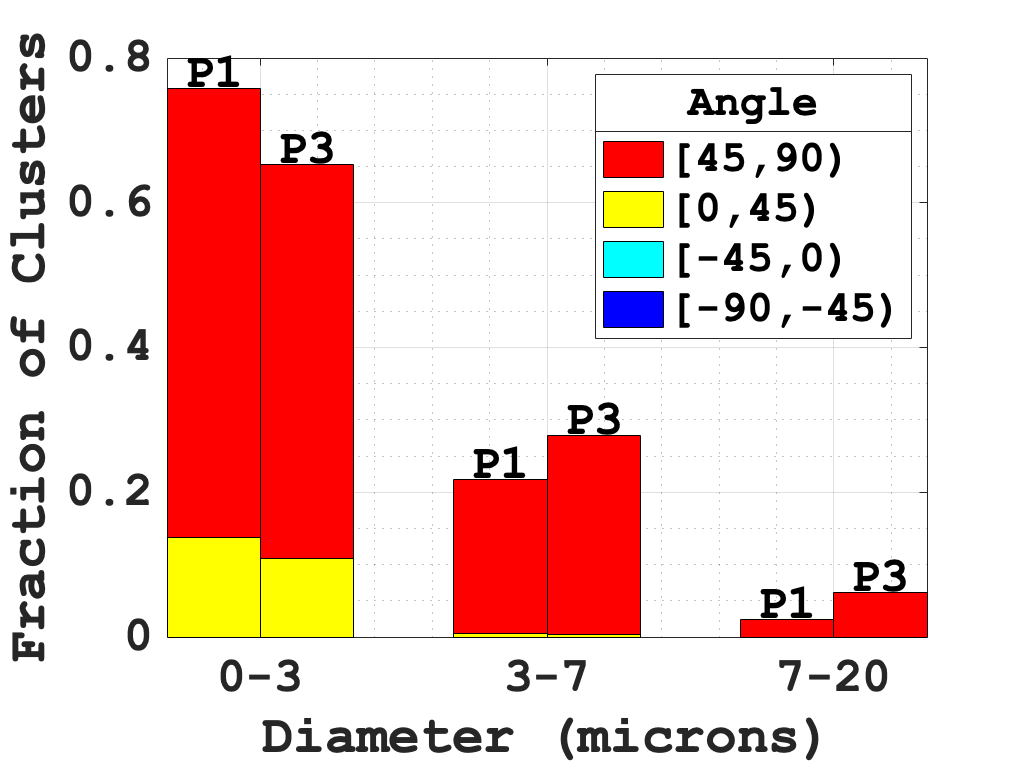}} 
    \subfloat[][Prolateness]{\includegraphics[width=0.5\linewidth]{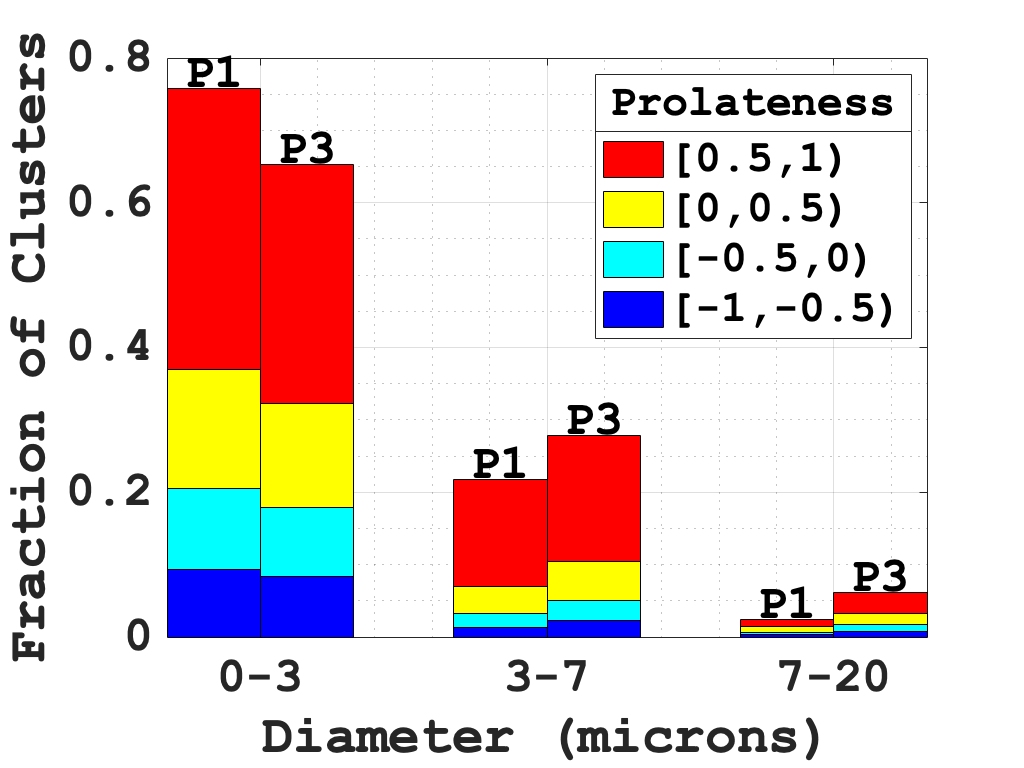}} 
    \caption{Cluster properties of 3-d biofilms using Parameter Sets 1 (P1) and 3 (P3). (a) Average biofilm height when clusters detach over time, where $\square$ is 0\%, \begin{Large}$\diamond$\end{Large} is 20\%, and $\triangle$ is 65\% EPS disruption. The (b) shape anisotropy, (c) angle relative to the $y$-axis, and (d) prolateness under 65\% EPS disruption are grouped by small, medium, and large clusters (0-3, 3-7, and 7-20 $\SI{}{\micro\meter}$).}
  \label{fig:3d_detach}
\end{figure}

As previously discussed, we have extended our modeling framework and are able to simulate cluster detachment from 3-d biofilms. Most calculations remain the same and only minor updates are required when moving from 2-d to 3-d, though the increased complexity greatly adds to simulation time. Given this constraint, we present results only for Parameter Sets 1 and 3, which are summarized in Fig. \ref{fig:3d_detach}. We examine the same characteristics from 2-d cases to portray the clusters that detach in 3-d cases, though some characteristics require augmentations to account for the added dimension. First, we consider average biofilm height at time of detachment in Fig. \ref{fig:3d_detach}(a); as with 2-d, height decreases over time. 

Next, we regard linearity. The 3-d analog of linearity is shape anisotropy, defined as
\begin{equation}\label{eq:shapeanisotropy}
\mathcal{S}^2=\frac{3}{2}\frac{\lambda_1^4+\lambda_2^4+\lambda_3^4}{(\lambda_1^2+\lambda_2^2+\lambda_3^2)^2}-\frac{1}{2},
\end{equation}
where eigenvalues of the covariance matrix of agent locations are given by $\lambda_1^2$, $\lambda_2^2$, and $\lambda_3^2$ (in descending order). Anisotropy is bounded $0\leq \mathcal{S}^2\leq 1$, with $\mathcal{S}^2=0$ for a cluster of spherically symmetric agents and $\mathcal{S}^2=1$ when the agents lie on a line. We observe all clusters are generally more spherical (Fig. \ref{fig:3d_detach}b). Further, smaller clusters (0-3 $\SI{}{\micro\meter}$ diameter) have 15 to 20\% of the clusters that are more linear. In 3-d, the angle relative to the $y$-axis is strongly skewed towards the $[45^{\circ}, 90^{\circ})$ range across all cluster size groups (Fig. \ref{fig:3d_detach}c). 

Finally, we further study the shape of non-spherical clusters. In 3-d, these clusters range from oblate 
to prolate. 
To capture this difference, we calculate the prolateness \cite{Rawdon},
\begin{equation}
\mathcal{P}=\frac{(2a-b-c)(2b-a-c)(2c-a-b)}{2(a^2+b^2+c^2-ab-ac-bc)^{3/2}},
\end{equation}
for major axis $a$ and minor axes $b,c$. Here, $\mathcal{P}>0$ corresponds to a prolate ellipsoid and $\mathcal{P}<0$ corresponds to an oblate ellipsoid.  We find that most clusters that detach are prolate. Oblate clusters, of which there are few, are observed more frequently for smaller clusters.

\section{Discussion}
\label{sec:discussion}

Towards the ultimate goal of predicting detachment of clusters from \textit{S. epidermidis} biofilms, we develop a mathematical framework (Fig. \ref{fig:framework}) that grows 24-hour-old \textit{in silico} biofilms and then simulates detachment over 1 hour. 
We benchmark the 24-hour-old biofilm 
with structural features from experiments \cite{Jonblat2024,packard2026biophysical,Cerca2012,Stewart13,Ganesan13}. We present results from different detachment scenarios, simulated via our clustering algorithm and detachment
model, and find that most detached clusters are small, with a diameter between 0-3 $\SI{}{\micro\meter}$. 

Our modeling framework exhibits novelty in three
key ways.  First, we utilize experimentally measured biofilm structural features including biofilm height, cellular number density, cellular clustering, and EPS content to benchmark the initial 24-hour-old biofilm. Second, our model explicitly reflects the connection between the spatial positioning of EPS and bacteria locations with EPS stickiness (and level of EPS disruption) when making detachment decisions. Third, our detachment model provides insight into the temporal dynamics of detachment that are difficult, if not impossible, to determine through experiments. 
Our model is also readily adaptable to evaluating detachment from biofilms formed by other species of bacteria when updated with relevant, species-specific parameters. 
We further discuss characteristics of detached clusters, novel modeling aspects, limitations, and model extensions in the following sections.

\subsection{Cluster Detachment Characteristics}
\label{sec:disc_cluster_detach_char}

The size and shape of detached clusters critically affect their biological fate—smaller, spherical clusters are more readily cleared by the immune system, whereas larger or oblate aggregates ($>\SI{10}{\micro\meter}$) can evade immune clearance and are more likely to adhere to vessel walls than smaller or more spherical clusters \cite{Alhede20,Vahidkhah15}. Our results show that more than 80\% of clusters that detached from our \textit{in silico} biofilms are $\leq \SI{7}{\micro\meter}$ in diameter (Fig. \ref{fig:PercentDiamLin}); this is true across all simulated EPS disruption levels. For comparison, experiments have reported the size of \textit{S. epidermidis} biofilm-detached clusters and classified nearly all (98.7\%) clusters as small or medium  and less than $\SI{3.4}{\micro\meter}$ on average \cite{packard2026biophysical}; only clusters in the large category (1.3\%) had diameters above $\SI{7}{\micro\meter}$. Further, our simulations show that average cluster diameter decreases over time (Fig.~\ref{fig:TimeCluster}b) and levels off at or below $\SI{3}{\micro\meter}$. 
We do observe some larger detaching clusters, a greater percentage of which are linear (Figs.~\ref{fig:TimeCluster}, \ref{fig:PercentDiamLin}).  
This means that many of these larger non-spherical clusters of bacteria and EPS could evade immune clearance and would have an increased adherence to walls. Cluster shape also strongly impacts the detachment decision as highly spherical particles are less likely to detach \cite{ting2023detachment}. We study shape via linearity (2-d case) and shape anisotropy (3-d case), and find that in both cases larger detached clusters tended to have high linearity (Fig. \ref{fig:2dparamcomp}c) or shape anisotropy (Fig. \ref{fig:3d_detach}b), whereas smaller clusters exhibited the full range of possible values.  We also observe that most of our clusters are prolate (Fig.~\ref{fig:3d_detach}d), contradicting an experimental study that found the majority of \textit{S. epidermidis} biofilm-detached cell clusters (79.8\%) to be oblate \cite{packard2026biophysical}. Perhaps this discrepancy is due in part to the fact that our 2-d geometric framework is  applied to 3-d, and a full treatment of the third dimension could yield different results. Another possibility is that specific biofilm matrix interactions within clusters may be required to accurately determine  the morphology of the biofilm-detached clusters. 

As mentioned, a novel component of our model is its consideration of geometry to determine clusters and make detachment decisions; these results yield insights not available in experiments. Ting et al. \cite{ting2023detachment} found that spheroidal particles are less likely to detach (from flat substrates) as orientation angle increases to 90$^{\circ}$ as measured from the vertical (parallel to the flow direction). Surprisingly, we find that more clusters  detach at a large angle relative to the $y$-axis (Fig. \ref{fig:PercentDiamLin}); smaller angles should promote detachment according to our theory (Section \ref{sec:detach_model}, in part based on Ting's work). Based on our observations (e.g., Fig. \ref{fig:TimeStep}), our cluster formation algorithm (Alg. \ref{Alg:tag_add_screen}) often determines clusters as having large angles with respect to the $y$-axis, so we may simply have a response bias. From our 3-d simulations and analysis, we find that detachment occurs mostly within orientation angle ranges of $[45^{\circ}, 90^{\circ})$, with a few smaller clusters detaching at angles between $[0^{\circ}, 45^{\circ})$. This is in contrast to our finding that 2-d detachment occurs across a range of orientation angles, although preference skews towards large angles. There is no reason to believe there is a bias to one direction in 3-d, and so perhaps this result is an artifact of how we apply our 2-d methods to the 3-d biofilms. 

Lastly, we comment on our findings in relation to the effect of EPS disruption, as this may guide efforts to dislodge biofilm clusters from medical devices using chemical treatment. We observe that the height of the biofilm when clusters detach negatively correlates with the amount of EPS disruption (Fig.~\ref{fig:2dSummary}a). Deeper study of this relationship provides rich future work. Interestingly, the time course of average biofilm height decreases roughly linearly for disruption levels of 0\%, 10\%, and 20\%; for 50\% and above, the rate of decrease leans exponential and steadies to zero after approximately 30 minutes. We also detect a positive correlation between cluster size at detachment and EPS disruption level for our 2-d biofilm simulations (Fig.~\ref{fig:2dSummary}b). Yet even at 95\% disruption, the average detached cluster size of $\SI{5.5}{\micro\meter}$ does not exceed the $\SI{10}{\micro\meter}$ threshold above which immune system evasion or vessel wall adherence is expected \cite{Alhede20,Vahidkhah15}. Cluster linearity also weakly correlates with EPS disruption level on average (Fig.~\ref{fig:2dSummary}c). 
We also emphasize that our models provide important insight into undisrupted biofilms (a disruption level of 0\%) as a baseline case, as few experimental studies have reported distributions of clusters detaching in this control scenario. Our results show that significant detachment still occurs in the zero disruption setting.

\subsection{Limitations}

Inherently, models require simplifying assumptions and here is no exception. Firstly, we assume that capsule EPS and independent EPS operate in similar ways. It has been shown that the role of capsule production is minor and species-specific \cite{Nunez2023}, however it remains unclear whether types of EPS should have different properties or model mechanisms. This factor could be explored in the future or may be more relevant for other models. Although we use adhesive moment model parameters from \cite{hartmann2019emergence} when possible, we extend the upper bound of the repulsion strength range and increase the attraction strength in our simulations to achieve a realistic balance of detachment decisions. While this is reasonable because \textit{S. epidermidis} bacteria cells have a radii that is approximately double of that studied by \cite{hartmann2019emergence}, this is a subjective element of our modeling that could be improved in the future with the availability of additional data. Further, model selection for the coefficient due to drag $C_d$ changes depending on the combination of elongation and flatness of clusters \cite{ganser1993rational,leith1987drag}. The biofilm clusters we identify are of varied aspect ratios, so we may not be able to select a single most-appropriate coefficient of drag model based on cluster shape. Also, during the one hour of detachment that we study, we are not accounting for bacteria and EPS dynamics, assuming that there is minimal additional bacteria cell division and EPS synthesis. This is an avenue for future work and comparison to experimental data. Finally, we assume clusters are perfectly elliptical or ellipsoidal. This allows us to use existing theory \cite{ting2021impact,ting2023detachment,maramizonouz2022drag,hartmann2019emergence} to efficiently determine drag and adhesive moments of clusters. Additionally, this assumption allows us to easily compute detached aggregate characteristics such as cluster size, linearity, and orientation angle. Naturally, this adds an amount of error to our measured quantities; although similar assumptions have also been made in experimental contexts \cite{packard2026biophysical}, leading to reasonable approximations.

\subsection{Extensions}

The current work standardizes many parameters (see Tables \ref{tab:cluster_form_params}- \ref{table:adhesion_force} and \ref{tab:iDynomics_parameters}-\ref{tab:iDynomics_parameters_2}) and makes assumptions to provide a tractable first model of biofilm cluster disruption and detachment that considers geometric factors. We discuss two possible extensions in detail.

\subsubsection{Matrix Disruption}

We note that the biofilm matrix composition can change over time and that compromising different components of the matrix may change our results, including which agents are tagged, which clusters are selected for detachment, etc. The current model does not differentiate which part of the matrix the disruptor targets. An immediate next step is to look at fractional components of the matrix  (e.g., measured percentages of components in a 24-hour-old biofilm have been found to comprise 52\% polysaccharides, 18.9\% protein, and around 11.6\% lipids \cite{Jonblat2024}) and connect various disruptors that target these components. A starting point could be using a matrix composition-based detachment amplitude \cite{pechaud2024modelling} that distinguishes polysaccharides and proteins.

Our current model assumes a constant proportion of healthy and compromised EPS, whose effect  we study by varying the percentage of EPS disruption. Following \cite{xavier2005biofilm}, assuming Michaelis-Menten kinetics, we could define a decay rate $r^*$ that acts on healthy EPS. In this way, the mass balance of healthy and compromised EPS becomes time-dependent. Additionally, disruptor concentration can be represented as a function of height of the agent and time.

In our model, the compromised ratio term within attachment strength $\beta_{A}$ for agent $A$ (Eq. \eqref{eq:attachment_strength}) is raised to the first power, corresponding to a linear dependence of biofilm mechanical resistance on the relevant structural component. This choice is motivated by parsimony and the lack of quantitative experimental evidence supporting a high order or sublinear relationship. Previous modeling studies \cite{xavier2005biofilm} have shown that treatment outcomes can be highly sensitive to the exponent governing the dependence of biofilm cohesiveness on EPS content, with qualitatively distinct behaviors emerging for values above and below one. In particular, exponents greater than one imply a strongly nonlinear increase in cohesiveness with structural fraction. This can lead to rapid loss of integrity once the EPS content is sufficiently reduced, thus potentially overestimating treatment effectiveness and predicting complete biofilm removal. Conversely, exponents below one correspond to a weaker dependence of cohesiveness on matrix composition, which may result in incomplete removal and even post-treatment regrowth following an initial decrease in biofilm mass. 
Our model avoids imposing either of these extreme nonlinear regimes in the absence of experimental constraints. Nevertheless, if the true biological relationship is superlinear 
or sublinear, 
the model may incorrectly estimate biofilm susceptibility or its capacity for recovery. Direct experimental measurements linking biofilm mechanical properties to structural composition would therefore be crucial for constraining this parameter and improving the predictive reliability of the model.

\subsubsection{Detachment Model}

Future work could further investigate drag, mobility, and ability of detached clusters to reattach based on different downstream forces. Drag forces on different ellipsoidal shapes have been previously studied \cite{MARAMIZONOUZ2022117964}; this may be extended to detached clusters from biofilms.

An important addendum is a full consideration of detachment in the 3-d setting. Our current method is to project clusters onto the 2-d plane that represents a cross-section through the flow. Ting et al. \cite{ting2021impact,ting2023detachment} consider either detachment in the 3-d setting or clusters oriented at an angle to the flow direction, but not both. To extend our ideas to 3-d clusters, we will need to measure either an azimuthal angle in the plane defined by the surface on which the biofilm sits, or rotational angles for roll, pitch, and yaw from the three axes. Then, multi-angle-dependent drag and adhesive shape factors will need to be designed with aspect ratio, orientation angle, and azimuthal angle (or roll, pitch, yaw) as inputs. These may be achieved by a combination of deriving analytical expressions from fluid mechanics principles and validation with experiment. It is also true that an ellipsoidal cluster boundary is not easily defined by an equation in 3-d, 
which is necessary for testing which agents are part of the cluster and which are neighbors; this can be addressed by reversing the rotations and shifting the ellipsoid back to the origin before examining the standard equation of an ellipsoid.

\section*{Code Availability}

Code is available at this \href{https://drive.google.com/drive/folders/19o7uWDcUBHtMvN2crrNxq-4I8LeCFqJi?usp=share_link}{Google Drive folder}. 

\section*{Acknowledgments}
This material is based upon work supported by the National Science Foundation under Grant No. DMS-1929284 while the authors were in residence at the Institute for Computational and Experimental Research in Mathematics (ICERM) in Providence, RI, during the Women in Mathematical Computational Biology program held in January 2025 and as a Collaborate@ICERM group in June 2025.  JAFK was supported by Los Alamos National Laboratory LDRD 20230853PRD2.

\bibliographystyle{spphys}
\bibliography{reference}

@article{Alhede20,
author={Alhede, M and Lorenz, M and Fritz, BG and Jensen, PO and Ring, HC and Bay, L and Bjarnsholt, T}, 
title={Bacterial aggregate size determines phagocytosis efficiency of polymorphonuclear leukocytes}, 
journal={Med Microbiol Immunol}, 
year=2020, 
volume=209,
pages="669--680",
}

@article{Bjarnsholt13,
title={The in vivo biofilm},
author={Bjarnsholt, T and Alhede, M and Eickhardt-Sørensen, SR and Moser, C and Kuhl, M and Jensen, PO and Hoiby, N},
journal={Trends Microbiol},
year=2013,
volume=21,
number=9,
}

@article{Rawdon,
author={Rawdon, EJ and Kern, JC and Piatek, M and Plunkett, P and Stasiak, A and Millett, KC},
title={Effect of knotting on the shape of polymers},
journal={Macromolecules},
year=2008,
volume=41,
pages="8281--8287",
}

@article{Stewart20,
author={Stewart, PS}, 
title={Antimicrobial Tolerance in Biofilms},
journal={Microbiol Spectr},
year=2015, 
volume=3,
number=3,
doi={10.1128/microbiolspec.MB-0010-2014},
}

@article{Qin,
author={Qin, Z and Ou, Y and Yang, L and Zhu, Y and Tolker-Nielsen, T and Molin, S and Qu, D}, 
title={Role of autolysin-mediated DNA release in biofilm formation of Staphylococcus epidermidis},
journal={Microbiol}, 
year=2007, 
volume=153,
pages="2083--2092", 
doi={10.1099/mic.0.2007/006031-0},
}

@article{Hammond14,
title={Variable Viscosity and Density Biofilm Simulations using an Immersed Boundary Method, {P}art {I}: {N}umerical Scheme and Convergence Results},
author={Hammond, JF and Stewart, EJ and Younger, JG and Solomon, MJ and Bortz, DM},
volume=98, 
issue=3,  
journal={J Comp Phys},
year=2014,
pages="295--340",
doi={10.3970/cmes.2014.098.295},
}

@Article{Li24,
author ="Li, C and Nijjer, J and Feng, L and Zhang, Q and Yan, J and Zhang, S",
title  ="Agent-based modeling of stress anisotropy driven nematic ordering in growing biofilms",
journal  ="Soft Matter",
year  ="2024",
volume  ="20",
issue  ="16",
pages  ="3401-3410",
doi  ="10.1039/D3SM01535A",
}

@article{Franca16,
author={Franca, A and Carvalhais, V and Vilanova, M and  Pier, GB and Cerca, N},
title={Characterization of an in vitro fed-batch model to obtain cells released from S. epidermidis biofilms},
journal={AMB Express},
volume=6, 
number=23,
year=2016,
doi={10.1186/s13568-016-0197-9},
}

@article{Ganesan13,
author = {Ganesan, Mahesh and Stewart, Elizabeth J. and Szafranski, Jacob and Satorius, Ashley E. and Younger, John G. and Solomon, Michael J.},
title = {Molar Mass, Entanglement, and Associations of the Biofilm Polysaccharide of Staphylococcus epidermidis},
journal = {Biomacromolecules},
volume = {14},
number = {5},
pages = {1474-1481},
year = {2013},
doi = {10.1021/bm400149a},
}

@article{ganser1993rational,
  title={A rational approach to drag prediction of spherical and nonspherical particles},
  author={Ganser, Gary H},
  journal={Powder Technol},
  volume={77},
  number={2},
  pages={143--152},
  year={1993},
  publisher={Elsevier}
}

@article{leith1987drag,
  title={Drag on nonspherical objects},
  author={Leith, David},
  journal={Aerosol Sci Tech},
  volume={6},
  number={2},
  pages={153--161},
  year={1987},
  publisher={Taylor \& Francis}
}

@article{Karygianni20,
title = {Biofilm Matrixome: Extracellular Components in Structured Microbial Communities},
journal = {Trends Microbiol},
volume = {28},
number = {8},
pages = {668-681},
year = {2020},
author = {Karygianni, L and Ren, Z and Koo, H and Thurnheer, T},
}

@article{Cockx24,
    author = {Cockx, BJR AND Foster, T AND Clegg, RJ AND Alden, K AND Arya, S AND Stekel, DJ AND Smets, BF AND Kreft, JU},
    journal = {PLOS Comp Biol},
    title = {Is it selfish to be filamentous in biofilms? Individual-based modeling links microbial growth strategies with morphology using the new and modular iDynoMiCS 2.0},
    year = {2024},
    volume = {20},
    pages = {1-32},
}

@article{Lardon11,
author={Lardon, LA and Merkey, BV and Martins, S and Dotsch, A and Picioreanu, C and Kreft, JU and Smets, BF}, 
title={iDynoMiCS: next-generation individual-based modelling of biofilms}, 
journal={Environ Microbiol}, 
year=2011,
volume=13,
pages="2416--34",
}

@article{Dzianach,
author={Dzianach, P.A. and Dykes, G.A. and Strachan, N.J. and Forbes, K.J. and Pérez-Reche, F.J.}, 
year=2019, 
title={Challenges of biofilm control and utilization: lessons from mathematical modelling}, 
journal={J R Soc Interface}, 
volume=16,
number=155,
pages=20190042,
}

@article{Picioreanu04, 
title={Advances in mathematical modeling of biofilm structure}, 
volume={1}, 
journal={Biofilms}, 
author={Picioreanu, C and Xavier, JB and van Loosdrecht, MCM}, 
year={2004}, 
pages={337–-349},
}

@article{xavier2005biofilm,
  title={Biofilm-control strategies based on enzymic disruption of the extracellular polymeric substance matrix--a modelling study},
  author={Xavier, Joao B and Picioreanu, Cristian and Rani, Suriani Abdul and van Loosdrecht, Mark CM and Stewart, Philip S},
  journal={Microbiol},
  volume={151},
  number={12},
  pages={3817--3832},
  year={2005},
  publisher={Microbiology Society}
}

@article{xavier2005framework,
  title={A framework for multidimensional modelling of activity and structure of multispecies biofilms},
  author={Xavier, Joao B and Picioreanu, Cristian and Van Loosdrecht, Mark CM},
  journal={Environ Microbiol},
  volume={7},
  number={8},
  pages={1085--1103},
  year={2005},
  publisher={Wiley Online Library}
}

@article{ting2021impact,
  title={Impact of shape on particle detachment in linear shear flows},
  author={Ting, Heng Zheng and Bedrikovetsky, Pavel and Tian, Zhao Feng and Carageorgos, Themis},
  journal={Chem Eng Sci},
  volume={241},
  pages={116658},
  year={2021},
  publisher={Elsevier}
}

@article{ting2023detachment,
  title={Detachment of inclined spheroidal particles from flat substrates},
  author={Ting, Heng Zheng and Yang, Yutong and Tian, Zhao Feng and Carageorgos, Themis and Bedrikovetsky, Pavel},
  journal={Powder Technol},
  volume={427},
  pages={118754},
  year={2023},
  publisher={Elsevier}
}

@article{ting2022image,
  title={Image interpretation for kaolinite detachment from solid substrate: Type curves, stochastic model},
  author={Ting, Heng Zheng and Yang, Yutong and Tian, Zhao Feng and Carageorgos, Themis and Bedrikovetsky, Pavel},
  journal={Colloids Surf A: Physicochem Eng Asp},
  volume={650},
  pages={129451},
  year={2022},
  publisher={Elsevier}
}

@article{hartmann2019emergence,
  title={Emergence of three-dimensional order and structure in growing biofilms},
  author={Hartmann, Raimo and Singh, Praveen K and Pearce, Philip and Mok, Rachel and Song, Boya and D{\'\i}az-Pascual, Francisco and Dunkel, J{\"o}rn and Drescher, Knut},
  journal={Nature Phys},
  volume={15},
  number={3},
  pages={251--256},
  year={2019},
  publisher={Nature Publishing Group UK London}
}

@article{sudarsan2016simulating,
  title={Simulating biofilm deformation and detachment with the immersed boundary method},
  author={Sudarsan, Rangarajan and Ghosh, Sudeshna and Stockie, John M and Eberl, Hermann J},
  journal={Commun Comput Phys},
  volume={19},
  number={3},
  pages={682--732},
  year={2016},
  publisher={Cambridge University Press}
}

@article{cleaver1996extension,
  title={Extension and generalization of the Gay-Berne potential},
  author={Cleaver, Douglas J and Care, Christopher M and Allen, Michael P and Neal, Maureen P},
  journal={Phys Rev E},
  volume={54},
  number={1},
  pages={559},
  year={1996},
  publisher={APS}
}

@article{maramizonouz2022drag,
  title={Drag force acting on ellipsoidal particles with different shape characteristics},
  author={Maramizonouz, Sadaf and Nadimi, Sadegh},
  journal={Powder Technol},
  volume={412},
  pages={117964},
  year={2022},
  publisher={Elsevier}
}

@article{NUNEZ2023,
title = {A comprehensive comparison of biofilm formation and capsule production for bacterial survival on hospital surfaces},
journal = {Biofilm},
volume = {5},
pages = {100105},
year = {2023},
issn = {2590-2075},
doi = {https://doi.org/10.1016/j.bioflm.2023.100105},
url = {https://www.sciencedirect.com/science/article/pii/S2590207523000023},
author = {Charles Nunez and Xenia Kostoulias and Anton Y. Peleg and Francesca Short and Yue Qu},
}

@article{Stewart17,
author = {Elizabeth J. Stewart and David E. Payne and Tianhui Maria Ma and J. Scott VanEpps and Blaise R. Boles and John G. Younger and Michael J. Solomon},
title = {Effect of Antimicrobial and Physical Treatments on Growth of Multispecies Staphylococcal Biofilms},
journal = {Applied and Environmental Microbiology},
volume = {83},
number = {12},
pages = {e03483-16},
year = {2017},
doi = {10.1128/AEM.03483-16},
}

@article{Chambless,
author = {Chambless, Jason D. and Stewart, Philip S.},
title = {A three-dimensional computer model analysis of three hypothetical biofilm detachment mechanisms},
journal = {Biotechnol Bioeng},
volume = {97},
number = {6},
pages = {1573-1584},
doi = {10.1002/bit.21363},
year = {2007},
}

@article{Clegg17,
author={Clegg, RJ and Kreft, JU}, 
title={Reducing discrepancies between 3D and 2D simulations due to cell packing density}, 
journal={J Theor Biol},
year=2017, 
volume=21,
number=423,
pages="26--30", 
doi={10.1016/j.jtbi.2017.04.016},
}

@article{Stewart13,
author = {Stewart, Elizabeth J. and Satorius, Ashley E. and Younger, John G. and Solomon, Michael J.},
title = {Role of Environmental and Antibiotic Stress on Staphylococcus epidermidis Biofilm Microstructure},
journal = {Langmuir},
volume = {29},
number = {23},
pages = {7017--7024},
year = {2013},
doi = {10.1021/la401322k},
}

@article{pechaud2024modelling,
  title={Modelling biofilm development: The importance of considering the link between EPS distribution, detachment mechanisms and physical properties},
  author={Pechaud, Y and Derlon, N and Queinnec, Isabelle and Bessiere, Y and Paul, E},
  journal={Water Res},
  volume={250},
  pages={120985},
  year={2024},
  publisher={Elsevier}
}

@article{packard2026biophysical,
	author = {Packard, Sydney R. and Bulacan, Gabriel and Peiris, Buddika and Paffenroth, Randy and Stewart, Elizabeth J.},
	title = {Biophysical properties and phenotypes of cell clusters detached from Staphylococcus epidermidis biofilms after matrix-targeted disruption},
	elocation-id = {2026.01.28.701379},
	year = {2026},
	doi = {10.64898/2026.01.28.701379},
	publisher = {Cold Spring Harbor Laboratory},
	URL = {https://www.biorxiv.org/content/early/2026/01/28/2026.01.28.701379},
	eprint = {https://www.biorxiv.org/content/early/2026/01/28/2026.01.28.701379.full.pdf},
	journal = {bioRxiv}
}

@article{Xavier05,
author={Xavier, JdeB and Picioreanu, C and van Loosdrecht, MC}, 
title={A general description of detachment for multidimensional modelling of biofilms}, journal={Biotechnol Bioeng}, 
year=2005, 
volume=91,
number=6,
pages="651--669",
}

@article{MARAMIZONOUZ2022117964,
title = {Drag force acting on ellipsoidal particles with different shape characteristics},
journal = {Powder Technol},
volume = {412},
pages = {117964},
year = {2022},
issn = {0032-5910},
doi = {10.1016/j.powtec.2022.117964},
author = {Sadaf Maramizonouz and Sadegh Nadimi},
}

@article{Kaplan,
author={Kaplan, JB}, 
title={Biofilm dispersal: mechanisms, clinical implications, and potential therapeutic uses}, 
journal={J Dent Res}, 
year=2010, 
volume=89,
number=3,
pages="205--18", 
doi={10.1177/0022034509359403},
}

@article{Vahidkhah15,
author ={Vahidkhah, K and Bagchi, P},
title  ={Microparticle shape effects on margination, near-wall dynamics and adhesion in a three-dimensional simulation of red blood cell suspension},
journal  ={Soft Matter},
year  ={2015},
volume  ={11},
issue  ={11},
pages  ={2097-2109},
}

@article{Petrova,
title = {Escaping the biofilm in more than one way: desorption, detachment or dispersion},
journal = {Current Opinion in Microbiology},
volume = {30},
pages = {67-78},
year = {2016},
doi = {10.1016/j.mib.2016.01.004},
author = {Olga E Petrova and Karin Sauer},
}

@article{Jonblat2024,
  author = {Jonblat, S. and As-Sadi, F. and Zibara, K. and Sabban, M.E. and Dermesrobian, V. and Khoury, A.E. and Kallassy, M. and Chokr, A.},
  title = {Staphylococcus epidermidis biofilm assembly and self-dispersion: bacteria and matrix dynamics},
  journal = {Int Microbiol},
  year = {2024},
  volume = {27},
  number = {3},
  pages = {831-844},
  month = {June},
  doi = {10.1007/s10123-023-00433-2},
  pmid = {37824024},
  note = {Epub 2023 Oct 12}
}

@article{Erskine18,
 author={Erskine, E. and MacPhee, C.E. and Stanley-Wall, N.R.}, 
title={Functional amyloid and other protein fibers in the biofilm matrix}, 
journal={J Mol Biol}, 
volume=430, 
pages="3642--3656",
year=2018,
}

@article{Panlilio,
author={Panlilio, H. and Rice, C.V.}, 
title={The role of extracellular DNA in the formation, architecture, stability, and treatment of bacterial biofilms}, 
journal={Biotechnol Bioeng},
volume=118, 
pages="2129--2141", 
year=2021,
}

@article{Okshevsky,
author={Okshevsky, M. and Meyer, R. L.}, 
title={The role of extracellular DNA in the establishment, maintenance and perpetuation of bacterial biofilms}, 
journal={Crit Rev Microbiol}, 
volume={41}, 
pages={341--352}, 
year={2015},
}

@misc{dbscan,
year = {2025},
title = {Density-based spatial clustering of applications with noise (DBSCAN) (R2025a)},
note = {The MathWorks Inc.},
address = {Natick, Massachusetts, United States},
url = {https://www.mathworks.com}
}

@misc{alphashape,
year = {2025},
title = {Polygons and polyhedra from points in 2-D and 3-D (alphaShape) (R2025a)},
note = {The MathWorks Inc.;
Natick, Massachusetts, United States},
url = {https://www.mathworks.com}
}

@misc{grabit,
author={Doke, J},
year=2016,
title={GRABIT},
url={https://www.mathworks.com/matlabcentral/fileexchange/7173-grabit}, 
note={MATLAB Central File Exchange}, 
}

@misc{rng,
year = {2025},
note = {The MathWorks Inc.; Natick, Massachusetts, United States},
title = {Control random number generator (rng)
 (R2025a)},
url = {https://www.mathworks.com}
}

@article{Alden13,
author={Alden, K and Read, M and Timmis, J and Andrews, PS and Veiga-Fernandes, H and Coles, M},
year=2013,
title={Spartan: a comprehensive tool for understanding uncertainty in simulations of biological systems},
journal={PLoS Comput Biol},
volume=9,
number=2,
pages="e1002916",
doi={10.1371/journal.pcbi.1002916},
}

@article{Cerca2012,
  author    = {Cerca, N. and Gomes, F. and Pereira, S. and others},
  title     = {Confocal laser scanning microscopy analysis of \textit{S. epidermidis} biofilms exposed to farnesol, vancomycin and rifampicin},
  journal   = {BMC Res Notes},
  volume    = {5},
  pages     = {244},
  year      = {2012}
}

@article{Mattei18,
author={Mattei, MR and Frunzo, L and D’Acunto, B and Pechaud, Y and Pirozzi, F and Esposito, G}, title={Continuum and discrete approach in modeling biofilm development and structure: a review}, 
journal={J Math Biol},
volume=76, 
pages="945--1003", 
year=2018,
}

@article{Klapper,
author = {Klapper, Isaac and Dockery, Jack},
title = {Mathematical Description of Microbial Biofilms},
journal = {SIAM Rev},
volume = {52},
number = {2},
pages = {221-265},
year = {2010},
doi = {10.1137/080739720},
}

@article{WANG10,
title = {Review of mathematical models for biofilms},
journal = {Solid State Commun},
volume = {150},
number = {21},
pages = {1009-1022},
year = {2010},
author = {Wang, Q and Zhang, T},
}

@article{Wang22,
author = {Shuai Wang and Huiyan Zhu and Gexi Zheng and Feng Dong and Chongxuan Liu},
title = {Dynamic Changes in Biofilm Structures under Dynamic Flow Conditions},
journal = {Appl Environ Microbiol},
volume = {88},
number = {22},
pages = {e01072-22},
year = {2022},
doi = {10.1128/aem.01072-22},
}

@article{Jo22,
author={Jo, J and Price-Whelan, A and Dietrich, LEP}, 
title={Gradients and consequences of heterogeneity in biofilms}, 
journal={Nat Rev Microbiol},
year=2022, 
volume=20,
pages="593--607", 
doi={10.1038/s41579-022-00692-2},
}

@article{Otto2009,
  author    = {Otto, M.},
  title     = {\textit{Staphylococcus epidermidis} — the `accidental' pathogen},
  journal   = {Nat Rev Microbiol},
  year      = {2009},
  volume    = {7},
  pages     = {555--567}
}

@article{WILLE20,
title = {Biofilm dispersion: The key to biofilm eradication or opening Pandora’s box?},
journal = {Biofilm},
volume = {2},
pages = {100027},
year = {2020},
issn = {2590-2075},
doi = {10.1016/j.bioflm.2020.100027},
author = {Jasper Wille and Tom Coenye},
}

@article{Fleming18,
author={Fleming, D and Rumbaugh, K}, 
title={The Consequences of Biofilm Dispersal on the Host}, 
journal={Sci Rep},
volume={8}, 
pages={10738},
year={2018},
doi={doi.org/10.1038/s41598-018-29121-2},
}

@article{Fleming01,
author={Flemming, HC and Wingender, J}, 
title={Relevance of microbial extracellular polymeric substances ({EPS}s)--{P}art {I}: {S}tructural and ecological aspects},
journal={Water Sci Technol}, 
year=2001,
volume=43,
number=6,
pages="1-8",
}

@article{Assefa,
author={Assefa, M. and Amare, A}, 
title={Biofilm-Associated Multi-Drug Resistance in Hospital-Acquired Infections: A Review},
journal={Infect Drug Resis}, 
year=2020,
volume=15,
pages="5061-5068",
}

@article{Wang23,
author={Wang, S. and Zhao, Y. and Breslawec, A.P. and Liang, T. and Deng, Z. and Kuperman, L.L. and Yu, Q.},
title={Strategy to combat biofilms: a focus on biofilm dispersal enzymes},
journal={NPJ Biofilms Microbiomes},
volume=9, 
number=63, 
year={2023}, 
doi={10.1038/s41522-023-00427-y},
}

@article{Schilcher2020,
  author    = {Schilcher, K. and Horswill, A. R.},
  title     = {Staphylococcal Biofilm Development: Structure, Regulation, and Treatment Strategies},
  journal   = {Microbiol Mol Biol Rev},
  year      = {2020},
  volume    = {84},
  number    = {e00026-19}
}
\section*{Appendix}

\setcounter{equation}{0}
\setcounter{figure}{0}
\setcounter{table}{0}
\setcounter{section}{0}
\appendix
\renewcommand{\thesection}{S}
\renewcommand{\theequation}{A\arabic{equation}}
\renewcommand{\thefigure}{A\arabic{figure}}
\renewcommand{\thetable}{A\arabic{table}}

A detailed list of all parameters utilized in iDynoMiCS are given in Tables \ref{tab:iDynomics_parameters} and \ref{tab:iDynomics_parameters_2}. Several parameters were updated to be representative of an \textit{S. epidermidis} biofilm.
\addcontentsline{toc}{section}{Appendix}
\begin{table}[h]
\renewcommand{\arraystretch}{1.15}
\begin{tabular}{p{2.5cm}p{5.1cm}p{3.8cm}}
\hline\noalign{\smallskip}
Parameter & Description & Value  \\
\noalign{\smallskip}\svhline\noalign{\smallskip}
        \texttt{endOfSimulation}  & Duration of biofilm growth (days)  & 1 (EXP)\\
        \texttt{timeStepIni} & Baseline or initial time step for solutes (h) & 0.1 \\
        \texttt{timeStepMin}  & Minimum adaptive time step (h) & 0.05 \\
        \texttt{timeStepMax} & Maximum adaptive time step (h) & 0.2 \\
        \texttt{agentTimeStep} & Fixed agent time step (h) & 0.05 \\
        \hline 
        \texttt{x,y,z} & Dimensions of the domain ($\SI{}{\micro\meter}$)& 66$\times$2$\times$33 for 2-d; 66$\times$66$\times$33 for 3-d  (EXP) \\
        \texttt{resolution}  & For agent grid ($\SI{}{\micro\meter}$) & 1 (EXP) \\
         & For solute grid ($\SI{}{\micro\meter}$)& 2  \\
         \texttt{boundaryLayer} & Area between biofilm and bulk ($\SI{}{\micro\meter}$)  & 13 (EXP) \\
         \texttt{maxTh} & Maximum biofilm height allowed ($\SI{}{\micro\meter}$) & 20 \\
         \texttt{specificArea} & Ratio between substratum and bulk compartment volume (m$^2$/m$^3$) & 0.015 (EXP) \\
         \hline  \noalign{\smallskip}
        $N_B^0$  & Initial number of $B$ & 10 in 2-d; 160 in 3-d (EXP)\\
        $N_E^0$  & Initial number of $E$ & 0 in 2-d and  3-d \\ \noalign{\smallskip} \hline 
        \texttt{shovingMaxNodes} & Maximum number of agents to apply shoving algorithm & 2$\times10^6$ \\
        \texttt{shovingFraction} & Equilibrium accepted when moving/shoved agents below this threshold & 0.025 \\
        \texttt{shovingMaxIter} & Maximum number of iterations to reach final position & 250 \\
        \texttt{shovingMutual} & Shoving applied to all overlapping agents  & true \\
        \texttt{shovingMaxNodes} & Maximum number of agents to apply shoving algorithm & 2$\times10^6$ \\ \hline
        \texttt{divRadius\_E} & Radius at which $E$ divides & 2 $\SI{}{\micro\meter}$ \\
        \texttt{divRadiusCV\_E} & Stochasticity for $E$ division & 0.05 \hspace{3cm} \texttt{divRadius\_E}$\in[1.95,2.05]$ \\
        \texttt{babyMassFrac\_E} & Fraction of $E$ mass to each $E$ agent when dividing & 0.5 \\
        \texttt{babyMassFracCV\_E} & Stochasticity for mass in $E$ division & 0.1 \hspace{3cm} \texttt{babyMassFrac\_E}$\in[0.4,0.6]$ \\
        \texttt{shoveFactor\_E} & Multiple of $E$ radius for neighborhood in shoving algorithm & 1 \\
        \texttt{shoveLimit\_E} & If overlap of agents with $E$ is greater than this distance, shoving will occur &  0 $\SI{}{\micro\meter}$ \\ \hline
        \texttt{divRadius\_B} & Radius at which $B$ divides & 0.75 $\SI{}{\micro\meter}$ \\
        \texttt{divRadiusCV\_B} & Degree of stochasticity for $B$ division & 0.1 \hspace{3cm} \texttt{divRadius\_B}$\in[0.65,0.85]$ \\
        \texttt{babyMassFrac\_B} & Fraction of $B$ mass to each $B$ when dividing & 0.5 \\
        \texttt{babyMassFracCV\_B} & Stochasticity for mass in $B$ division & 0.05 \hspace{3cm} \texttt{babyMassFrac\_B}$\in[0.45,0.55]$ \\
        \texttt{shoveFactor\_B} & Multiple of $B$ radius for neighborhood in shoving algorithm & 0.75 \\
        \texttt{shoveLimit\_B} & If overlap of agents with $B$ is greater than this distance, shoving will occur &  0 $\SI{}{\micro\meter}$ \\ 
        \texttt{epsMax} & Maximum volume fraction of $B$ that can be taken up by the EPS capsule before EPS excretion will occur & 0.1 \\ 
\noalign{\smallskip}\hline\noalign{\smallskip}
\end{tabular}
    \caption{iDynoMiCS model parameters to create a 24-hour-old \textit{S. epidermidis} biofilm.  Parameter  baseline values were taken from the iDynoMiCS tutorial  \cite{Lardon11} and EXP refers to those parameters modified to have biofilm growth representative of several experimental studies \cite{Jonblat2024,packard2026biophysical,Stewart13}.  Here, $B$ is a representative bacteria agent and $E$ is a representative EPS agent. 
    }
    \label{tab:iDynomics_parameters}
\end{table}

\begin{table}[h]
\renewcommand{\arraystretch}{1.15}
\begin{tabular}{p{3cm}p{5.1cm}p{3.2cm}}
\hline\noalign{\smallskip}
Parameter & Description & Value  \\
\noalign{\smallskip}\svhline\noalign{\smallskip}
 \texttt{density} & How tight biomass is packed into a region (g/L) & 150 for $B$ (cell and capsule); 75 for $E$ \\ \hline
 \texttt{diffusivity} & Rate of movement in water (m$^2$/day) & 2$\times 10^{-4}$ for o2d,  1.7$\times 10^{-4}$ for NO3, and 1$\times 10^{-4}$ for COD 
 \\
       \texttt{biofilmDiffusivity} & Multiplicative factor to diffusivity in water & 0.8 \\
      \texttt{Sbulk} & Solute concentration in bulk (g/L) & 1$\times 10^{-2}$ for o2d, 0 for NO3, and 1$\times 10^{-2}$ for COD  
      \\ \hline
      
       \texttt{muMaxAer} & Maximum reaction rate utilized in Monod kinetics for aerobic growth of biomass in $B$ (1/h) & 0.25 for o2d and COD \\
       \texttt{KsAer} & Half-maximum solute concentration utilized in Monod kinetics for aerobic growth of biomass in $B$ & 4$\times 10^{-3}$ for COD, 0.2$\times 10^{-3}$ for o2d \\ 
       \texttt{yieldAer} & From the stoichiometric equation, factor that ensures mass balance when utilizing solutes to aerobically create biomass in $B$ (unitless) & $-0.5873$ for o2d, $-1.5873$ for COD, 1.5 for $B$ biomass (cell) and 0.2 for capsule biomass around $B$\\ \hline
       \texttt{muMaxAnaer} & Maximum reaction rate utilized in Monod kinetics for anaerobic growth of biomass in $B$ (1/h) & 0.2 for COD and NO3 \\
       \texttt{KsAnaer} & Half-maximum solute concentration utilized in Monod kinetics for anaerobic growth of biomass in $B$ & 4$\times 10^{-3}$ for COD, 0.5$\times 10^{-3}$ for NO3 \\ 
       \texttt{yieldAnaer} & From the stoichiometric equation, factor that ensures mass balance when utilizing solutes to anaerobically create biomass in $B$ (unitless) & $-1.5873$ for COD, $-0.1285$ for NO3, 0.8 for $B$ biomass (cell) and 0.2 for capsule biomass around $B$\\ 
       \hline 
     
       \texttt{yieldE} & From the stoichiometric equation, factor that ensures mass balance when losing  biomass in capsule (unitless) & $-1$ for biomass capsule and $-1$ for COD \\ 
       
       \texttt{kHydE} & Rate at which capsular EPS is excreted to neighboring $E$ (1/h) & 0.007 \\
       
\noalign{\smallskip}\hline\noalign{\smallskip}
\end{tabular}
    \caption{iDynoMiCS model parameters to create a 24-hour-old \textit{S. epidermidis} biofilm.  Parameter  baseline values were taken from the iDynoMiCS tutorial  \cite{Lardon11} and an EXP refers to those parameters modified to have biofilm growth representative of several experimental studies \cite{Jonblat2024,packard2026biophysical,Stewart13}.  Here, $B$ is a representative bacteria agent and $E$ is a representative EPS agent. COD is organic carbon or chemical oxygen demand, 
    o2d is oxygen, NO3 is nitrate, and Aer is aerobic growth. 
    }
    \label{tab:iDynomics_parameters_2}
\end{table}

\setcounter{equation}{0}
\setcounter{figure}{0}
\setcounter{table}{0}
\setcounter{section}{0}
\appendix
\renewcommand{\thesection}{S}
\renewcommand{\theequation}{A\arabic{equation}}
\renewcommand{\thefigure}{A\arabic{figure}}
\renewcommand{\thetable}{A\arabic{table}}

\section*{Supplementary Files Information}

\setcounter{subsection}{0}
Videos are available in a Supplementary Videos folder in this \href{https://drive.google.com/drive/folders/19o7uWDcUBHtMvN2crrNxq-4I8LeCFqJi?usp=share_link}{google drive folder}. 

\subsection{Video of 3-d view of the clustered biofilm  \label{supp:vid_1}}

The visualization provides additional geometric insight into 3-d cluster formation that are not apparent from fixed-angle renderings in Fig. \ref{fig:3d_cluster_after}.

\subsection{Video of 2-d biofilm break off under different disruptor ratios}\label{vid:2dprocess}
This video shows three simulated detachment processes over 60 minutes for disruptor ratios of 0\%, 20\%, and 65\%. Detached clusters and single agents are tagged using the same legend as in Fig.~\ref{fig:TimeStep}: orange for detached bacterial agents, red for detached EPS agents, and red ellipses for detached clusters. The top video (disruptor ratio 0) exhibits the slowest detachment dynamics, with break off occurring approximately every 3--5 minutes due to stochastic fluctuations in the model. This behavior is consistent with experimental observations in which detachment occurs continuously over time  \cite{packard2026biophysical}. Across all three simulations, higher disruptor ratios result in faster break off and yield a lower final biofilm height.

\subsection{Video of 3-d biofilm break off under different disruptor ratios}\label{vid:3dprocess}
This video presents the corresponding 3-d simulations of biofilm detachment under the same conditions as Video~\ref{vid:2dprocess}, with disruptor ratios of 0, 20\%, and 65\%. An example of a 3-d biofilm snapshot can be found in Fig. \ref{fig:3d_cluster_after}. The visualization tracks biofilm morphology and detachment dynamics throughout a 60-minute simulation. As in the 2-d case, increasing the disruptor ratio accelerates break off events and results in a lower final biofilm height. Compared with the lower-ratio simulations, the 65\% disruptor condition produces markedly larger and more cohesive detached clusters. This reflects increased structural weakening of the biofilm matrix at higher disruptor concentrations, leading to more extensive fragmentation and thicker cluster-scale break off events.

\end{document}